\documentclass[aps,12pt,a4paper]{revtex4-2}
\usepackage{epsfig}
\usepackage{graphicx}
\usepackage{amsmath,amssymb,color}
\usepackage[english]{babel}

\parskip=\medskipamount

\newcommand{\eq}[1]{(\ref{#1})}
\newcommand{\fig}[1]{Fig.~\ref{#1}}

\newcommand{\be}{\begin{equation}}
\newcommand{\ee}{\end{equation}}

\newcommand\disp{\displaystyle}

\newcommand{\la}{\left<}
\newcommand{\ra}{\right>}
\newcommand{\eps}{\varepsilon}

\newcommand{\re}{\textrm{Re}\,}
\newcommand{\im}{\textrm{Im}\,}

\begin{document}

\title{Generalized Devil's staircase and RG flows}

\author{A. Flack$^{1}$, A. Gorsky$^{2}$, and S. Nechaev$^{1,3}$}

\affiliation{$^{1}$LPTMS, CNRS--Universit\'e Paris-Saclay, 91405 Orsay Cedex, France, \\ $^{2}$Institute for Information Transmission Problems RAS, 127051 Moscow, Russia, \\ $^{3}$Laboratory of Complex Networks, Center for Neurophysics and Neuromorphyc Technologies, Moscow, Russia}

\date{\today}

\begin{abstract}

We discuss a two-parameter renormalization group (RG) flow when parameters are organized in a single complex variable, $\tau$, with modular properties. Throughout the work we consider a special limit when the imaginary part of $\tau$ characterizing the disorder strength tends to zero. We argue that generalized Riemann-Thomae (gRT) function and the corresponding generalized Devil's staircase emerge naturally in a variety of physical models providing a universal behavior. In 1D we study the Anderson-like probe hopping in a weakly disordered lattice, recognize the origin of the gRT function in the spectral density of the probe and formulate specific RG procedure which gets mapped onto the discrete flow in the fundamental domain of the modular group $SL(2,Z)$. In 2D we consider the generalization of the phyllotaxis crystal model proposed by L. Levitov and suggest the explicit form of the effective potential for the probe particle propagating in the symmetric and asymmetric 2D lattice of defects. Analyzing the structure of RG flow equations in the vicinity of saddle points we claim emergence of BKT-like transitions at $\im\tau\to 0$. We show that the RG-like dynamics in the fundamental domain of $SL(2,Z)$ for asymmetric lattices asymptotically approaches the ``Silver ratio''. For a Hubbard model of particles on a ring interacting via long-ranged potentials we investigate the dependence of the ground state energy on the potential and demonstrate by combining numerical and analytical tools the emergence of the generalized Devil's staircase. Also we conjecture a bridge between a Hubbard model and a phyllotaxis. 

\end{abstract}

\maketitle

\tableofcontents

\section{Introduction}

The Devil's staircase is the fingerprint of the ``incommensurability phenomena'' in a variety of physical systems \cite{aubry1983devil, bak1982commensurate}. The geometric signature of the incommensurability is the Riemann-Thomae (RT) function which often emerges in spectra of sparse systems of various physical origin.  Meanwhile, the Riemann-Thomae function also appears in a plethora of fundamental problems, such as stability diagram in fractional quantum Hall effect \cite{Tao1, Tao2}, interactions of non-relativistic ideal anyons with rational statistics in the ``magnetic gauge'' approach \cite{lundholm}, quantum $1/f$ noise and Frenel-Landau shift \cite{planat}, distribution of quotients of reads in DNA sequencing experiment \cite{dna}, frequency of specific subgraphs counting in the protein-protein network of a \textit{Drosophilla} \cite{drosophilla}. Though the degree of similarity with the original RT function could vary, and experimental profiles may drastically depend on the peculiarities of each particular physical system, a general probabilistic scheme resulting in emergence of the fractal hierarchical distribution can be considered as the manifestation of number-theoretic laws in nature.

One possible pattern behind the Riemann-Thomae function and the Devil's staircase is as follows. Consider a physical problem, for example the fractional quantum Hall effect (FQHE), and push the system into the particular limit in the parameter space. For FQHE this is the so-called ``thin torus limit'' -- see for example \cite{Tao2}. The system hosts some defects, and in the limit under consideration defects form a lattice which is a Wigner crystal in the thin torus limit of FQHE. Consider now the propagation of a probe particle through the sample which can be studied, for instance, by analysing the spectral density. The modular $SL(2,Z)$ group acts in the parameter space of this system. The imaginary part $\im \tau$ of the modular parameter $\tau$ gets identified with some function of disorder, while the real part $\re \tau$ corresponds to the chemical potential for the topological charge relevant for the studied problem. The motion of the probe particle in the crystal of defects can be mapped onto the motion in the fundamental domain of $SL(2,Z)$, and the rearrangements of the lattice can be treated by analyzing the RG flow in the vicinity of transition points which are identified with points of lattice bifurcations. Generally speaking, from the probe particle perspective, the rearrangement of the lattice can be studied by varying the chemical potential of defects (or of their number).

Another view on the Devil's staircase deals with a general classification of quantum systems spectra which usually involve discrete, continuous, as well as more tricky singular-continuous supports. Recent discussion of the latter case can be found in \cite{altshuler2023random} and it incorporates the Devil's staircase as an intrinsic ingredient. The mechanism behind the appearance of the Devil's staircase in the spectrum can be illustrated in the context of the Peierls model \cite{dzyaloshinskij1982commensurability} of electrons interacting with the lattice of ions. Starting with the integrable version of the Peierls model described by the Toda lattice, authors of \cite{brazovskii1982exactly} have shown that the spectral curve of the Toda system can be identified with the dispersion law of fermions. At particular values of Toda integrals of motion there very tiny bands in the fermion spectrum emerge and if one adds an arbitrary small non-integrable perturbation, the Devil's staircase structure gets formed nearby these bands at rational fermionic densities indicating the emergence of incommensurability transitions. In our work the similar viewpoint is used for studying the spectral statistics of a probe particle propagating along a weakly disordered crystal, which we interpret as a two-parametric renormalization group (RG) flows in a particular limit (as it is explained below).

The general classification of RG flows rhymes with the development of bifurcations (``catastrophes'') over time in the theory of dynamical systems -- see, for instance \cite{gukov2017rg}. In the catastrophe theory there are focuses, saddles, limits cycles and other attributes of the singularity theory, with corresponding fixed points, RG cycles and more exotic RG behavior. For instance, recently the RG counterparts of homoclinic orbits in the theory of dynamical systems have been found in the field theory \cite{jepsen2021homoclinic}, they also provide examples of chaotic RG flows \cite{bosschaert2022chaotic}. The incommensurability phenomenon is also known in the theory of dynamical systems. Hence, following the same logic, one could expect the existence of RG counterpart of the incommensurability. Indeed, the RG approach was successful in describing the Devil's staircase pattern in a Harper equation for the electron in a crystal in presence of a magnetic field \cite{wilkinson1984critical, wilkinson1987exact} where it was argued that the tunneling in the phase space is the crucial ingredient.

In many situations it is convenient to combine two real parameters of a 2D RG flow into the single complex parameter, $\tau$, which can be interpreted as the modulus of the complex structure for an auxiliary elliptic curve. The familiar examples are: the Anderson localization problem with the time symmetry breaking (TSB) term \cite{Altland2015topology}, the integer quantum Hall effect (IQHE) \cite{pruisken1984localization, levine1984theory}, and the Yang-Mills theory with the TSB $\theta$-term \cite{montonen1977magnetic, cardy1982phase}. In all these examples the real part of the complex parameter is the TSB parameter. We suggest a bit more general perspective and propose to consider the following generic complex (modular) parameter:
\be
\tau= {\rm [topological ~ term]} + i\,{\rm [disorder]},
\label{eq:tau}
\ee
hence the RG flow unites the topology and the disorder. Let us provide some known examples supporting this perspective: 
\begin{enumerate}
\item[(i)] In the integer quantum Hall effect (IQHE) the complex parameter $\tau$ is built of two conductivities, $\sigma_{xx}$ and $\sigma_{xy}$ \cite{pruisken1984localization}:
\be
\tau = \sigma_{xy} + i\sigma_{xx};
\ee
\item[(ii)] In gauge (Yang-Mills) theories the modular parameter $\tau$ involves the coupling constant, $g_{YM}$, and the $\theta$-term in the following combination \cite{montonen1977magnetic, cardy1982phase}:
\be
\tau = \frac{\theta}{2\pi} + \frac{4i\pi}{g_{YM}^2};
\ee
\item[(iii)] In the case of three-diagonal random matrices with the off-diagonal disorder the parameter $\tau$ enters in the combination (see \cite{krapiv,polov} and Section III for detail):
\be
\tau = \epsilon + i f
\ee
where $\epsilon$ is the function of the spectral parameter, $\lambda$, and $f$ depends on the strength of the disorder; 
\item[(iv)] In the Anderson model with the TSB term the parameter $\tau$ reads \cite{Altland2015topology}:
\be
\tau = \theta + iD
\ee
where $\theta$ counts windings and $D$ is the diffusion coefficient;
\item[(v)] For the polymer propagating in the lattice of obstacles, the entanglement complexity is fully characterized by the parameter $\tau$ known as a ''primitive path configuration'' (see for detail \cite{primitive1, nech1, nechaev-houches}:
\be
\tau = {\rm (chemical ~ potential ~ of ~ winding)} + i\,{\rm (complexity ~ of ~ entanglement)}.
\ee
where the ``primitive path'' was defined in \cite{primitive1, primitive2} and has the meaning of a geodesic path on some hyperbolic manifold \cite{nechaev-houches, nech-UFN}.
\end{enumerate}

At any fixed value $\tau=\chi+i\xi$ the partition functions of considered systems fully enjoy symmetries of the $SL(2,Z)$ modular group and hence are the modular functions. However when $\chi$ and/or $\xi$ run over time and depend on a scale, $\mu$, the situation is more subtle. It general, the RG flow involves two $\beta$-functions and is described by the set of equations
\be
\begin{cases}
\disp \frac{d\chi}{d \ln\mu}= \beta_{\chi}(\chi,\xi) \medskip \\
\disp \frac{d\xi}{d \ln\mu}= \beta_{\xi}(\chi,\xi)
\label{eq:beta}
\end{cases}
\ee
Typically, the disorder term enjoys both the perturbative and non-perturbative remormalizations, while the topological parameter is renormalized only non-perturbatively. 

There are some known patterns of $\beta$-functions with familiar properties: 
\begin{itemize}
\item For the Integer Quantum Hall Effect (IQHE) case the $\beta$-function can be expressed via the Grassmanian $\frac{U(n+m)}{U(n)\times U(m)}$ $\sigma$-model:
\be
\begin{cases}
\disp \beta_{\sigma_{xx}} = \beta^0_{xx} - C_{m,n}\sigma_{xx}^{m+n+2} e^{-2\pi\sigma_{xx}} \cos (2\pi\sigma_{xy}) \\ \medskip
\disp \beta_{\sigma_{xy}}= -D_{m,n}\sigma_{xx}^{m+n+2} e^{-2\pi\sigma_{xx}} \cos (2\pi\sigma_{xy}) \\ \medskip
\disp  \beta^0_{\sigma_{xx}}= -\frac{m+n}{2\pi} - \frac{mn+1}{2}\sigma_{xx}^{-1}
\end{cases}
\ee
where $D_{m,n}$ and $C_{m,n}$ are not completely universal. 
\item For the Russian Doll model which is the toy example of the system with the cyclic RG flows (see, \cite{bulycheva2014limit} for review), the RG flow is discrete
\be
\begin{cases}
\disp g_{N+1}= g_N + \frac{g_N^2 +\theta_N^2}{N} \\ \medskip
\disp \theta_{N+1}=\theta_N
\label{eq:beta1}
\end{cases}
\ee
\item For the Berezinskii-Kosterlitz-Thouless (BKT) transition one has: 
\be
\begin{cases}
\disp \beta_u= -c_1 uv \\
\disp \beta_v=-c_2 u^2
\label{eq:beta2}
\end{cases}
\ee
\end{itemize}

We focus our attention on a specific limit of RG flows when the non-perturbative renormalization coming from instanton-like contributions dominates -- see, for example, \cite{wilkinson1987exact}. This happens in all examples when $\xi=\im{\tau}\to 0$, which means that we are looking at the limit of a weak disorder in some frame, and the modular parameter is mainly governed by the ``$\theta$'' (i.e. winding-like) terms. The details are model-dependent, however in all cases the $\theta$-term has one and the same physical sense: it serves for counting topological defects. The weak disorder limit ($\xi \to 0$) in some cases could mean the strong coupling. For example, in the Yang-Mills theory, since $\xi=\frac{4\pi}{g_{YM}^2}$, in the limit $\xi\to 0$ the theory is strongly coupled in the ``magnetic frame'', while in the ``electric frame'' the theory is weakly coupled. We are searching for some universality in the $\im{\tau}\to 0$ regime. Throughout the work we argue that gRT function and generalized Devil's staircase emerge naturally in the $\xi\to 0$ limit and are universal.  

In all cases (i)--(v) mentioned above, the $\beta$-functions can be expressed in terms of elliptic functions on some Riemann surface. However, when couplings $\chi$ and $\xi$ run in time, the construction of corresponding Riemann surfaces is a nontrivial issue. The Riemann surface is bundled over some manifold and the RG flow is identified with the dependence of the modular parameter on the point of the fibration base. The benchmark example is provided by the Seiberg-Witten solution for the low-energy effective action of the $N=2$ SYM theory \cite{seiberg1994electric}. The renormalization of both $(\re{\tau}, \im{\tau})$ can be found directly from the partition function of ensemble of instantons which can be evaluated via the localization approach \cite{nekrasov2003seiberg}.

Remarkably, two more ways to handle RG flows are available. The first one can be formulated quite generally. We add the probe object into the ensemble of defects, study its dynamics and derive from the induced dynamics the behavior of $\beta$-functions. This logic works well in the Seiberg-Witten solution when the surface defect inserted into the instanton ensemble plays the role of a probe. The prepotential of the low-energy $N=2$ super Yang-Mills (SYM) theory, $\cal{F}$, yields the exact $\beta$-function \cite{seiberg1994electric}. This $\beta$-function can be derived from the semiclassical wave function of the probe whose dynamics is governed by the integrable system of Calogero-Toda type \cite{gorsky1995integrability, martinec1996integrable, donagi1996supersymmetric}. The derivation of the potential governing the dynamics of the probe from the first principle is not a simple issue and only recently this problem has been fully elucidated \cite{nekrasov2019bps}. In our study we use a similar probe analysis.

The second way to treat RG flows is more sophisticated and involves the nontrivial dynamics rooted in the so-called ``vertex realization'' of the infinite-dimensional algebras \cite{alday2010liouville}. In brief, the partition function of the instanton ensemble gets mapped onto the particular conformal block in the Liouville theory, or its extension to higher spins (the so-called ``$W_n$ theory''). The modular parameter becomes the position of the vertex operator insertion on the sphere. Hence the flow in the modular domain in the asymptotic regime gets reduced to the investigation of a particular asymptotic regime of a conformal block. 
%We shall comment this viewpoint on the example of the phyllotaxis model.

The paper is organized as follows. In Section II we recall the main facts concerning the Riemann-Thomae (RT) function and provide its analytic regularization in terms of the Dedekind $\eta$-function. In Section III we consider the non-perturbative problem of a probe particle propagating in a weakly disordered 1D lattice and pay  attention to the spectral density of the probe whose Hamiltonian is given by the tridiagonal matrix with the off-diagonal bimodal disorder. In the same section we formulate the version of the RG flow in the fundamental domain of the modular group yielding the RT function. In Section IV we consider the propagation of the probe in 2D lattice of defects and investigate the dependence of probe dynamics on the lattice pattern. We show that a lattice with a specific asymmetry yields the Silver ratio as the asymptotic value of the flow in the fundamental domain of the group $SL(2,Z)$ in the weak disorder limit. In the same Section we discuss the RG flows in vicinity of bifurcation (transition) points and conjecture the BKT-like divergence of the correlation length. In Section V we discuss the spin system with long-range interactions exhibiting the Devil's staircase behavior and focus on its dependence on the form of the long-range potential. We find that gRT emerges in considered systems and exponents involved in gRT function provide a kind of universality. Using duality properties between the integrable systems we comment on the similarity between the incommensurability observed in the long-range Hubbard model and in FQHE in the thin torus limit. In Discussion we overview related issues which require more profound analysis. In Conclusion we summarize our findings and formulate open questions. 

\section{Riemann-Thomae function, Euclid Orchard and Devil's staircase}

\subsection{Reminder on the standard Riemann-Thomae function}

\begin{figure}[ht]
\centerline{\includegraphics[width=16cm]{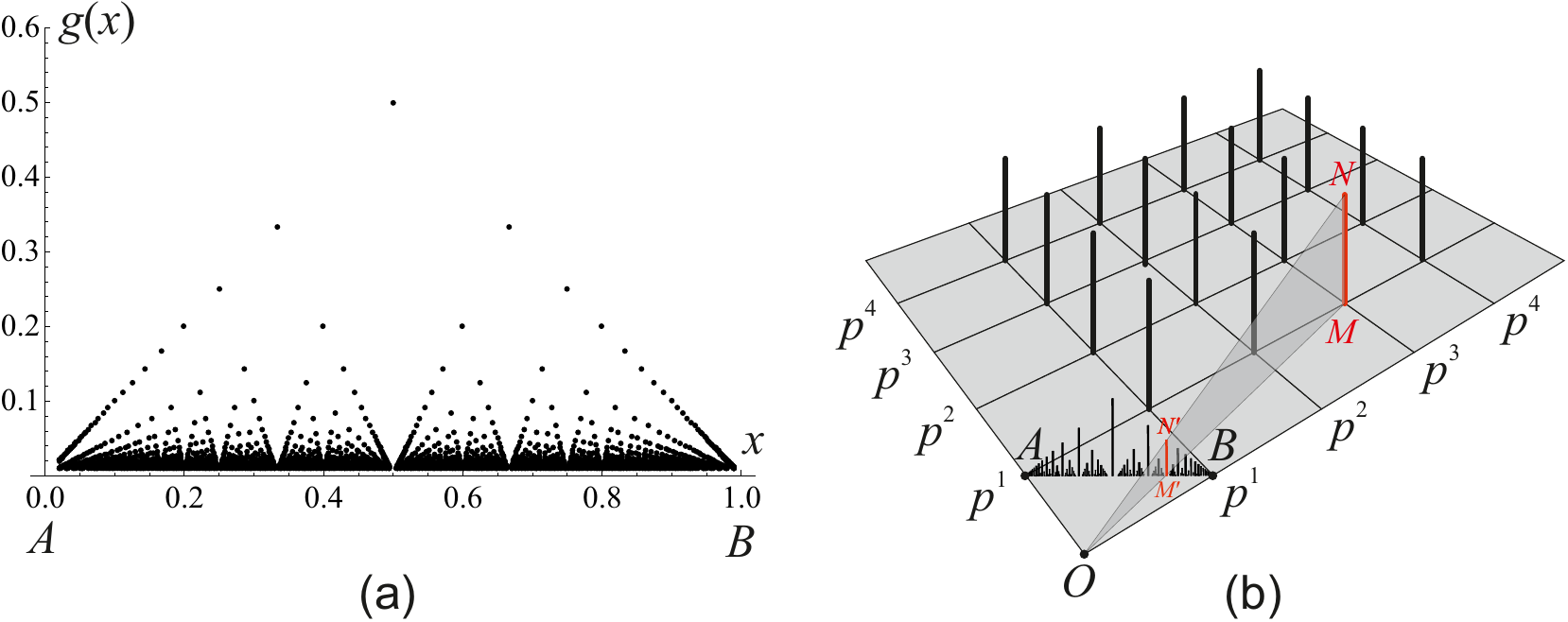}}
\caption{(a) Riemann-Thomae ``raindrop'' function, (b) Riemann-Thomae function constructed by the Euclid Orchard.}
\label{fig:001}
\end{figure}

The Riemann function, $g(x)$, also known as the Thomae function \cite{RT}, has many other names: the popcorn function, the raindrop function, the countable cloud function, the ruler function, the modified Dirichlet function. It is one of the simplest number-theoretic functions possessing a nontrivial fractal structure (another famous example is the everywhere continuous but nowhere differentiable Weierstrass function). The Riemann-Thomae (RT) function is defined in the open interval $x \in (0,1)$ according to the following rule:
\be
g(x) = \begin{cases} \frac{1}{n} & \mbox{if $x=\frac{m}{n}$, and $(m,n)$ coprime} \medskip \\
0 & \mbox{if $x$ is irrational} \end{cases}
\label{eq:001}
\ee
The function $g$ is discontinuous at every rational point: irrationals, where $g$ vanishes, come infinitely close to any rational number. At the same time, $g$ is continuous at irrationals -- see \fig{fig:001}a.

To provide basic properties of the RT function, consider some irrational number, $t$, at which $g(t)=0$, and take some $0 < \eps < 1$. Without the loss of generality, $\eps$ is assumed to be rational, otherwise one may replace $\eps$ with any smaller rational $\eps'=\frac{k}{s}<\eps$ such that $\mathrm{gcd}(k,s) = 1$. Thus, there is a finite set of rational numbers $\Omega_\eps = \{n= \frac{i}{j}, 1<j \le s, 1 \le i < j\}$, that are not smaller than $g(\eps)$. Now assign $\delta(\eps) = \inf\{|n - t|, n \in \Omega_\eps\}$ that defines the vicinity of $t$, in which the values of $g$ are smaller than $\eps$: $|t-y|<\delta(\eps) \to |g(t)-g(y)|=|g(y)|<\eps$. These inequalities prove the continuity of the Riemann function $g(x)$ at any irrational value of $x$.

An elegant representation of the Riemann-Thomae function emerges in a so-called ``Euclid orchard'' construction -- see \fig{fig:001}b \cite{euclid}. Consider an orchard of trees of unit heights located at every point $(am, ak)$ of the two-dimensional square lattice, where $m$ and $k$ are non-negative integers defining the lattice, and $a$ is the lattice spacing, which is convenient to choose as $a = \frac{1}{\sqrt{2}}$. Suppose that the observer stays on the line $m = 1 - k$ between the points $A(0,a)$ and $B(a,0)$, and watches the trees in the first quadrant along the rays emitted from the origin $O(0,0)$. Along these rays, we see only the first (non-shadowed) tree with coprime coordinates, $M(am,ak)$, while all other trees are shadowed. Introduce the rotated coordinate system $(x,y)$ with the axis $0x$ along the segment $AB$ and the axis $0y$ normal to the orchard's plane, as shown in \fig{fig:001}a. We set the origin of the $0x$ axis at the point $A$, then the point $B$ has the coordinate $x = a$. Having the focus located at the origin, the tree $MN$ at the point $M(am,ak)$ is projected to the tree $M'N'$ located at $M'(x,y=0)$, where $x = \frac{m}{k+m}$ and the height $|M'N'|$ of this tree is $\frac{1}{k+m}$ -- see \fig{fig:001}b. Denoting $k+m$ by $n$, we immediately conclude that the ``visibility diagram'' in the Euclid orchard is exactly the Riemann-Thomae function for the variable $x=\frac{m}{n}$.

The RT function arises in the Euclid orchard problem as a purely geometrical object. However, the Riemann-Thomae function also deserves a transparent probabilistic interpretation. Suppose two random integers, $\phi$ and $\psi$, are taken independently from an exponential probability distribution, $Q_n = (1-p)p^n$, where $0<p<1$. If $\mathrm{gcd}(\phi,\psi) = 1$, the variable $\nu = \frac{\phi}{\phi+\psi}$ has the distribution $P(\nu)\sim g(\nu)$ in the asymptotic limit $q=1-p \to 0$, namely:
\be
P\left(\nu = \frac{\phi}{\phi+\psi}\right) = \begin{cases} \disp \sum_{n=1}^{\infty} p^{n(\phi+\psi)} = \frac{(1-q)^{\phi+\psi}}{1 - (1-q)^{\phi+\psi}}\bigg|_{q\to 0} \approx \frac{1}{q}\, \frac{1}{\phi+\psi} & \quad \mbox{$(\phi,\psi)$ coprime} \medskip \\ 0 & \quad \mbox{otherwise} \end{cases}
\label{eq:002}
\ee
Thus, $P\left(\frac{\phi}{\phi+\psi}\right)$ coincides with the Riemann-Thomae function $g\left(\frac{\phi}{\phi+\psi}\right)$ defined in \eq{eq:001} up to the scaling factor (amplitude) $\frac{1}{q}$.

The emergence of the distribution \eq{eq:002} can be understood on the basis of the generalized Euclid orchard construction if one considers a $(1+1)$-dimensional directed walk on the lattice starting at the origin and making $\phi$ steps along one axis, followed by $\psi$ steps along the other axis. At every lattice site, the walk survives with the probability $p$ and dies with the probability $q=1-p$. Having an ensemble of such walks, one arrives at the model of ``hooked walks'' in the Euclid orchard. Thus, for some point, $\nu$, and at a small death probability, $q$, ($q\to 0$), a fraction of survived walkers, $P(\nu)$, computed in \eq{eq:002} is described by the Riemann function \eq{eq:001}.

\subsection{Relation of the Riemann-Thomae function to Eisenstein series and Dedekind $\eta$-function}

\begin{figure}[ht]
\centerline{\includegraphics[width=16cm]{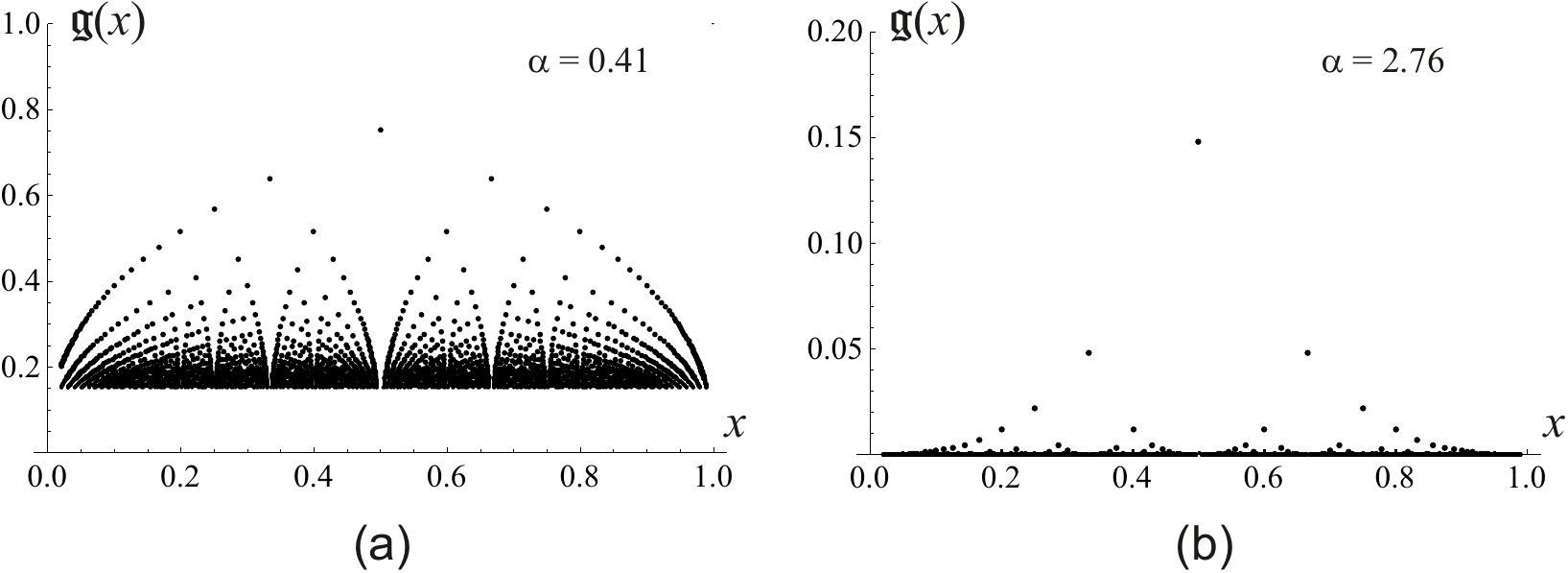}}
\caption{Powers of the Riemann-Thomae function, $\mathfrak{g}(x)$, for two sample values ($n=100$): (a) $\alpha=0.41$ and (b) $\alpha=2.76$.}
\label{fig:002}
\end{figure}

Define the generalized Riemann-Thomae (gRT) function, $\mathfrak{g}(x)$, as follows:
\be
\mathfrak{g}(x) = \begin{cases} h(n) & \mbox{if $x=\frac{m}{n}$, and $(m,n)$ coprime} \medskip \\
0 & \mbox{if $x$ is irrational} \end{cases}
\label{eq:003}
\ee
where $h(n)=n^{-\alpha}$ ($\alpha>0$). The sample plots of $\mathfrak{g}(x)$ for two arbitrary chosen values, $\alpha=0.41$ and $\alpha=2.76$ (for $n=100$) are shown in \fig{fig:002}a,b.

A natural, physically justified analytic regularization of the Riemann-Thomae function is highly demanded. Below, we briefly describe such a regularization for gRT function $\mathfrak{g}(x)$ where $h(n)=n^{-2}$ considered firstly in \cite{polov} and extend our construction to any $f(n)=n^{-\alpha}$ ($\alpha>0$) and even to non-algebraically decaying potentials. Namely, we demonstrate that the analytic approximation of the function $\mathfrak{g}(x)$ for $h(n)=n^{-2}$ involves the Dedekind $\eta$-function, $\eta(x+iy)$, defined in the halfplane $y>0$.

Let us rewrite $\mathfrak{g}(x)$ for $h(n)=n^{-2}$ on the interval $x \in (0,1)$ as follows:
\be
\mathfrak{g}(x) = \lim_{N\to\infty} \sum_{n=1}^{N} \sum_{k=1}^{n} \frac{1}{2n^{2}}\delta\left(x - \frac{k}{n}\right)
\label{eq:004}
\ee
The function $\mathfrak{g}(x)$ assigns zero to all irrational points. Now, using the identity:
\be
\delta(x) = \frac{1}{\pi} \lim_{\eps\to+0} \im \frac{1}{x - i\eps},
\label{eq:007}
\ee
the function $\mathfrak{g}(x)$ in \eq{eq:004} can be regularized by $0<\eps \ll 1$ as follows:
\be
\mathfrak{g}(x) = \bar{C}(\eps) \bar{\mathfrak{g}}(x,\eps)
\label{eq:007a}
\ee
where
\begin{multline}
\bar{\mathfrak{g}}(x,\eps) = \frac{1}{2\pi} \lim_{N\to\infty} \sum_{n=1}^{N} \sum_{m=1}^{N} \frac{\eps}{\left(x n - m\right)^2 + \eps^2 n^2} \\ = \frac{1}{2\pi}\sum_{\{m,n\} \in \mathbb{Z}^2 \backslash \{0, 0\}} \frac{\eps}{\left(x n - m\right)^2 + \eps^2 n^2} - \frac{\eps \pi}{12}\left(1 + \frac{1}{x^2 + \eps^2}\right)
\label{eq:008}
\end{multline}
The latter term in \eq{eq:008} is proportional to $\delta(x)$ and can be neglected. The coefficient $\bar{C}(\eps)$ in \eq{eq:007a} can be computed from the normalization condition and it is instructive to fix it at the end of derivation. Since $n$ and $m$ in \eq{eq:008} run over all integer points except $0$, it is convenient to change a sign in front of $m$: $m\to -m$ and rewrite \eq{eq:008} as follows
\be
\bar{\mathfrak{g}}(x,\eps) = \frac{1}{2\pi}\lim_{N\to\infty} \sum_{\{m,n\} \in \mathbb{Z}^2 \backslash \{0, 0\}} \frac{\eps}{\left(x n + m\right)^2 + \eps^2 n^2}
\label{eq:009}
\ee

Recall now the definition of the non-holomorphic Eisenstein series, $E(z,s)$, \cite{eisen}:
\be
E(z,s) = \sum_{\{m,n\} \in \mathbb{Z}^2 \backslash \{0, 0\}} \frac{y^s}{|n z + m|^{2s}};
\qquad z = x+iy
\label{eq:010}
\ee
where $E(z,s)$ is a function of $z=x+iy$ and is defined in the upper half-plane $y>0$ for all $\re(s)>1$. Comparing \eq{eq:009} and \eq{eq:010}, we can straightforwardly conclude that the function $\bar{\mathfrak{g}}(x,\eps)$ matches the Eisenstein series $E(z, s)$ at $s=1$ upon the identification $\eps = y$, i.e. $x+i\eps = z$. Thus,
\be
\bar{\mathfrak{g}}(x,\eps) = \frac{1}{2\pi}E(z,s = 1)
\label{eq:011}
\ee

The non-holomorphic Eisenstein series of weight 0 and level 1 can be analytically continued to the whole complex $s$-plane with one simple pole at $s=1$. Notably $E(z,s)$, as function of $z$, is the $SL(2,\mathbb{Z})$--automorphic solution of the hyperbolic Laplace equation:
\be
-y^2 \left(\frac{\partial^2}{\partial x^2}+\frac{\partial^2}{\partial y^2}\right) E(x,y, s) = s(1-s)\; E(x,y, s)
\label{eq:012}
\ee
The residue of $E(z, s)$ at $s=1$ is known as the first Kronecker limit formula \cite{epstein, siegel, motohashi}. Explicitly, it reads at $s\to 1$:
\be
E(z, s\to 1) = \frac{\pi}{s-1} + 2\pi\left(\gamma + \ln 2 - \ln \left(y^{1/2}|\eta(z)|^2\right)\right) + O(s-1)
\label{eq:013}
\ee
where $\gamma$ is the Euler constant and $\eta(z)$ is the Dedekind $\eta$-function. Equation \eq{eq:013} establishes the important connection between the Eisenstein series and the Dedekind $\eta$-function, which we exploit below. The Dedekind $\eta$-function is defined as follows:
\be
\eta(z)=e^{\pi i z/12}\prod_{n=0}^{\infty}(1-e^{2\pi i n z})
\label{eq:014}
\ee
The argument $z=x+iy$ is called the modular parameter, and $\eta(z)$ is defined for all $y>0$. The function $\eta(z)$ is invariant with respect to the action of the modular group $SL(2,\mathbb{Z})$:
\be
\begin{cases}
\disp \eta (z+1)=e^{\pi i z/12}\;\eta(z) \medskip \\
\disp \eta\left(-\frac{1}{z}\right) = \sqrt{-i}\; \eta(z)
\end{cases}
\label{eq:015}
\ee
In general,
\be
\eta\left(\frac{az+b}{cz+d}\right) = \omega(a,b,c,d)\; \sqrt{cz + d}\; \eta(z)
\label{eq:016}
\ee
where $ad-bc=1$ and $\omega(a,b,c,d)$ is a 24th degree root of unity \cite{dedekind}.

\begin{figure}[ht]
\centerline{\includegraphics[width=15cm]{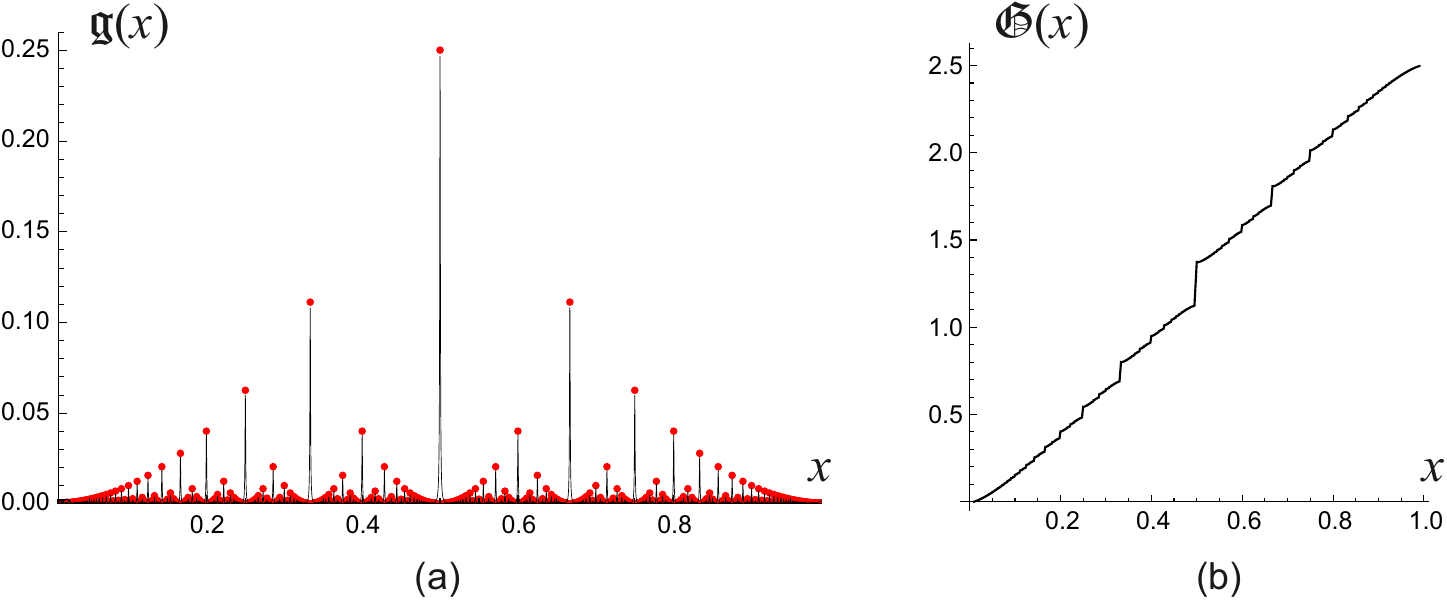}}
\caption{(a) Plots of the discrete generalized Riemann-Thomae function $\mathfrak{g}(x)$ for $h(n)=n^{-2}$ (red points) at rational points in $0<x<1$, and of everywhere continuous function $-\frac{12y}{\pi}\ln|y^{1/4}\eta(x+iy)|$ (black curve) taken at $y = 5\times 10^{-4}$; (b) Integrated function, $\mathfrak{G}(x)(x)$ (see \eq{eq:019}) has a Devil's staircase structure.}
\label{fig:003}
\end{figure}

Collecting together \eq{eq:011} and \eq{eq:013}, taking into account that $\eps=y$, and omitting the divergent constant at $s\to 1$, we get the following asymptotic analytic expression for the Riemann-Thomae function $\mathfrak{g}(x)$ for $h(n)=n^{-2}$:
\be
\mathfrak{g}(x) = \bar{C}(y) \bar{\mathfrak{g}}(x,y) = - C(y)\,\ln f(x,y)
\label{eq:017}
\ee
where 
\be
f(x,y) = y^{1/4}|\eta(x+iy)|
\label{eq:018}
\ee
The coefficient $C(y)$ in \eq{eq:017} we compute from the condition 
$$
\mathfrak{g}\left(x=\frac{1}{2}\right) = -\lim_{y\to 0} \left[C(y)\,\ln f\left(x=\frac{1}{2},y\right)\right] = \frac{1}{4}
$$
(see the Appendix \ref{app:1} for more detail). Specifically, the value of $C(y)$ up to logarithmic corrections reads
\be 
C(y) = \frac{12y}{\pi}
\label{eq:018a}
\ee
In \fig{fig:003}a we have depicted the function $\mathfrak{g}(x)$ for $h(n)=n^{-2}$ with the maximal denominator $n_{\rm max}=10^{2}$ together with the function $-\frac{12y}{\pi}\ln f(x,y)$ at the fixed value $y=5\times 10^{-4}$, where $f(x,y)$ is defined in \eq{eq:018}--\eq{eq:018a}, while in \fig{fig:003}b we have plotted the cumulative (integrated) function $\mathfrak{G}(x)$, where
\be
\mathfrak{G}(x) = \int_0^x \mathfrak{g}(x')dx'
\label{eq:019}
\ee

The function $\mathfrak{G}(x)$ has a typical ``Devil's staircase'' shape \cite{devil} reflected horizontally and rotated by $\pi/2$. Later on we shall consider the ``generalized Devil's staircase'', $G_{\alpha}(x)$ defined as in \eq{eq:019} for $h(n) = n^{-\alpha}$ (see \eq{eq:003}) at any values of $\alpha>0$ ($\mathfrak{G}(x)\equiv G_2(x)$).

\section{Riemann-Thomae function and the spectral density of random tridiagonal operators}

Consider an ensemble of random operators represented by tridiagonal $N\times N$ ($N\gg 1$) random symmetric matrices $A_N$ with the bimodal (Bernoulli) distribution of sub-diagonal matrix elements:
\be
A_N = \left(\begin{array}{ccccc}
0 & x_1 & 0 & \cdots & 0 \smallskip \\  x_1 & 0 & x_2 & & \smallskip \\  0 & x_2 & 0 & & \smallskip \\ \vdots &  &  &  & \smallskip \\ & & & & x_{N-1} \smallskip \\ 0 & & & x_{N-1} & 0
\end{array} \right);
\qquad x_k=\left\{\begin{array}{ll} 1 & \mbox{with probability $p$} \medskip \\
0 & \mbox{with probability $q=1-p$} \end{array} \right.
\label{eq:022}
\ee
We are interested in spectral properties of an ensemble of such matrices. Namely, we compute the density of eigenvalues,  $\rho(\lambda)$, in the limit $N\to\infty$ and demonstrate its connection to the Riemann-Thomae function $g(x)$ defined in \eq{eq:001}. For the first time this question was addressed in \cite{krapiv} and below we present slightly more extended version of our construction. 

\begin{figure}[ht]
\centerline{\includegraphics[width=16cm]{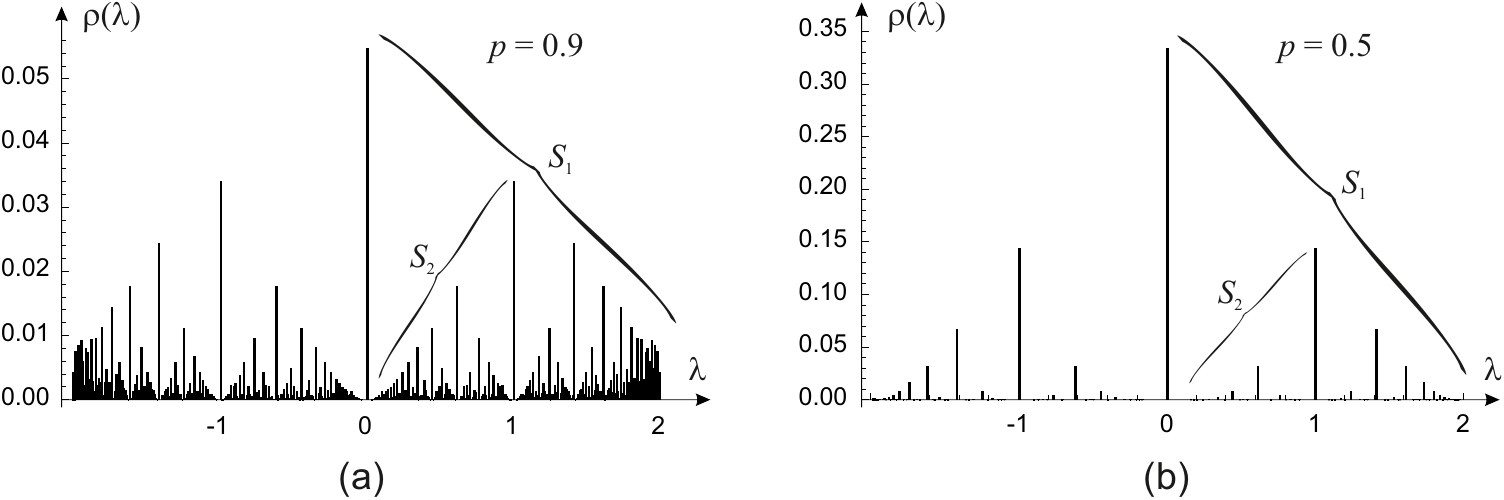}}
\caption{The spectral density $\rho(\lambda)$ for the ensemble of 500 three-diagonal random operators of size $N=500$ at $p=0.9$ (a) and $p=0.5$ (b).}
\label{fig:004}
\end{figure}

To proceed, note that at any $x_k=0$, one can split the matrix $A_N$ into independent blocks along the diagonal. So, it is instructive to consider subsequences of gapless sets of with $x_k=1$. The set of eigenvalues of a symmetric gapless $n\times n$ three-diagonal block $A_n$ with $x_k=1$ for all $k=1,...,n$, is 
\be
\lambda_{k,n} = -2\cos\frac{\pi k}{n+1}; \qquad (k=1,...,n)
\label{eq:023}
\ee
The probability of having a gapless subsequence with $n$ consecutive "1" is $Q_n=p^n$, since all $x_k$ are independently distributed Bernoulli variables. The sample plots $\rho(\lambda)$ for two different values of $p$, namely for $p=0.9$ and $p=0.5$ computed numerically for $N=500$ over 500 different matrix realizations, are shown in \fig{fig:004}.

The spectral density $\rho(\lambda)$ can be written in the limit $N\to\infty$ in terms of the Riemann-Thomae function $g(\tau)=g(x+iy)$. The resulting expression reads:
\be
\rho(\lambda) = \frac{p^{1/g(x+iy)}}{1-p^{1/g(x+iy)+1}}
\label{eq:025}
\ee
where 
\be
g(x+iy)=\sqrt{-\frac{12y}{\pi} \ln \left(y^{1/4} |\eta(x+iy)|\right)}, \qquad x=\frac{1}{\pi}\arccos\frac{\lambda}{2}, \quad y=1-p
\label{eq:025a}
\ee
The parameter $y$ in \eq{eq:025} has a sense of a ``resolution cutoff'' of the Dedekind relief -- see \fig{fig:set} in Section IIA. The relation between the strength of the disorder, $p$, and the cutoff, $y$, can be established using the following qualitative arguments. On one hand, the maximal denominator, $n_{max}$, of the Thomae function (see \eq{eq:003}) defines the total number of peaks that can be resolved upon $n_{max} \ln p \sim 1$,  as it follows from \eq{eq:025}. On the other hand, the cutoff $y$ can be estimated as $y\sim 1/n_{max}$. Thus, in the limit $p\to 1$ one has $y\approx 1-p$. In next Section we provide the outline of the derivation of \eq{eq:025}. Let us emphasize that throughout the paper we pay attention to the {\it non-perturbative} limit $p\to 1$ of a weak disorder.

\subsection{Analogy with the Peierls model}

Let us discuss how this toy 1D example matches the general perspective on the Devil's staircase. To this aim, we compare our example with the Peierls model of 1D superconductivity where the Devil's staircase has been recognized as well \cite{dzyaloshinskij1982commensurability}. One starts with the integrable version of the Peierls model \cite{brazovskii1982exactly} in which the Hamiltonian can be written as follows 
\be
H= \Psi^{\dagger}L_{f}\Psi + \sum_{i=1}^{k} t_k H_k
\label{eq:peierls}
\ee
The Hamiltonian \eq{eq:peierls} describes the interactions between phonon-like degrees of freedom, $\theta_n$, of fluctuations around the regular lattice $x_n= na + \theta_n, \quad n= 1\dots N$ and fermions propagating at the top of the lattice. The Hamiltonian of phonons in the integrable case involves a few lowest Toda chain Hamiltonians, $H_k = {\rm Tr}\, L_f^k$, where the $N\times N$ Lax operator for the Toda chain 
\be
L_{f}=\left(\begin{array}{cccccc}
p_1 & c_1 & 0 & \cdots & & \eta \\  c_1 & p_2 & c_2 & & & \\  0 & c_2 & p_3 &  & & \\ \vdots &  &  &   &  & \\ & &  &  &  & \\ & & & & p_{N-1} & c_{N-1} \\ \eta & & & & c_{N-1} & p_N
\end{array} \right)
\ee
plays the role of the Hamiltonian for the fermions. The quasimomentum, $\eta$, corresponds to the periodic lattice, $p_k$ are the momenta of phonons and $c_k= \exp (\theta_{k+1} - \theta_k)$. The Lax representation for the Toda chain provides the consistency of the phonon and fermion dynamics. The solution at the generic values of the Toda Hamiltonians and some value of the fermionic density, $\rho=\frac{q}{N}$, is expressed in terms of the hyperelliptic Riemann surface whose moduli are defined by the Hamiltonians and fermionic density. The Riemann surface simultaneously plays the role of the dispersion law for fermions.

The transition from commensurability to incommensurability in the fermionic spectrum occurs if two conditions are fulfilled simultaneously:
\begin{itemize}
\item Select the values of Toda Hamiltonians in such way that some bands in the spectrum became very tiny;
\item Select the rational value of the fermionic density and add a very weak non-integrable deformation of the Hamiltonian.
\end{itemize}
Upon these two conditions the Devil's staircase emerges \cite{dzyaloshinskij1982commensurability}. It can also be reformulated in terms of the Whitham dynamics for the Toda chain describing the flow of the integrable systems solutions in the moduli space. Whitham dynamics was interpreted as a version of a RG flow for many systems -- see, for instance, discussion in \cite{gorsky2015rg}.

Let us compare the Peierls model setup and the toy model described in the previous Section. We have the 1D lattice of $N$ sites and the particle hopping on this lattice, which can be identified with the particle Hamiltonian \eq{eq:022}. The two-step procedure similar to the one described above for the Peirles system is as follows:
\begin{itemize}
\item We see that in our random operator \eq{eq:022} all diagonal matrix elements are equal to zero. In terms of the Peierls model it means that $p_i=0$ and the lattice is frozen. Freezing the lattice indeed means a deep degeneration of the Riemann surface;
\item Instead of adding a weak non-integrable deformation to the Peierls Hamiltonian, we introduce the weak randomness in the lattice which does the same job.
\end{itemize}
Hence, the emergence of the Devil's staircase in the spectrum of a tridiagonal matrix and in the Peierls model is of similar origin. The probe particle in our model plays the same role as Lax fermions in the Peierls model for the effective Whitham RG dynamics. The key point ensuring the full matching of the Peierls model with our scheme is the formulation of the RG procedure that can be translated into the flow in the fundamental domain of $SL(2,Z)$. This will be explained in detail below.

\subsection{Riemann-Thomae function and hyperbolic geometry}

\begin{figure}[ht]
\centerline{\includegraphics[width=16cm]{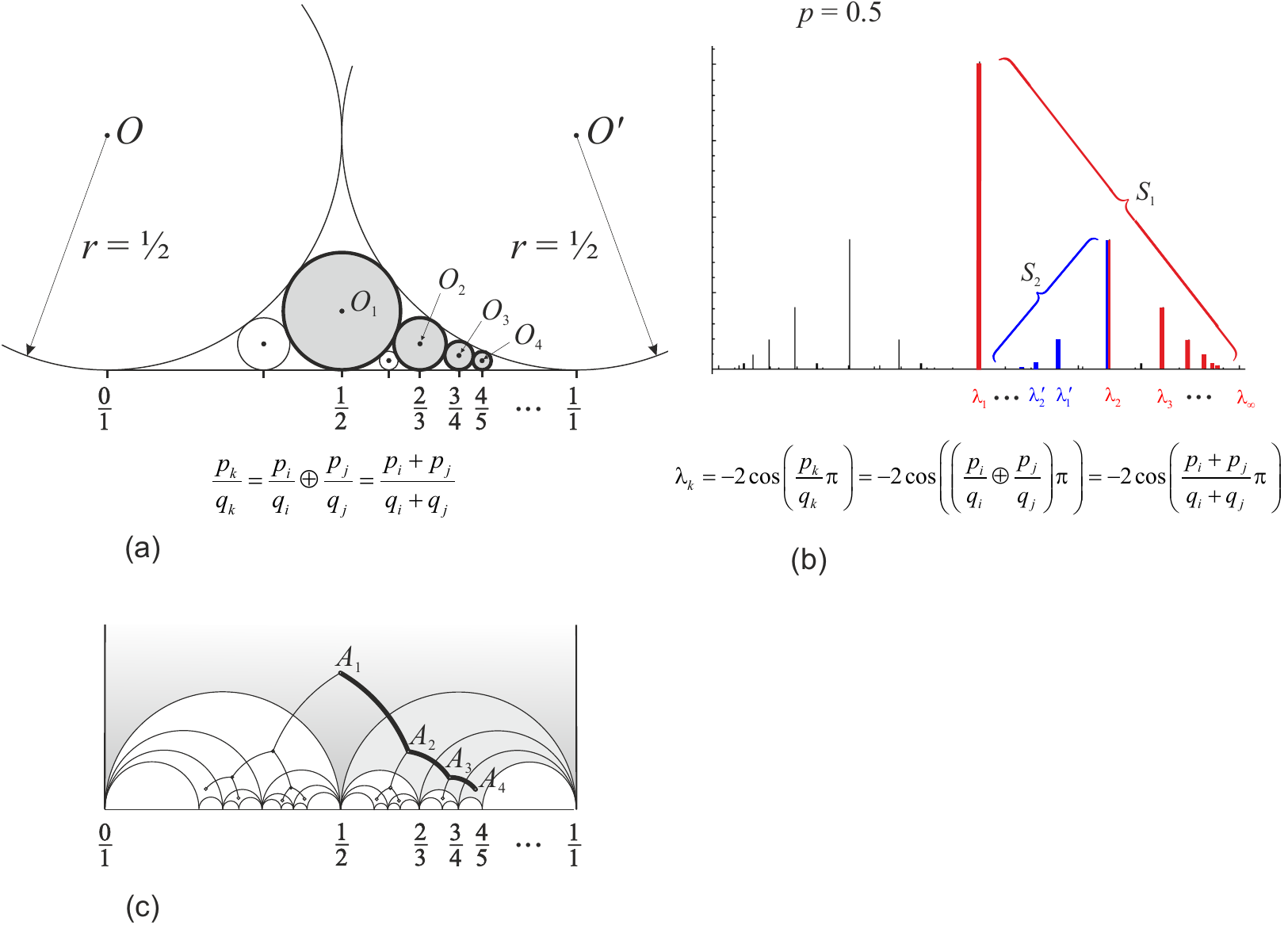}}
\caption{(a) The Ford circles as illustration of the Farey sequence construction: each circle touches two neighbors (right and left circles) and the segment. The position of newly generated circle is determined via the $\oplus$ addition: $\frac{p_{i}}{q_{i}} \oplus \frac{p_{j}}{q_{j}}=\frac{p_{i}+p_{j}}{q_{i}+q_{j}}$; (b) The spectral density $\rho_{\eps}(\lambda)$ for the ensemble of tridiagonal matrices of size $N=10^3$ at $p=0.5$ and its relation to Farey numbers; (c) The same Farey sequence generated by recursive fractional-linear transformations of the fundamental domain of the modular group $SL(2,Z)$.}
\label{fig:005}
\end{figure}

Let us remind a textbook definition: sequences of coprime fractions constructed via the $\oplus$ addition constitute the ``Farey sequence'' \cite{farey}
\be
\frac{p_{i}}{q_{i}} \oplus \frac{p_{j}}{q_{j}} = \frac{p_{i}+p_{j}}{q_{i}+q_{j}}.
\label{eq:030}
\ee
A simple geometric model behind the Farey sequence, known as the Ford circles \cite{ford}, is shown in \fig{fig:005}a. The corresponding generic recursive algorithm constitutes the Farey sequence construction.  Schematically, the construction goes as follows: take the segment $[0,1]$ and draw two circles $O$ and $O'$ both of radius $r=\frac{1}{2}$ touching each other, and the segment at the terminal points 0 and 1. Now inscribe a new circle $O_3$ touching $O_1$, $O'$ and $[0,1]$. Where is the position of the new circle along the segment? The position of the newly generated circle projected to the segment $[0,1]$ is determined via the $\oplus$ operation \eq{eq:030}. For example, the centers $x_{O_2}$ of the circle $O_2$, $x_{O_3}$ of the circle $O_3$ and $x_{O_4}$ of the circle $O_4$ are located correspondingly at the points:
\be
\begin{cases}
\disp x_{O_2}=\left(\frac{p}{q}\right)_{O_1}\oplus \left(\frac{p}{q}\right)_{O'}=\frac{1}{2} \oplus \frac{1}{1}=\frac{1+1}{2+1}=\frac{2}{3} \medskip \\
\disp x_{O_3}=\left(\frac{p}{q}\right)_{O_2}\oplus \left(\frac{p}{q}\right)_{O'}=\frac{2}{3} \oplus \frac{1}{1}=\frac{2+1}{3+1}=\frac{3}{4} \medskip \\
\disp x_{O_4}=\left(\frac{p}{q}\right)_{O_3}\oplus \left(\frac{p}{q}\right)_{O'}=\frac{3}{4} \oplus \frac{1}{1}=\frac{3+1}{4+1}=\frac{4}{5}
\end{cases}
\label{eq:ford}
\ee
etc. 

The connection of the Ford construction with the spectrum of the ensemble of random operators \eq{eq:022} goes as follows. Write the position of the central peak at $\lambda_1=0$ as $\lambda_1=0=-2\cos \left(\frac{1}{2}\pi\right)$. Write the spectral edge at $\lambda_{\infty}=2$ as $\lambda_{\infty}=-2\cos \left(\frac{1}{1}\pi\right)$. Now consider the ``enveloping'' sequence of monotonically decreasing peaks which is denoted in \fig{fig:004} as $S_1$. One can show that positions of eigenvalues $\lambda_2, \lambda_3,...$ shown in red in \fig{fig:005}b are determined similarly to \eq{eq:ford}, namely
\be
\begin{cases}
\disp \lambda_2=-2\cos\left(\left(\frac{1}{2}\oplus \frac{1}{1}\right)\pi\right)= -2\cos\left(\frac{1+1}{2+1}\pi\right)=-2\cos\left(\frac{2}{3}\pi\right) \medskip \\
\disp \lambda_3=-2\cos\left(\left(\frac{2}{3}\oplus \frac{1}{1}\right)\pi\right)= -2\cos\left(\frac{2+1}{3+1}\pi\right)=-2\cos\left(\frac{3}{4}\pi\right) \medskip \\
\disp \lambda_4=-2\cos\left(\left(\frac{3}{4}\oplus \frac{1}{1}\right)\pi\right)= -2\cos\left(\frac{3+1}{4+1}\pi\right)=-2\cos\left(\frac{4}{5}\pi\right)
\end{cases}
\label{eq:031}
\ee
etc.

Positions of resonances in other monotonic sequences, for example in the sequence $S_2$ in \fig{fig:004}, one can again find recursively using the Farey construction \eq{eq:030}. Corresponding eigenvalues for the sequence $S_2$ are denoted as $\lambda_2, \lambda_1', \lambda_2', ...$ and they are shown in blue in \fig{fig:005}b. Their positions are:
\be
\begin{cases}
\disp \lambda_1'= -2\cos\left(\left(\frac{1}{2}\oplus \frac{2}{3}\right)\pi\right)= -2\cos\left(\frac{1+2}{2+3}\pi\right)=-2\cos\left(\frac{3}{5}\pi\right) \medskip \\
\disp \lambda_2'= -2\cos\left(\left(\frac{1}{2}\oplus \frac{3}{5}\right)\pi\right)= -2\cos\left(\frac{1+3}{2+5}\pi\right)=-2\cos\left(\frac{4}{7}\pi\right) \medskip \\
\disp \lambda_3'= -2\cos\left(\left(\frac{1}{2}\oplus \frac{4}{7}\right)\pi\right)= -2\cos\left(\frac{1+4}{2+7}\pi\right)=-2\cos\left(\frac{5}{9}\pi\right)
\end{cases}
\label{eq:032}
\ee
etc. The same Farey sequences can be sequentially generated by fractional-linear transformations (reflections with respect to the arcs) of the fundamental domain of the modular group $SL(2,Z)$ -- the triangle lying in the upper halfplane $\im z>0$ of the complex plane $z$ as shown in \fig{fig:005}c.

\subsection{Spectral density $\rho(\lambda)$ and the ``visibility diagram''}

The spectral density $\rho(\lambda)$ of the ensemble of $N\times N$ random matrices $A_N$ with the bimodal Bernullian distribution of matrix elements can be written in a form of a resolvent:
\be
\rho(\lambda) = \lim_{N\to\infty}\frac{1}{N}\la \sum_{n=1}^N\sum_{k=1}^{n} \delta(\lambda-\lambda_{k,n}) \ra =  \lim_{\stackrel{N\to\infty}{\eps\to 0}} \frac{\eps}{\pi N} \sum_{n=1}^N Q_n \sum_{k=1}^n \im\, \frac{1}{\lambda-\lambda_{k,n} - i\eps}
\label{eq:033}
\ee
where $\la ...\ra$ means averaging over the distribution $Q_n=p^n$, and the identity \eq{eq:007} has been used to regularize the $\delta$-function. Substituting \eq{eq:023} into \eq{eq:033}, we find the following expression for $\rho(\lambda)$:
\be
\rho(\lambda) = \lim_{\stackrel{N\to\infty}{\eps\to 0}} \frac{1}{\pi N} \sum_{n=1}^{N} p^n
\sum_{k=1}^n\frac{\eps}{\left(\lambda+2\cos\frac{\pi k}{n+1}\right)^2+\eps^2}
\label{eq:034}
\ee
The sum in \eq{eq:034} looks rather complicated, however by exchanging the orders of summations (first running the summation in $n$ and then -- in $k$) one can advance in understanding the structure of \eq{eq:034} and its relation to the Riemann-Thomae function $g(x)$ as stated in \eq{eq:025}.

\begin{figure}[ht]
\centerline{\includegraphics[width=16cm]{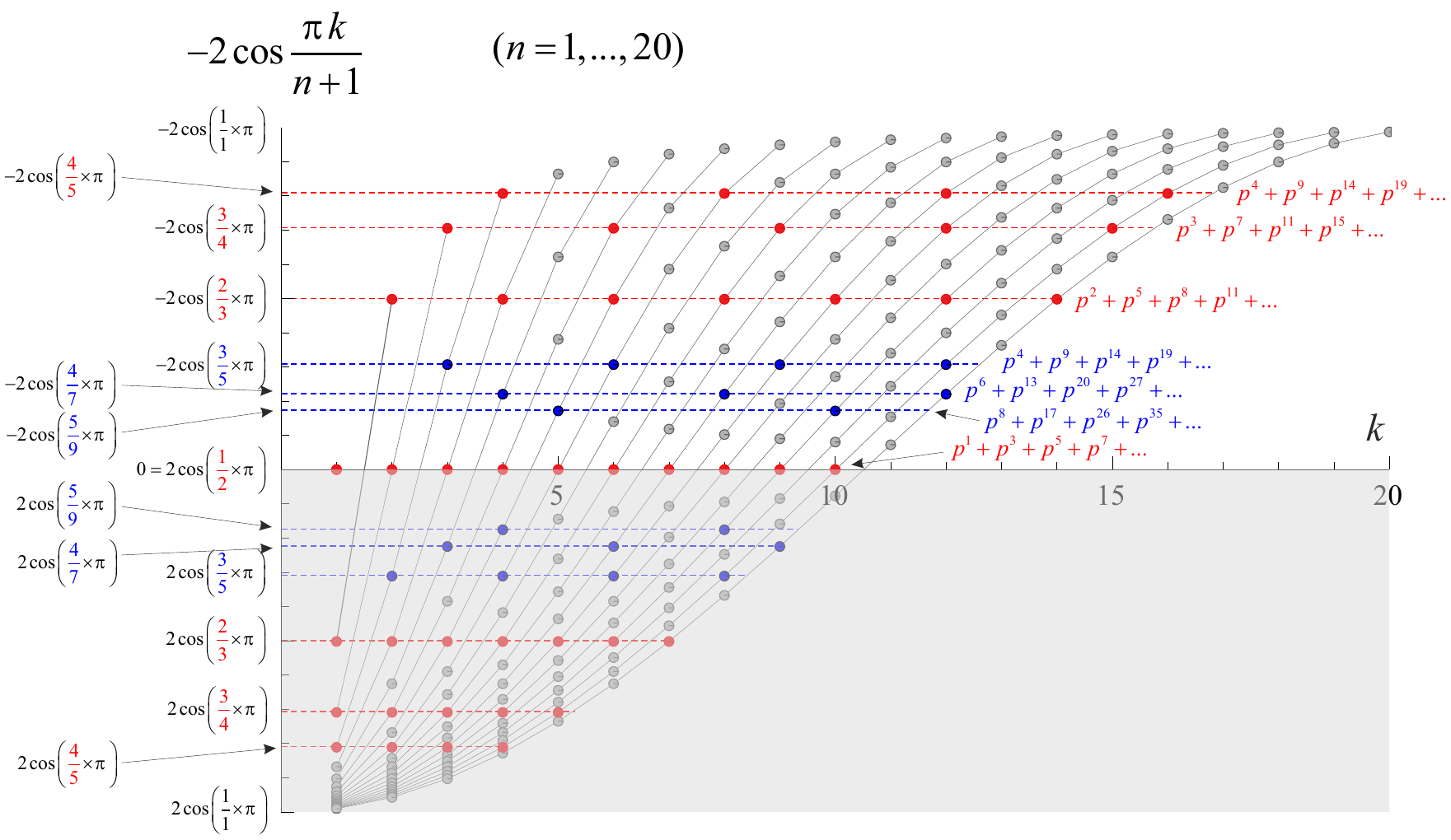}}
\caption{Family of 20 curves $f(k,n)=-2\cos\frac{\pi k}{n+1}$. Each curve corresponds to a particular $n=1,...,20$ (from left to right), points along each curve mark values $k=1,...,n$. Each horizontal dashed line corresponds to the multiplicity of the eigenvalue and contributes to the height of the peak in the spectral density.}
\label{fig:006}
\end{figure}

To proceed, note that the function $\rho(\lambda)$ is nonzero only at $\lambda_{k,n}$. In \fig{fig:006} we have plotted eigenvalues $\lambda_{k,n}=-2\cos\frac{\pi k}{n+1}$ as a function of $k$ for a set of 20 fixed values of $n$: $n=1,..., 20$. Every point in \fig{fig:005} designates some eigenvalue $\lambda_{k,n}$; points along each solid gray curve correspond to \emph{different} values of $k$ for \emph{one and the same} value of $n$. Each horizontal dashed line represents the set of \emph{the same} eigenvalues $\lambda$ coming from \emph{different} $n$. Let us exchange orders of summation in $k$ and in $n$ in \eq{eq:034} and first sum weights of points along each horizontal dashed line over all $n$. In such a way we account of  the degeneracy of a corresponding eigenvalue $\lambda_k$ and, respectively, the height of a peak in the spectral density $\rho(\lambda)$ at $\lambda_k$.

Thus, by summing corresponding powers of $p$ we get the height of any peak (i.e. eigenvalue). To compute $\rho(\lambda)$ we split the spectrum into monotonic sequences of peaks as it has been discussed in the previous section. Sequences $S_1$ and $S_2$ shown in \fig{fig:005}b are typical representatives. For them we have:
\be
S_1: \left\{\begin{array}{ll}
\disp \lambda_1=-2\cos\frac{\pi}{2} & \quad \rho_{S_1}(\lambda_1) = p^1+p^3+p^5+p^{7}+... \medskip \\
\disp \lambda_2=-2\cos\frac{2\pi}{3} & \quad \disp \rho_{S_1}(\lambda_2) = p^2+p^5+p^8+p^{11}+... \medskip \\
\disp \lambda_2=-2\cos\frac{3\pi}{4} & \quad \rho_{S_1}(\lambda_3) = p^3+p^7+p^{11}+p^{15}+... \medskip \\ & ... \\
\disp \lambda_n=-2\cos\frac{\pi n}{n+1} & \quad \disp \rho_{S_1}(\lambda_n) = \sum\limits_{s=1}^{\infty} p^{(n+1)s-1} = \frac{p^n}{1-p^{n+1}} \quad (n=1,2,...) \end{array} \right.
\label{eq:035}
\ee
and
\be
S_2: \left\{\begin{array}{ll}
\disp \lambda_2=-2\cos\frac{2\pi}{3} & \quad \rho_{S_2}(\lambda_2) = p^2+p^5+p^8+p^{11}+... \medskip \\
\disp \lambda_1'=-2\cos\frac{3\pi}{5} & \quad \rho_{S_2}(\lambda_1') = p^4+p^9+p^{14}+p^{19}+... \medskip \\
\disp \lambda_2'=-2\cos\frac{4\pi}{7} & \quad \rho_{S_2}(\lambda_2') = p^6+p^{13}+p^{20}+p^{27}+... \medskip \\ & ... \\
\disp \lambda_{n}'=-2\cos\frac{\pi n}{2n-1} & \quad \disp \rho_{S_2}(\lambda_{n}) = \sum\limits_{s=1}^{\infty} p^{(2n-1)s-1} = \frac{p^{2n-2}}{1-p^{2n-1}} \quad (n=2,3,...)
\end{array} \right.
\label{eq:036}
\ee

\begin{figure}[ht]
\centerline{\includegraphics[width=14cm]{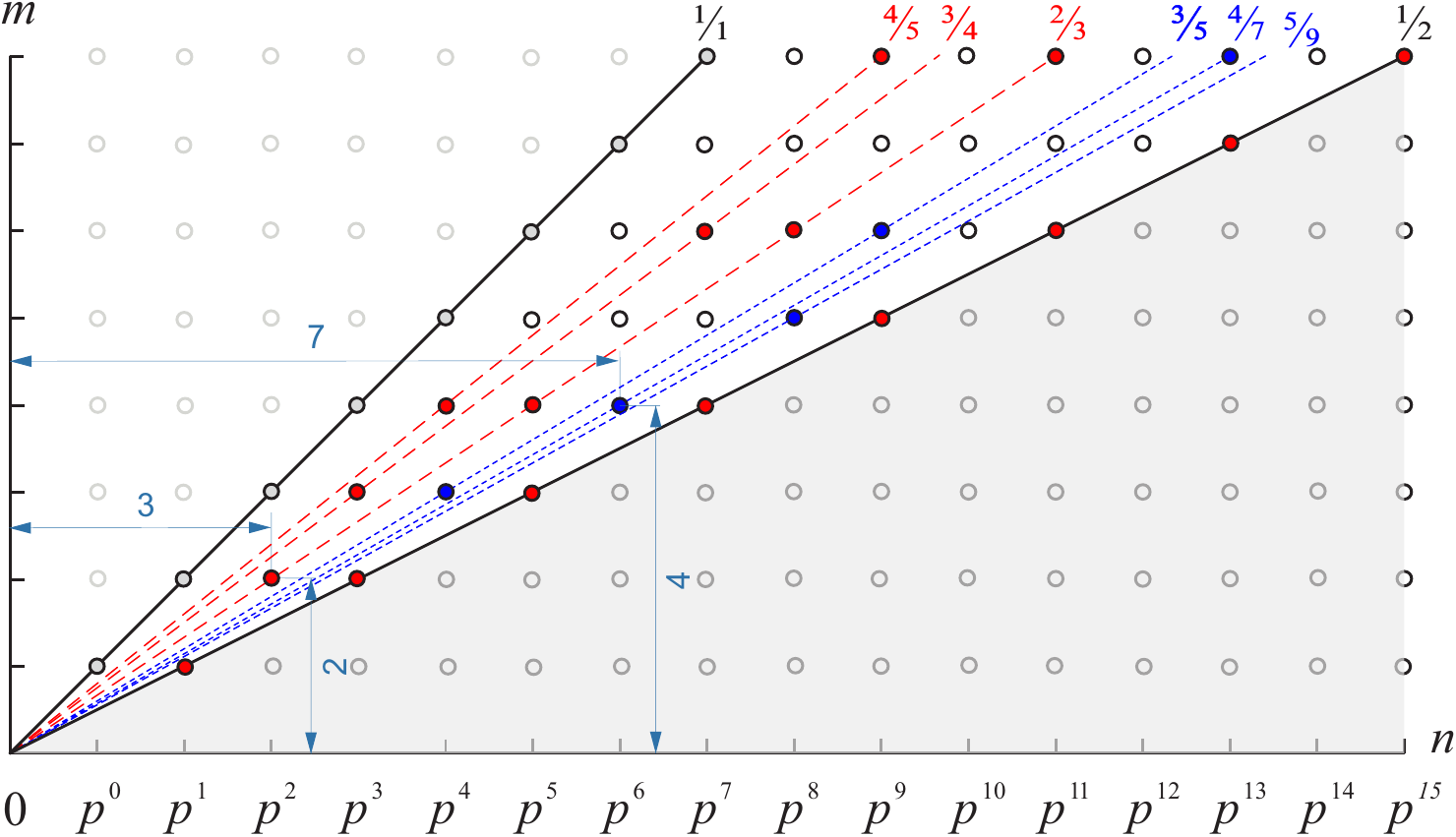}}
\caption{Visibility diagram. Each point in the vertical row $n$ carries a weight $p^{n-1}$. Integer points within the wedge $[\pi/4,\pi/2]$ are designated by circles. The dashed rays are emitted at rational tangents $\frac{m}{n}$, where $m$ and $n$ are coprimes. The weights corresponding to marked integer points, are summed up along the rays.}
\label{fig:007}
\end{figure}

Equations \eq{eq:035}--\eq{eq:036} are tightly linked to the so-called visibility diagram (our notations are slightly different with the definition of the visibility diagram defined in \cite{wallett}). Consider the square lattice of integer points $(m,n)$ and add a weight $p^{n-1}$ to each vertical row. Emit rays from the point $(0,0)$ at rational tangents of angles $\alpha_{m,n}=\arctan \frac{m}{n}$ ($m$ and $n$ are coprimes, $1\le m\le n-1$) within the wedge $[\pi/4,\pi/2]$ in the positive direction as shown in \fig{fig:007} and sum up the weights, $p^{n-1}$, of all integer points along each emitted ray.

Let us show that the visibility diagram provides a straightforward way of calculating the spectral density $\rho(\lambda)$. To see this, consider again the monotonic sequences of peaks $S_1$ and $S_2$ shown in \fig{fig:004} and in \fig{fig:006}. For better visualization we keep colors throughout the text: the sequence $S_1$ is shown in red and $S_2$ -- in blue. Summing rational points on the visibility diagram in \fig{fig:007} along a ray with the angle $\alpha_{m,n}=\arctan\frac{m}{n}$ we get exactly the same results as given in \eq{eq:035}--\eq{eq:036}. We demonstrate that on two examples:
\begin{itemize}
\item Pick up the ray with $\tan \alpha_{2,3}=\frac{2}{3}$, which corresponds to $\lambda=-2\cos\left(\frac{2}{3}\pi\right)$. Summing rational points (weighted with the corresponding power of $p$) along this ray we get the height: $p^2+p^5+p^8+p^{11}+... = \sum_{s=1}^{\infty}p^{3s-1}=\frac{p^2}{1-p^3}$, which coincides with the value for $\lambda_2$ (at $n=2$) in \eq{eq:035}.
\item Pick up the ray with $\tan \alpha_{4,7}=\frac{4}{7}$ corresponding to the eigenvalue $\lambda=-2\cos\left(\frac{4}{7}\pi\right)$. Summing points along this ray we get: $p^6+p^{13}+p^{20}+p^{27}+... = \sum_{s=1}^{\infty}p^{7s-1}=\frac{p^6}{1-p^7}$, which coincides with the value for $\lambda_2'$ (at $n=4$) in \eq{eq:036}.
\end{itemize}

The analysis of the visibility diagram allows to formulate the following general prescription for the evaluation of the spectral density $\rho(\lambda)$ at $\lambda = -2\cos\left(\frac{m}{n}\pi\right)$
\be
\rho(n) = \sum_{s=1}^{\infty} p^{(n+1)s-1} = \frac{p^{n}}{1-p^{n+1}}
\label{eq:037}
\ee
where $n$ is the denominator of the fraction $\frac{m}{n}$ ($m$ and $n$ are coprimes). The angle $\alpha_{m,n} = \arctan\frac{m}{n}$ uniquely defines the eigenvalue
\be
\lambda = -2 \cos\left(\frac{m}{n}\pi\right)
\label{eq:038}
\ee
Comparing \eq{eq:037} with the definition of the Riemann-Thomae function $g(x)$ in \eq{eq:001} we can immediately identify $n$ with the denominator of the corresponding rational fraction $x=\frac{m}{n}$. So, we have $n=1/g(x+iy)$ where the cutoff $y$ defines the maximal denominator, $n_{max}$. At irrational $x$ the function $g(x)$ is 0, which provides $n=\infty$, thus giving $\rho(n)=0$ according to \eq{eq:037}. From \eq{eq:038} we have $\lambda=-2\cos(\pi x)$. Inverting this expression and taking into account the symmetry of the spectrum, we get $x=\frac{1}{\pi}\arccos\frac{\lambda}{2}$ which together with \eq{eq:037} leads to \eq{eq:025}.

\section{Riemann-Thomae function and phyllotaxis}

Amazing connection of cell packing with Fibonacci sequences, known as phyllotaxis \cite{phyllotaxis} was observed a long time ago in the works of naturalists and remains till now one of the most known manifestations of number theory in natural science. The generic description of growing plants based on symmetry arguments allowed researchers to uncover the role of Farey sequences in the plant's structure formation (see, for example, \cite{phyllo2, phyllo3, phyllo4}), however, the question why the nature selects the Fibonacci sequence, among other possible Farey ones, was hidden until modern time. A tantalizing answer to this question has been given by L. Levitov in 1990 in \cite{levitov}, who proposed an ``energetic'' approach to the phyllotaxis, suggesting that the development of a plant is connected with an effective motion along the optimal path on a Riemann surface associated with the energetic relief of growing tissue.

The energetic mechanism suggested in \cite{levitov1} was applied later in \cite{levitov3} to the investigation of the geometry of flux lattices pinned by layered superconductors. It has been shown that under the variation of a magnetic field, the structure of the flux lattice can undergo a sequence of rearrangements encoded by the Farey numbers. However, lattices emerging in sequential rearrangements are characterized by the specific subsequence of the Farey set, namely, by the Fibonacci numbers. Very illuminating experiments have been provided in \cite{rotating} for lattice formed by drops in rotating liquid, and in \cite{cactus} for the equilibrium structure of a ``magnetic cactus''.

\begin{figure}[ht]
\centerline{\includegraphics[width=14cm]{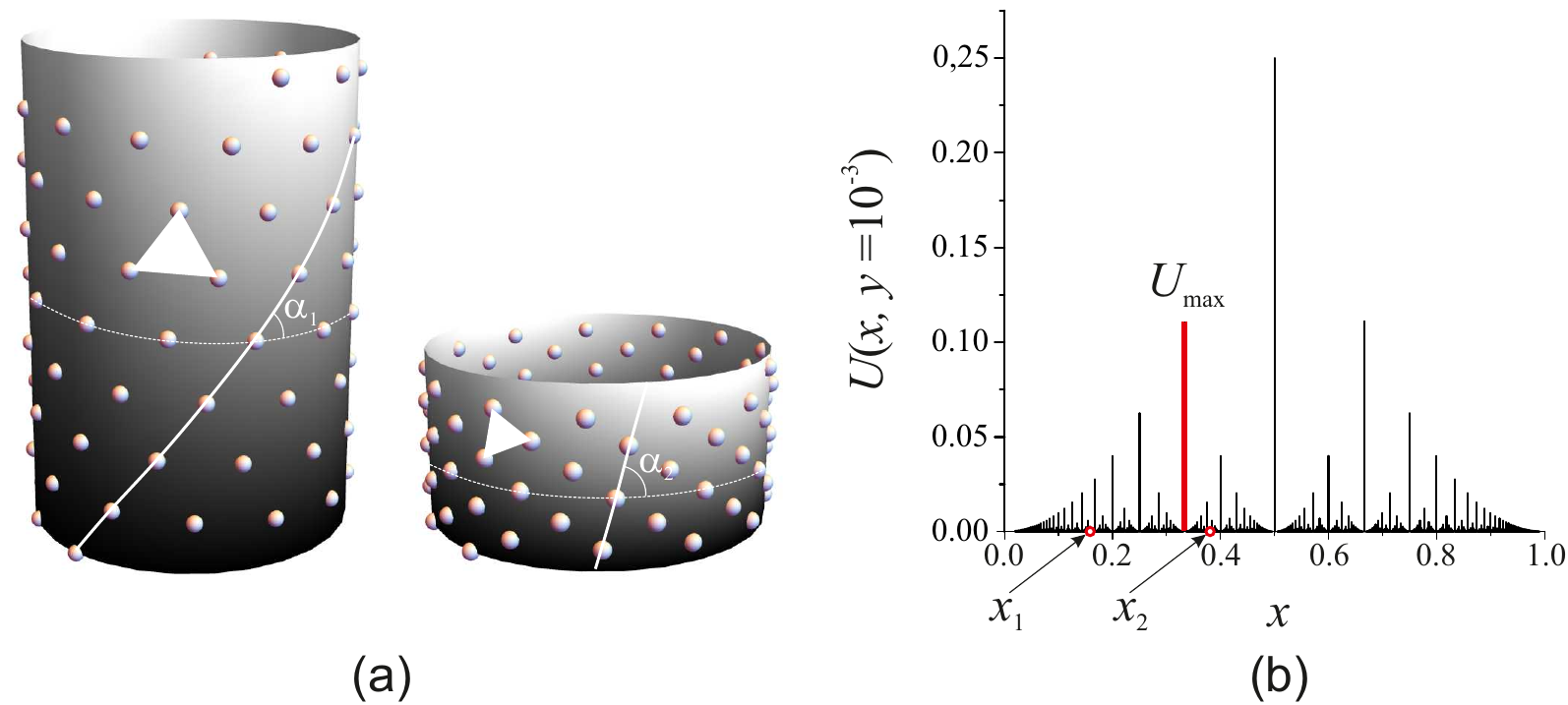}}
\caption{(a) Repulsive points distributed on the surface of the cylinder form a lattice, characterized by the parameter $\alpha$, with a minimal energy. The lattice is rearranged when the cylinder is compressed along its vertical axis; (b) Dependence $U(x, y={\rm const})$ defined in \eq{eq:044} for the compressed lattice ($y=0.001 \ll 1$ and $\beta=1$) as a function of $x$.}
\label{fig:008}
\end{figure}

Here we consider, following L. Levitov, the model system of $N$ strongly repulsive particles disposed and equilibrated on the surface of a cylinder of fixed diameter, $D$, and height, $H$ and look at the rearrangement of these particles when the cylinder is compressed along its height under the condition that $N$ and $D$ remain unchanged -- see \fig{fig:008}a. At the continuous compression, for each height, particles form a triangular ``Abrikosov'' lattice with minimal energy \cite{abrikosov}. Various lattice topologies parametrized by the modular parameter, $\tau=D+iH$, represent the valleys separated by energetic barriers on the manifold $\Gamma$ with the non-archimedean \emph{ultrametric} structure \cite{ultra1}.

The notion of ultrametricity deals with the concept of hierarchical organization of energy landscapes \cite{mez}. A complex system is assumed to have a large number of metastable states corresponding to local minima in a complex potential energy landscape. These minima are clustered in hierarchically nested basins: larger basins consist of smaller basins, each of these consists of even smaller ones, \emph{etc}. The basins of local energy minima are separated by a hierarchically arranged set of barriers: large basins are separated by high barriers, and smaller basins within each larger one are separated by lower barriers. Ultrametric geometry fixes taxonomic (i.e. hierarchical) tree-like relationships between elements and, speaking figuratively, is much closer to Lobachevsky geometry, rather to the Euclidean one.

We provide an explicit construction of the energetic relief in a phase space of all possible patterns of compressed lattices for symmetric and asymmetric interaction potentials and demonstrate that the ground state is related to the deepest valley in $\Gamma$ constructed via the Dedekind $\eta$-function. The lattice rearrangement caused by the compression of the cylinder along its axis is associated with the adiabatic flow along the geodesic in the energetic relief $\Gamma$, which can be understood as an RG flow. 

At each height, $H$, particles on the cylindric surface form a lattice with a minimal energy. Different lattice topologies, parameterized by the modular parameter $\tau=D+iH$, have valleys separated by barriers on the manifold $\Gamma$ with the non-archimedean \emph{ultrametric} structure. For strongly compressed lattices ($y \ll 1$) the energy $U(x,y)$ as a function of $x$ has a peak at every rational point, $x = \frac{m}{n}$ as it is shown in \fig{fig:008}b. One sees that $U(x,y)$ shares the hierarchical (ultrametric) behavior, which should be understood as follows: the transitions between two arbitrary local minima at $x_1$ and $x_2$, are determined by the passage over the highest barrier $U_{\rm max}(x_1,x_2)$, separating the points $x_1$ and $x_2$.

It is instructive to remind the definition of the ultrametric space. The ultrametric space, ${\cal M}$ is a set of elements supplied with the \emph{metric}, i.e. the pairwise distance, $d(x_1,x_2)$ between elements $x_1$ and $x_2$ which meets three requirements:
\begin{enumerate}
\item[(i)] non-negativity, $d(x_1,x_2)>0$ for $x_1\neq x_2$, and $d(x_1,x_2)=0$ for $x_1=x_2$,
\item[(ii)] symmetry, $d(x_1,x_2)=d(x_2,x_1)$,
\item[(iii)] the \emph{strong triangle inequality}, $d(x_1,x_2) \le \max\{d(x_1,x_3), d(x_3,x_2)\}$ (instead of the ordinary triangle inequality $d(x_1,x_2)\le d(x_1,x_3) + d(x_3,x_2)$ in the Euclidean metric space).
\end{enumerate}
The ultrametric organization of the energy relief means that we identify the energy with the metric. Namely considering $U(x,y={\rm const})$ as a function of $x$, we may set $d(x_1,x_2)=U_{\rm max}(x_1,x_2)$ as it is depicted in \fig{fig:008}b.

This Section is organized as follows. We begin with the derivation of the symmetric potential $U(x,y)$ separating valleys between different equilibrium configurations of particles on the cylinder when the cylinder is compressed along its height, $H$, under the condition that $N$ and $D$ remain unchanged -- see \fig{fig:008}. The corresponding analytic expression for potential barriers separating the valleys matches the generalized Riemann-Thomae function $\mathfrak{g}(x)$ defined in \eq{eq:003} for $h(n)=n^{-2}$. We consider the RG flow of the minimum of the potential $U(x,y)$ when $y$ is tending to 0 (i.e. the lattice is strongly compressed) and propose the topological interpretation of the corresponding flow in terms of the diffusion of a particle in the triangular (equal-sided) lattice of obstacles tessellating the Euclidean plane. Finally, we propose the generalization of the model to non-symmetric potentials acting between particles on the cylinder and show that the corresponding RG flow might differ from the Fibonacci sequence which has a Golden ratio $\frac{1}{2}\left(\sqrt{5}-1\right)\approx 0.618$ as a fixed point. Specifically, we demonstrate that for some non-symmetric potentials which have different strengths along the cylinder axis and along its circumference, we find a set of ``metallic fixed points'' expressed in terms of the so-called ``metallic ratios'' among which the so-called ``Silver ratio'' \cite{silver} is one of the known representatives. The basin of attraction of a ``Silver ratio'' is lower than that of a Golden ratio, which could be a reason why the Golden ratio is distributed in nature much wider than the Silver ratio.

\subsection{Construction of the potential and RG flow}

\begin{figure}[ht]
\centerline{\includegraphics[width=12cm]{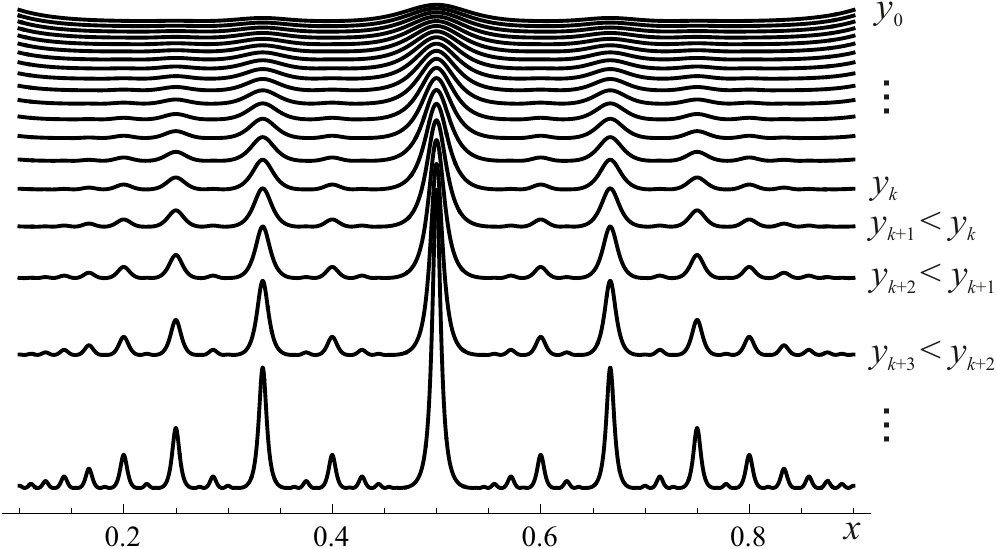}}
\caption{Set of plots of $U(x|y)$ on $x$ taken at different values of $y$ in the region $0.003<y<0.05$. For better view the curves for different $y$ are shifted in the vertical direction. As smaller $y$ as more generations of peaks are proliferated.}
\label{fig:set}
\end{figure}

Any particle on the cylinder can be parameterized by a pair $(z_n,\alpha_n\, \{\mathrm{mod}\, 2\pi\})$, where $n\in \mathbb{N}$, subject that all particles are arranged according to monotonic growth of $z_n$. Projecting the cylindrical surface conformally onto the plane, we get new coordinates, $\mathbf{r}_{n,m}(x,y)$, of particles on the planar lattice,
\be
\mathbf{r}_{n,m}(x,y) = \left(\frac{m + n x}{\sqrt{y}},\, n\sqrt{y} \right), \quad \{m,n\} \in \mathbb{Z}^2
\label{eq:039}
\ee
where the connection between cylindrical and planar lattices is set by the following change of variables:
\be
x=\frac{\alpha}{2\pi}, \quad y=\frac{h}{2\pi} \quad (y>0)
\label{eq:040}
\ee
Strong repulsive potential acting between particles can be approximated by the conformally-invariant $1/r^2$ potential. Consider two arbitrary particles one of which is located at the origin of the $(x,y)$-plane and the second -- at some point $(x_{m,n}, y_{m,n})$. Suppose that the potential $U({\bf r}_{m,n})$ has the following form:
\be
U({\bf r}_{m,n}) = \frac{q}{{\bf r}^2_{m,n}}
\label{eq:041}
\ee
where $q>0$ is some arbitrary parameter having sense of a charge. The energy of a whole lattice reads
\be
U(x,y) = \sum_{\{m,n\} \in \mathbb{Z}^2 \backslash \{0, 0\}} U(x_{m,n},y_{m,n}) = \sum_{\{m,n\} \in \mathbb{Z}^2 \backslash \{0, 0\}} \frac{q}{{\bf r}^2_{m,n}}
\label{eq:042}
\ee
Substituting \eq{eq:040} into \eq{eq:041}, we get:
\be
U(x,y) = \sum_{\{m,n\} \in \mathbb{Z}^2 \backslash \{0, 0\}} \frac{qy}{(m+n x)^2 + y^2 n^2} 
\label{eq:043}
\ee
Comparing \eq{eq:043} to \eq{eq:011} and \eq{eq:014}, we conclude that
\be
U(x,y) \approx q E(x+iy, s\to 1) \to 4\pi q \ln \left(y^{1/4}|\eta(x+iy)|\right) + \mathrm{const}
\label{eq:044}
\ee
where $E(x,s)$ is the non-holomorphic Eisenstein series (see \eq{eq:010}). Recall that here again we have exploited the 1st Kronecker limit formula, and dropped out the term divergent at $s\to 1$ (which is independent on $x,y$). The self-similarity of the function $U(x,y)$ is clearly seen in \fig{fig:set} where we have plotted a set of curves $U(x|y)\equiv U(x,y)$ taken at different values of $y$. For better visualization the curves at different $y$ ($0.003<y<0.05$) are shifted in the vertical direction as shown in \fig{fig:set}. 

\begin{figure}[ht]
\centerline{\includegraphics[width=16cm]{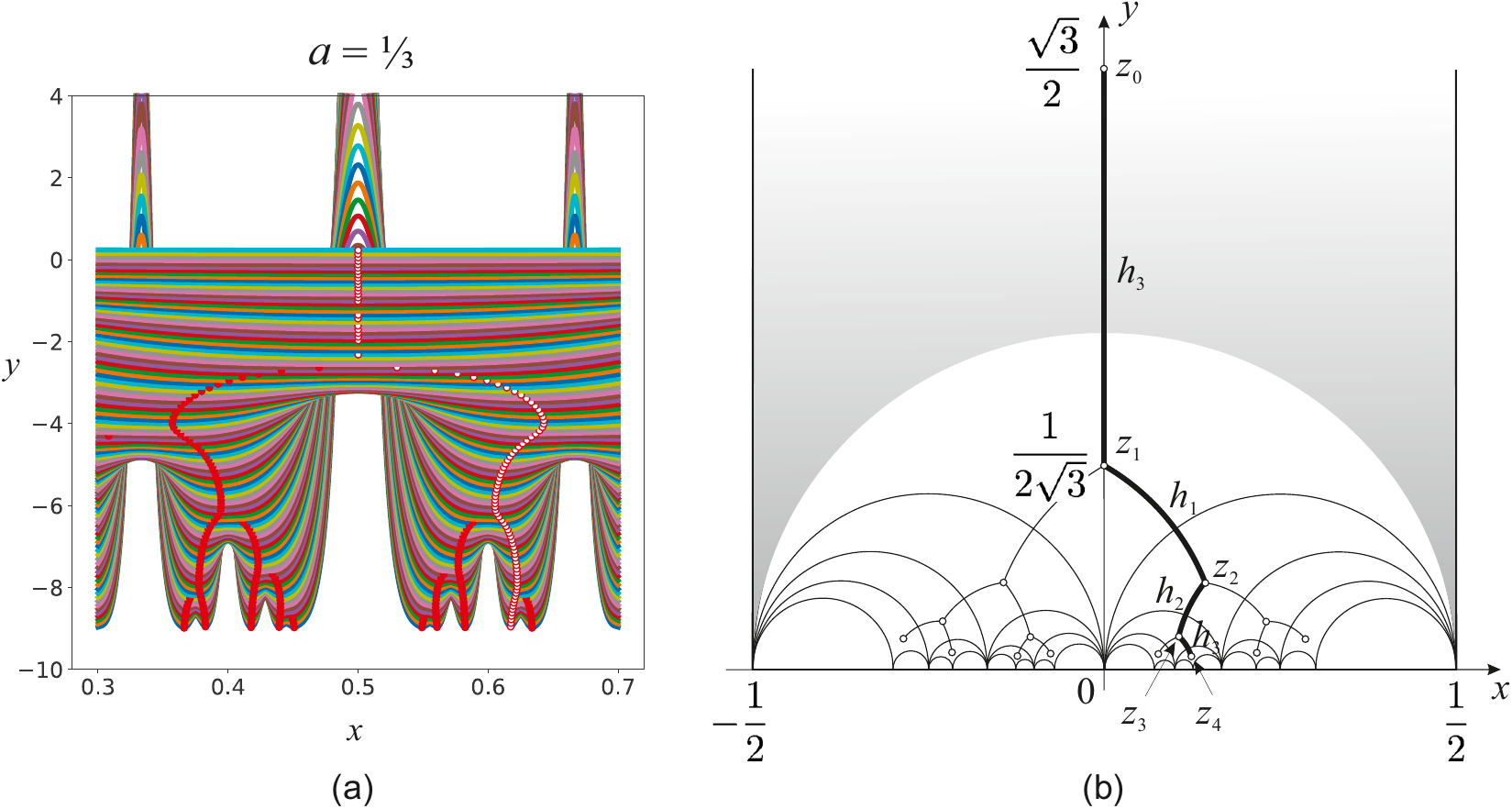}}
\caption{(a) Evolution of a minimum of a potential $U(x)$ when $y$ is continuously changing from $y_0=1.00$ towards 0. For better visualization each minimum at a given value of $y$ is marked by a white dot; (b) The fundamental domain of the modular group and few sequential reflections highlighted by the arcs of geodesics: $A_0 \to A_1 \to A_2 \to A_3 \to A_4...$, i.e. $(h_3 h_1 h_2)(h_3 h_1 h_2)(h_3 h_1 h_2)...$.}
\label{fig:009}
\end{figure}

Having the function $U(x|y)$ we may construct a trajectory that describes the continuous flow of the minimum of $U(x|y)$ as a function of $x$ (where $x\in[0,1]$) when $y$ is continuously changing from $+\infty$ down to 0. The corresponding flow is depicted in \fig{fig:009}a by a sequence of white dots for a family of plots $U(x|y_m)$ where $y_m = y_0 - c m$, and $m=0,1,2,...,M$. The parameters $c>0$ and $M$ are chosen such that $y_M>0$ (in our numeric computations $y_M=0.005$, $y_0=1.00$, $c=0.005$, $M=199$).

The flow of the minimum of the potential $U(x)$ when $y\to 0$, passes through the successive reflections of the fundamental domain of the free group $\Gamma_2$ is shown in \fig{fig:009}b. The corresponding Cayley graph is a 3-branching Cayley tree. Recall that the 3-branching Cayley tree can also be viewed as the Cayley graph of the group $\Lambda$, which has the free product structure: $\Lambda \sim \mathbb{Z}_2 \otimes \mathbb{Z}_2 \otimes \mathbb{Z}_2$, where $\mathbb{Z}_2$ is the cyclic group of second order. The matrix representation of generators $h_1, h_2, h_3$ of the group $\Lambda$ is well known: 
\be 
h_1 = \left(\begin{array}{cc} 1 & -1 \medskip \\ 0 & -1 \end{array}\right); \qquad 
h_2 = \left(\begin{array}{cc} 1 & 1 \medskip \\ 0 & -1 \end{array}\right); \qquad 
h_3 = \left(\begin{array}{cc} 0 & \tfrac{1}{2} \medskip \\ 2 & 0 \end{array}\right)
\label{eq:cyclic}
\ee
The optimal flow shown in \fig{fig:009}a by white dots corresponds to sequential reflections of the fundamental domain of the group $\Lambda$ depicted in \fig{fig:009}b by bold black arcs. Taking the point, $z_0=\tfrac{\sqrt{3}}{2}\,i$, we can find  its image, $z_N$, after $N$ recursive applications of generators from the set $\{h_1,h_2,h_3\}$ according to the following formula:
\be
z_N=\frac{1}{2}+\begin{cases} \disp \frac{a_N  \bar{z}_0 + b_N}{c_N\bar{z}_0 + d_N} & \mbox{for $N = 2k-1$, $k=1,2,...$} \medskip \\ 
\disp \frac{a_N  z_0 + b_N}{c_N z_0 + d_N} & \mbox{for $N = 2k$, $k=1,2,...$} 
\end{cases}
\label{eq:cyclic2}
\ee
where $\bar{z}$ means complex conjugation of $z$ and $\{a_N, b_N, c_N, d_N\}$ are the coefficients of the matrix 
\be
\left(\begin{array}{cc} a_N & b_N \\ c_N & d_N \end{array} \right) = \overbrace{h_3 h_2 h_1 h_3...}^{N~{\rm terms}} 
\label{eq:cyclic3}
\ee
Using \eq{eq:cyclic}-\eq{eq:cyclic3} we reproduce the coordinates of the points $A,B,C$ in \fig{fig:009}b. The sequence which converges to the Golden ratio is as follows:
\be
\left(\begin{array}{cc} a_{3M} & b_{3M} \\ c_{3M} & d_{3M} \end{array} \right) = \overbrace{\left(h_3 h_2 h_1\right)\left(h_3 h_2 h_1\right) ... \left(h_3 h_1 h_2\right)}^{N~{\rm terms}} = \left(h_3 h_2 h_1\right)^{3M}
\label{eq:image}
\ee
where $N=3M$, $M=1,2,3,...$.  The limiting value of $x_{\infty}=\re z_{N\to \infty}$ is the Golden ratio:
\be
x_{\infty} = \frac{1}{2} + \lim_{M\to\infty} \frac{a_{3M} c_{3M}+ b_{3M} d_{3M}}{c_{3M}^2+d_{3M}^2} = \frac{1}{2}\left(\sqrt{5}-1\right) \approx 0.618034...
\label{eq:18}
\ee
The sequence of ``zigzag'' reflections  is encoded in the continued fraction expansion of the Golden ratio, $\phi$:
\be 
\phi=\frac{1}{2}(\sqrt{5}-1)=\cfrac{1}{1+\cfrac{1}{1+\cfrac{1}{1+\cfrac{1}{1+\cdots}}}}
\label{eq:zigzag}
\ee
where interlacing odd and even ``1'' correspond to the left and right turns of a zigzag path.

\subsection{Topological view at the potential $U(x,y)$}

\begin{figure}[ht]
\centerline{\includegraphics[width=15cm]{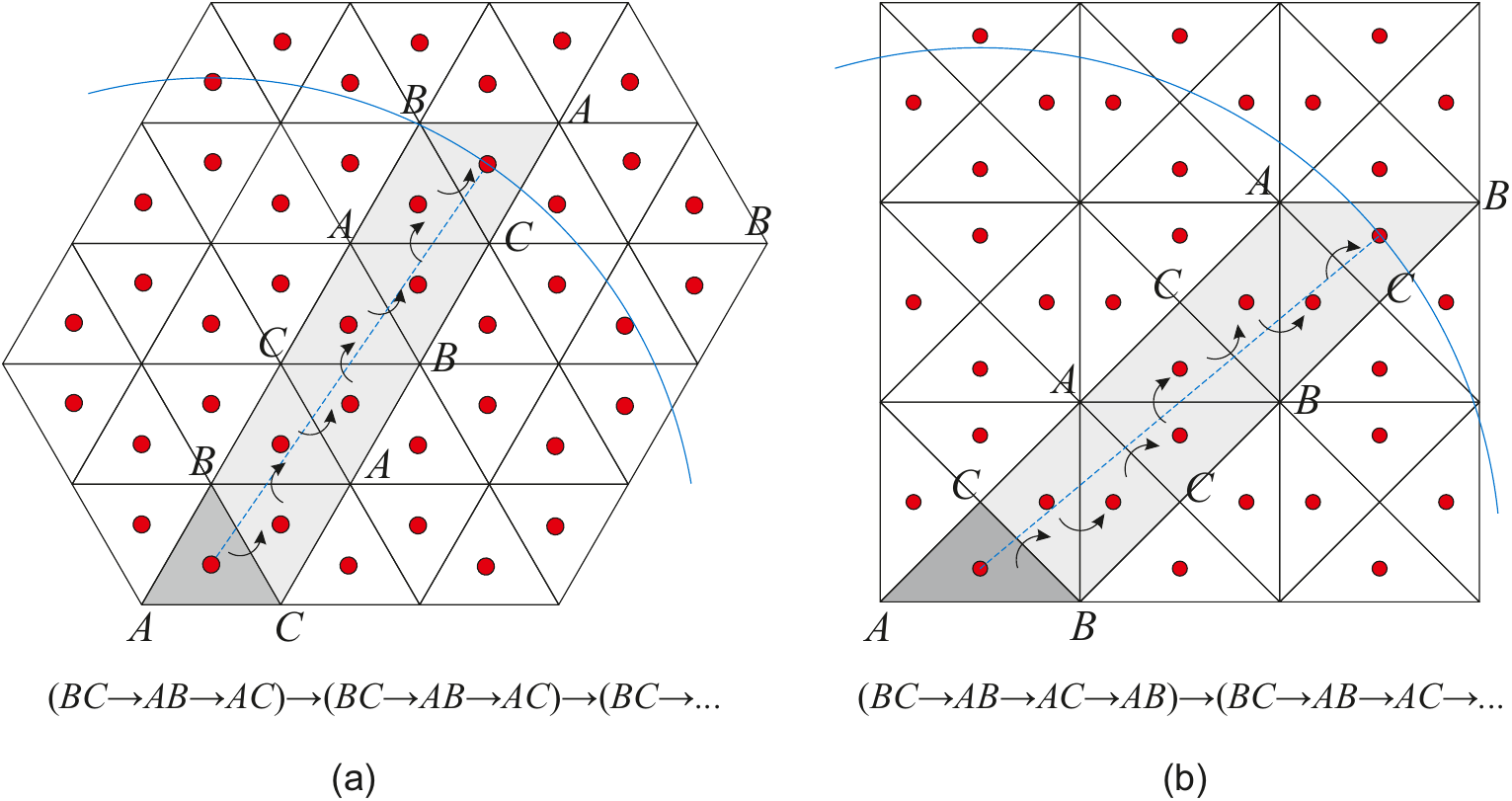}}
\caption{Tessellation of an Euclidean plane by reflections of an elementary triangle $ABC$ with respect to its sides: (a) $ABC$ is equal-sided triangle with angles $(\frac{\pi}{3},\frac{\pi}{3},\frac{\pi}{3})$; (b) $ABC$ is isosceles triangle with angles $(\frac{\pi}{2},\frac{\pi}{4},\frac{\pi}{4})$.}
\label{fig:010}
\end{figure}

The construction of the optimal path in the relief constructed on the basis of the Dedekind $\eta$-function allows for geometric interpretation. Take the triangle $ABC$ centered at $w_0$ in the complex plane $w$ with the Euclidean metric. Tessellate the plane $w$ by images of $ABC$ obtained by reflections of this triangle with respect to its sides. Consider two patterns of the elementary cell: (a) the triangle with angles $(\frac{\pi}{3},\frac{\pi}{3},\frac{\pi}{3})$, (b) the triangle with angles $(\frac{\pi}{2},\frac{\pi}{4},\frac{\pi}{4})$. The corresponding tessellations are schematically shown in \fig{fig:010} where red points designate the images of the point $w_0$ under reflections. Suppose that we have made $k$ reflections of the triangle $ABC$ with respect to its sides ($k=7$ in \fig{fig:010}). Let us address the following question: which sequence of $k$ successive reflections corresponds to the {\it maximal} Euclidean distance, $d(w_0,w_k)$, between the initial point $w_0$ and its image after $k$ reflections, $w_k$? The answer seems straightforward: $d(w_0,w_k)$ is the ``most aligned'' sequence which is encoded in the following set of reflections lying in a gray strip in \fig{fig:010}a,b:
\be
\left\{\begin{array}{ll}
(BC\to AB\to AC)\to (BC\to AB\to AC)\to (BC\to ... & \mbox{for the triangle $(\frac{\pi}{3},\frac{\pi}{3},\frac{\pi}{3})$} \medskip \\ (BC \to AB \to AC \to AB)\to (BC \to AB \to AC \to ... 
& \mbox{for the triangle $(\frac{\pi}{2},\frac{\pi}{4},\frac{\pi}{4})$} 
\end{array}
\right.
\label{eq:seq}
\ee

To see the connection of the longest Euclidean distance $d(w_0,w_k)$ in the plane $w$ covered by the triangular lattice, with the optimal flow in the modular domain, let us make the conformal mapping of the equal-sided triangle lying in $w=u+iv$ to the fundamental domain of the modular group -- the zero-angled triangle bounded by arcs (see \fig{fig:009}b). The corresponding mapping can be performed in two steps: (i) we map the triangle $ABC$ on $w$ with branching points at the corners onto the upper half plane of the complex plane $\zeta=\xi+i\chi$ with the branching points at $0,1, i\infty$, and then (ii) we map the upper half plane of $\zeta$ onto the fundamental domain of the modular group. Such a composite mapping has been described in  detail in \cite{nech1, conf}, so below we reproduce the final result only: 

\noindent (i) Conformal map $w \to \zeta$ is realized via the Kristoffel-Schwartz integral: 
\be 
w(\zeta) = \frac{2\pi}{\sqrt{2}\,\Gamma^3(1/3)}\int_0^{\zeta} {\xi}^{-2/3}(1-\xi)^{-2/3}d\xi
\label{eq:map1}
\ee 

\noindent (ii) Conformal mapping $\zeta \to z$ is realized via the $k^2(z)$ modular function: 
\be 
\zeta(z) = k^2(z) = \frac{\theta_2^4(0,e^{i\pi z})}{\theta_3^4(0,e^{i\pi z})}
\label{eq:map2}
\ee
where $\theta_i(0,q)$ ($i=1,...,4$) are the Jacobi elliptic $\theta$-functions. 

In what follows we will need only the Jacobian $J$ of the composite mapping $w(\zeta(z))$ which can be easily computed:
\begin{multline} 
J=\left|\frac{dw(z)}{dz}\right|^2=\left|\frac{dw(\zeta)}{d\zeta}\right|^2 \left|\frac{d\zeta(z)}{dz}\right|^2 = \\ \frac{4\pi^4}{3\Gamma^6(1/3)} \left|\theta_2(0,e^{i\pi z})\theta_3(0,e^{i\pi z})\theta_4(0,e^{i\pi z})\right|^{8/3} = \frac{ \pi^{4/3} 2^{14/3}}{3\Gamma^6(1/3)} |\eta(z)|^8
\label{eq:jacobian}
\end{multline}
where the following relations between Jacobi theta-functions have been used:
\be
\left\{\begin{array}{l}
\disp \frac{d}{dz}\ln \frac{\theta_2(0,e^{i\pi z})}{\theta_3(0,e^{i\pi z})} = i \frac{\pi}{4} \theta_4(0,e^{i\pi z}) \medskip \\ \disp  \theta_3^4(0,e^{i\pi z})-\theta_2^4(0,e^{i\pi z}) = \theta_4^4(0,e^{i\pi z}) \medskip \\ \disp \frac{d}{dz}\theta_1(0,e^{i\pi z}) = 2 \eta^3(z) \medskip \\
\disp \frac{d}{dz}\theta_1(0,e^{i\pi z}) = \pi \theta_2(0,e^{i\pi z})\theta_3(0,e^{i\pi z})\theta_4(0,e^{i\pi z})
\end{array}
\right.
\ee

Consider now the diffusion-like problem in the plane $w$ equipped with the triangular lattice of obstacles. The corresponding probability distribution of random paths, $P(w,t)$ obeys the parabolic equation:
\be 
\partial_t P(w,t) = D \partial^2_{w\bar{w}} P(w,t)
\label{eq:diff}
\ee
where $w=u+iv$ and $\bar{w}=u-iv$. To classify topological states of trajectories from the point of view of their entanglements with obstacles, it is instructive to pass to the covering space as it has been explained in \cite{nech1,nech-UFN}. Taking into account the conformal invariance of the Laplace operator and making use of the Laplace transform, $P(w,\lambda)=\int_0^{\infty} P(w,t)e^{-\lambda t}dt$, we perform the conformal mapping $w\to z$, and rewrite Eq.\eq{eq:diff} in a form of a stationary diffusion equation in $z$-plane in the effective ``potential'' $W(z)$:
\be
\lambda W(z) P(z,\lambda) = D \partial^2_{z\bar{z}} P(z,\lambda)
\label{eq:diff-conf}
\ee
where the potential $W(z)$ is defined by the Jacobian $J$ of the conformal mapping expressed again via the Dedekind $\eta$-function:
\be
W(z)=\lambda \left|\frac{dw(z)}{dz}\right|^2 = \lambda |\eta(z)|^8
\label{eq:conf-pot}
\ee

Eq. \eq{eq:conf-pot} shows that the effective potential emerging in the topological problem of diffusion in the array of obstacles has exactly the same structure of minima and maxima as the potential acting between repulsive particles located at the surface of the cylinder in the phyllotaxis problem. The topological meaning of coordinates $x$ and $y$ is as follows: $y$ describes the ``complexity'' of the entanglement, which in the polymer language is known as the ``length of the primitive path'' \cite{primitive1, primitive2} and has a meaning of the geodesic length in the covering space \cite{nech1, nech-UFN}, while $x$ describes the local winding of the path around obstacles. 

One can immediately see now that the ``most straight'' sequence \eq{eq:seq} for the tessellation of the plane by the equal-sided triangle $(\frac{\pi}{3},\frac{\pi}{3},\frac{\pi}{3})$ shown in \fig{fig:010}a exactly matches the ``zigzag'' paths coded by the cyclic sequence $\left(h_3 h_1 h_2\right)\left(h_3 h_1 h_2\right)\left(h_3 h_1 h_2\right)(h_3...$ of generators of the group $\Lambda$ in \fig{fig:009}. The extension to the asymmetric potential is considered in the next section.

\subsection{Optimal flows in asymmetric potentials}

\begin{figure}[ht]
\centerline{\includegraphics[width=12cm]{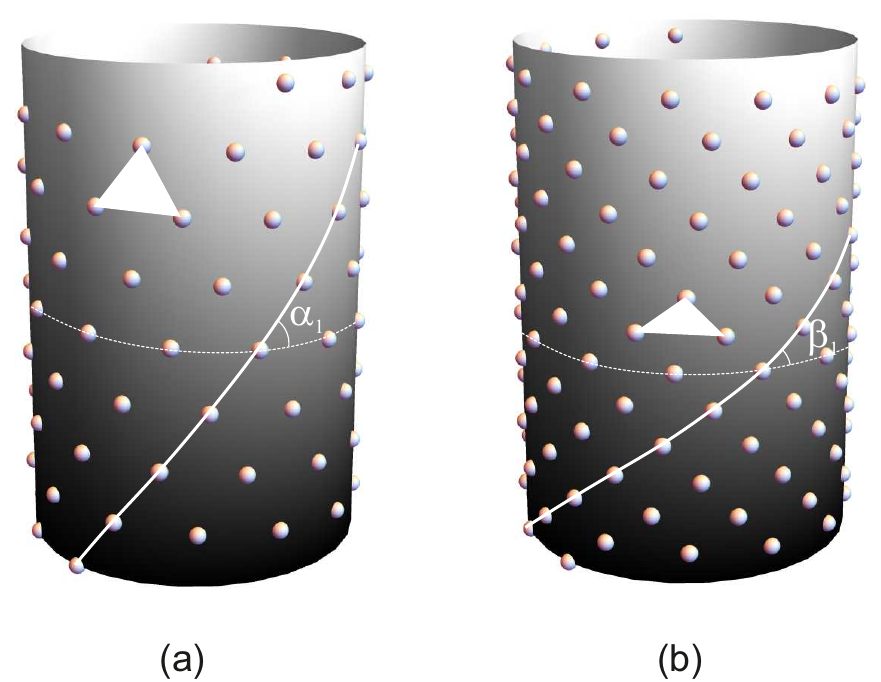}}
\caption{(a) Particles on the cylinder interacting with the symmetric potential form the equal-sided triangular ``Abrikosov lattice''; (b) Particles on the cylinder having stronger interaction along the circumference of the cylinder and weaker along its axis form the lattice with the elementary placket in a form of isosceles triangle.}
\label{fig:011}
\end{figure}

Let us return to the problem of finding an optimal lattice of repulsive particles on the cylinder. Suppose now that the potential acting between particles is asymmetric: it is stronger along the circumference of the cylinder and weaker along its axis. In that case, the equilibrium configuration of particles will not be anymore the Abrikosov triangular lattice depicted in \fig{fig:011}a, but rather the triangular lattice with the elementary placket in a form of an ``isosceles triangle'' as it is shown in \fig{fig:011}b.

The question which we address here is as follows. If we squeeze the cylinder along its axis, should the stable lattice patterns for particles interacting with the asymmetric potential follow again the Fibbonacci sequence, or the corresponding optimal flow will choose another minimal energy valley with a fixed point distinct from the Golden ratio? As we shall see, particular asymmetries of the potential force the flow to select an optimal path in the phase space distinct from the Fibonacci sequence. In considered examples of lattices with elementary triangles $(\frac{\pi}{2},\frac{\pi}{4},\frac{\pi}{4})$ and $(\frac{2\pi}{3},\frac{\pi}{6},\frac{\pi}{6})$ the flows follow so-called ``metallic ratios'' (in particular, ``Silver ratio'' for the triangle $(\frac{\pi}{2},\frac{\pi}{4},\frac{\pi}{4})$).

Taking into account the geometrical interpretation of the Jacobian of the conformal transform provided in the previous section, we conjecture that one can mimic the asymmetry in the interaction between particles on the cylinder by considering the {\it isosceles} triangle (instead of the equal-sided one) tessellating the plane $w$. The conformal mapping of the elementary cell in the form of an isosceles triangle with angles $(\alpha \pi,\beta \pi,\gamma \pi)=( a\pi, a\pi, (1-2a)\pi)$ to the fundamental triangle of the modular group can be constructed by a straightforward generalization of the conformal transform described by equations \eq{eq:map1}--\eq{eq:map2}. Namely, instead of \eq{eq:map1} one has
\begin{multline}
w(\zeta) = \frac{\pi}{\sin (\pi \gamma) \Gamma(\alpha) \Gamma(\beta) \Gamma(\gamma)} \int_{0}^{\zeta} \xi^{\alpha-1} (1-\xi)^{\beta-1} d\xi = \\ \frac{\pi}{\sin (\pi(1-2a)) \Gamma^2(a)\Gamma(1-2a)}\int_0^{\zeta} {\xi}^{a-1}(1-\xi)^{a-1} d\xi
\label{eq:map1a}
\end{multline}
where $\alpha = a$, $\beta = a, \gamma=1-2a$. Equation \eq{eq:map2} remains unchanged. Computing the Jacobian $J(z,a)$ of the composite conformal mapping, one gets
\be
J(z,a) = \left|\frac{dw(z)}{dz}\right|^2 = \frac{\pi^4}{\sin^2(2\pi a)\Gamma^4(a)\Gamma^2(1-2a)} \left|\theta_2^{a}(0,e^{i\pi z}) \theta_3^{1-2a}(0,e^{i\pi z}) \theta_4^{a}(0,e^{i\pi z})\right|^8
\label{eq:conf-a}
\ee
At $a=\frac{1}{3}$ we return to the expression \eq{eq:jacobian} for the Jacobian $J(z,\frac{1}{3})\equiv J$.

The isosceles triangle completely (without gaps and overlays) tessellates the plane by reflections with respect to its sides for values $a=\left\{\frac{1}{3};\, \frac{1}{4};\, \frac{1}{6}\right\}$ only. If we do not restrict ourselves by the condition to tessellate the plane by {\it isosceles} triangle, there is one with angles $(\frac{\pi}{3}, \frac{\pi}{6}, \frac{\pi}{2})$ which tessellates the Euclidean plane completely, however this case is not considered in our work because it does not correspond any physical choice of interaction potential acting between particles. All other triangles lead to an incomplete tessellation of the plane or to the tessellation with overlays.  

For $a=\frac{1}{4}$ and $a=\frac{1}{6}$ the potentials $J(z,a=\frac{1}{4})$ and $J(z,a=\frac{1}{6})$ have the following explicit expressions
\be
\begin{cases}
\disp J\left(z,\tfrac{1}{4}\right) = \frac{\pi^3}{\Gamma^4(1/4)} \theta_2^{2}(0,e^{i\pi z}) \theta_3^{4}(0,e^{i\pi z}) \theta_4^{2}(0,e^{i\pi z}) & \mbox{for $a=\frac{1}{4}$} \medskip \\
\disp J\left(z,\tfrac{1}{6}\right) = \frac{4\pi^4}{3\Gamma^4(1/6)\Gamma^2(2/3)} \theta_2^{4/3}(0,e^{i\pi z}) \theta_3^{16/3}(0,e^{i\pi z}) \theta_4^{4/3}(0,e^{i\pi z}) & \mbox{for $a=\frac{1}{6}$}
\end{cases}
\label{eq:conf-asymm}
\ee

\begin{figure}[ht]
\centerline{\includegraphics[width=15cm]{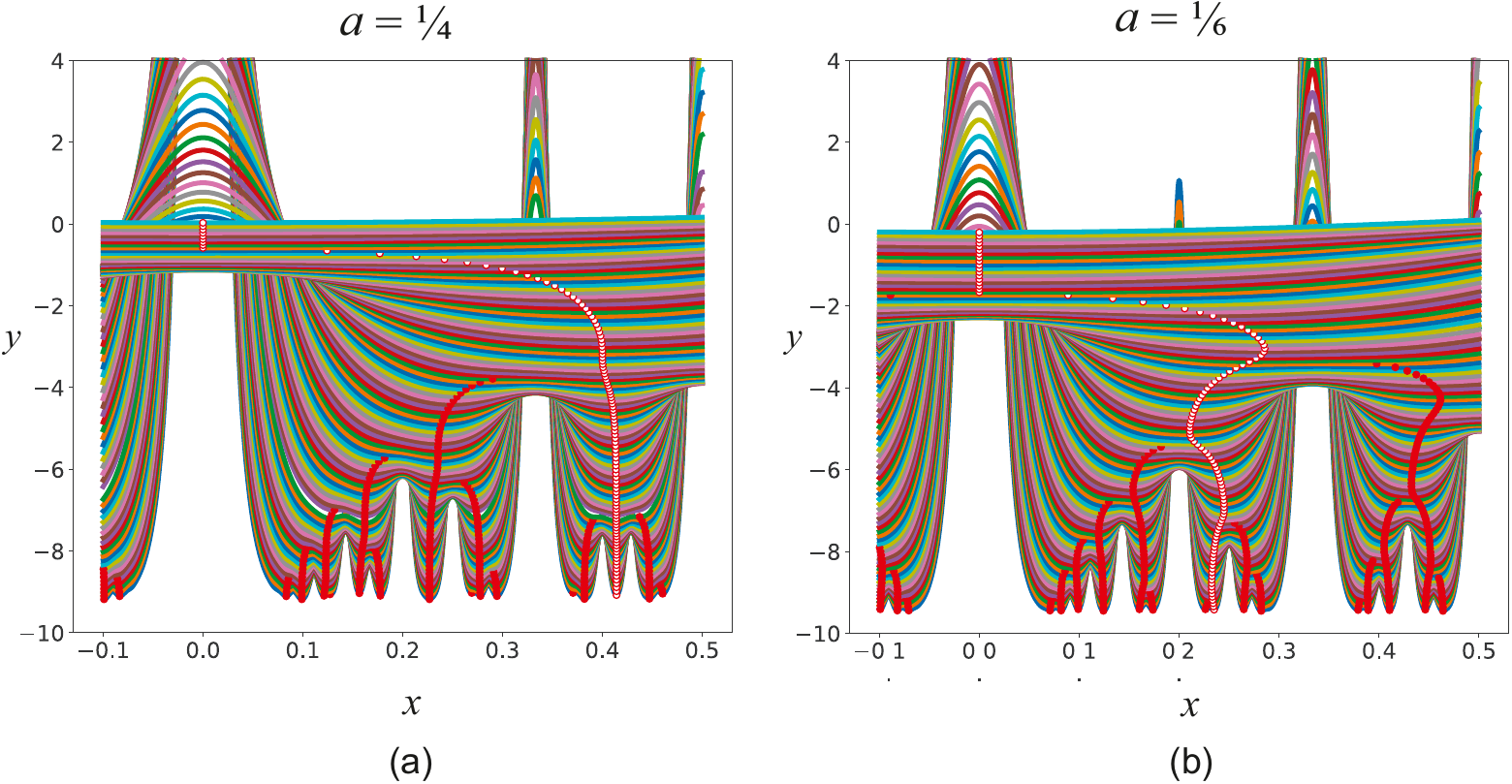}}
\caption{Evolution of a minimum of the potential $J(x|y,a)$ when $y$ is continuously changing from $y_0=1.00$ towards 0. For better visualization each minimum at a given value of $y$ is marked by a white dot; (a) the flow for $a=\frac{1}{4}$; (b) the flow for $a=\frac{1}{6}$.}
\label{fig:asymm}
\end{figure}

For triangles with $a=\frac{1}{4}$ and $a=\frac{1}{6}$ we construct the sets of sequential reflections of the fundamental domain of the modular group operating with generators $\{h_1, h_2, h_3\}$ as it has been done for the equal-sided triangle with $a=\frac{1}{3}$ (see \eq{eq:cyclic}--\eq{eq:image}). The flows of optimal paths and corresponding reflections are shown in \fig{fig:asymm}a,b.

Sets of reflections for $a=\frac{1}{4}$ and $a=\frac{1}{6}$ have the following explicit expression:
\be
\begin{cases}
\disp x_{\infty} = 2\lim_{M\to\infty} \frac{a_{4M} c_{4M}+ b_{4M} d_{4M}}{c_{4M}^2+d_{4M}^2} = \sqrt{2}-1 \approx 0.414214... & \mbox{for $a=\frac{1}{4}$} \medskip \\
\disp x_{\infty} = 2\lim_{M\to\infty} \frac{a_{3M} c_{3M}+ b_{3M} d_{3M}}{c_{3M}^2+d_{3M}^2} = \sqrt{5}-2 \approx 0.236068... & \mbox{for $a=\frac{1}{6}$}
\end{cases}
\label{eq:asymm}
\ee
where 
\be
\left(\begin{array}{cc} a_{4M} & b_{4M} \\ c_{4M} & d_{4M} \end{array} \right) = \left(h_3 h_1 h_3 h_2\right)^{4M}; \qquad \left(\begin{array}{cc} a_{3M} & b_{3M} \\ c_{3M} & d_{3M} \end{array} \right) = \left(h_3 h_1 h_2\right)^{3M} 
\label{eq:prod-asymm}
\ee
and the generators $h_1,h_2,h_3$ are defined in \eq{eq:cyclic}.

One sees from \fig{fig:asymm}a that the ``optimal'' sequence of reflections of the triangle with $a=\frac{1}{4}$ (i.e. the triangle $(\frac{\pi}{2},\frac{\pi}{4},\frac{\pi}{4})$) has the repeating cycle $(h_3 h_1 h_3 h_2)$ encoded in the longest sequence of reflections $(BC \to AB \to AC \to AB)$ in \fig{fig:010} and leads to the attraction point with the irrationality $\sqrt{2}$. This permits us to suggest that besides the Fibonacci sequence, there is a class of asymmetric potentials producing another stable pattern with the continued fraction expansion of the so-called ``Silver ratio'', $Ag$, given by the Pell-Lucas sequence:
\be 
Ag=\sqrt{2}-1=\cfrac{1}{2+\cfrac{1}{2+\cfrac{1}{2+\cfrac{1}{2+\cdots}}}}
\label{eq:zigzag2}
\ee
where ``2'' corresponds to two left turns followed by two right turns of a zigzag path on a Cayley tree of the modular group. 

The optimal flow for $a=\frac{1}{6}$ in the potential given by $J(z,a=\frac{1}{6})$ (see \fig{fig:asymm}b) is encoded in the repeating cycle $(h_3 h_1 h_2)$ -- see \eq{eq:prod-asymm}. The continued fraction expansion of the limiting value $M=\sqrt{5}-2$, which belongs to the generic set of so-called ``Metallic ratios'' is as follows
\be 
M_4=\sqrt{5}-2=\cfrac{1}{4+\cfrac{1}{4+\cfrac{1}{4+\cfrac{1}{4+\cdots}}}}
\label{eq:zigzag3}
\ee
where ``4'' corresponds to four left turns followed by four right turns of a zigzag path on a Cayley tree of the modular group. However since $\phi=(M_4-1)/2$, the reflections of the triangle $(\frac{\pi}{6},\frac{\pi}{6},\frac{2\pi}{6})$ do not lead to the new irrationality of the limiting point. In a forthcoming work we plan to study the basins of attractions of the stable points corresponding to the Golden and Silver ratios. Our preliminary computations allow us to conjecture that the basin of attraction of $\phi$ is bigger than that of $Ag$, which permits us to conjecture that this is might be a reason of wider spreading of the Golden ratio in nature than that of the Silver ratio.

\subsection{RG flow in modular domain for symmetric potential in vicinity of bifurcation points and Berezinsky-Kosterlitz-Thouless-like transitions}

\begin{figure}[ht]
\centerline{\includegraphics[width=16cm]{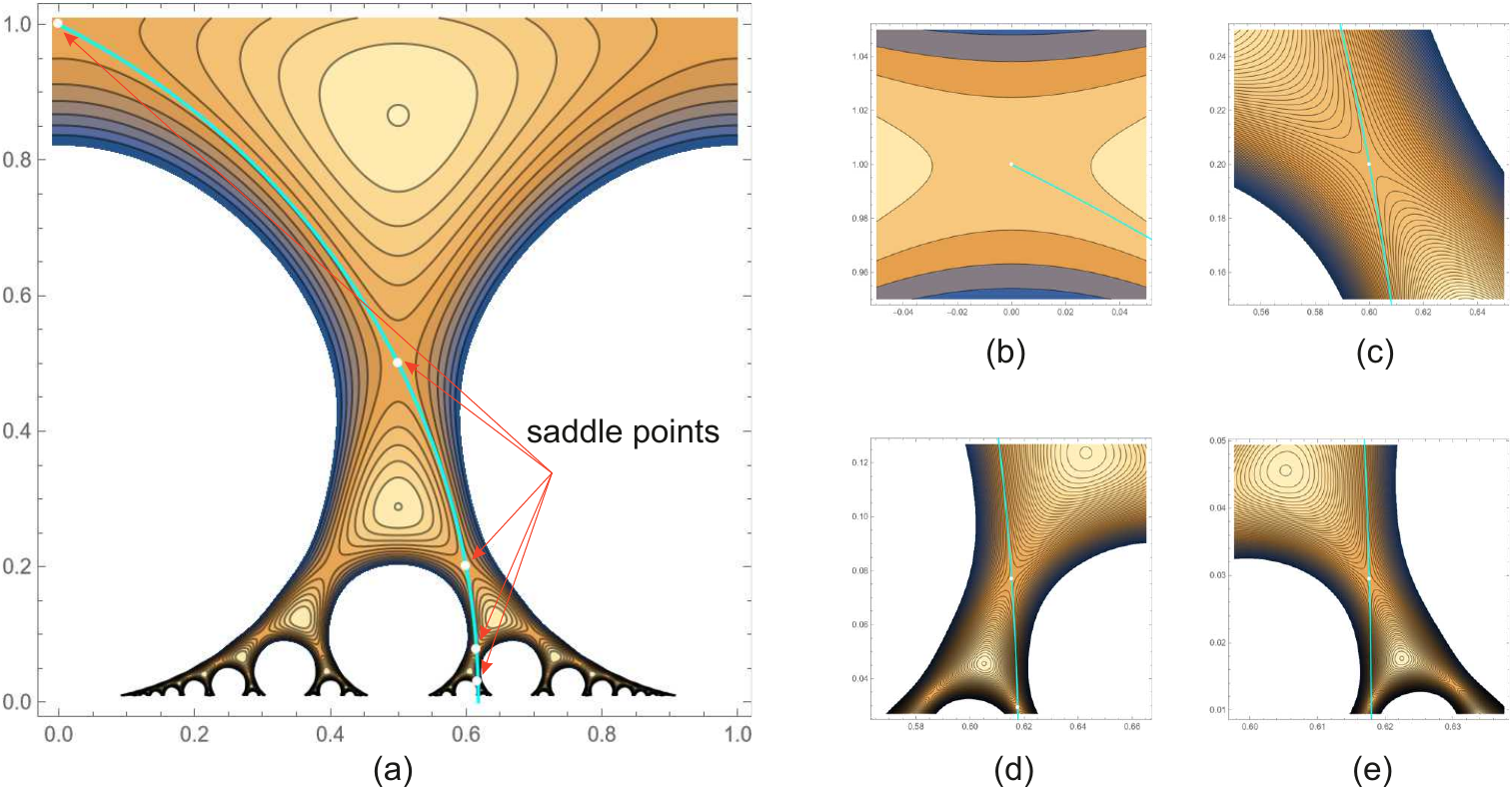}}
\caption{(a) Contour plot of the Riemann surface $U(x,y)=\ln \left(y^{1/4}|\eta(x+iy)|\right)$ in the region ($0.01<x<0.99$,\, $0.005<y<1$) within the bounding box $-0.27<U(x,y)<-0.25$. White points designate the bifurcation points of the RG flow, the cyan arc is the open geodesics passing through all saddles $(x_s(k),y_s(k))$ where $k=0,...,\infty$ -- see \eq{eq:saddles-all}; (b)--(e) are plot of the surface $U(u,v)$ in vicinity of four first saddles $(x_s(k),y_s(k))$ for $k=0,..,3$.}
\label{fig:saddle}
\end{figure}

Understanding RG flow as adiabatic particle's dynamics (APD) in a complex potential is very useful in studying the behavior of RG flows in the vicinity of critical points which can be regarded as bifurcation points in the APD problem. Here we derive the corresponding RG equation for the potential $U(x,y) = c \ln (y^{1/4} |\eta(x+iy)|)$ emerging throughout our study (see, for example \eq{eq:044}). The function $U(z)$, where $z=x+iy$, plays the role of a $\beta$-function which remains invariant under the action of the group $SL(2,Z)$, in particular when $y$ tends to 0. Recall, that in the phyllotaxis problem changing $y$ from $=+\infty$ down to 0 can be interpreted as the re-distribution of the system of repulsive particles (equilibrated at the surface of the cylinder) when the cylinder is squeezed along its axis. The contour plot of $U(x,y)$ for $c=1$ in the region $0.01<x<0.99$,\, $0.005<y<1$ within the bounding box $-0.27<U(x,y)<-0.25$ is shown in \fig{fig:saddle}a. The $(x_s,y_s)$ coordinates of saddle points have the generic expression:
\be
x_s = \frac{n_1 m_1 + n_2 m_2}{m_1^2 + m_2^2}; \qquad y_s = \frac{1}{m_1^2 + m_2^2}
\label{eq:saddle}
\ee
where $(m_1,m_2,n_1,n_2)$ are some integers. In particular, white dots in \fig{fig:saddle}a have the following  coordinates: $(\frac{1}{2}, \frac{1}{2}), (\frac{3}{5}, \frac{1}{5}), (\frac{8}{13}, \frac{1}{13}), (\frac{21}{34}, \frac{1}{34})$. From the topological point of view there is no difference between all these saddle points, however the orientation of saddles with respect to the $x$-axis is different and the geodesic (cyan line in \fig{fig:saddle}a) passes through different saddle points at different angles. The coordinates $(x_s,y_s)$ of saddles constituting the Fibonacci series are:
\be
\Big(x_s(k),y_s(k)\Big)= \left(\frac{G_1^{2k}-G_2^{2k}}{G_1^{2k+1}-G_2^{2k+1}}, \frac{\sqrt{5}}{G_1^{2k+1}-G_2^{2k+1}}\right); \qquad k=0,1,2,...,\infty
\label{eq:saddles-all}
\ee
where $G_1=\frac{1}{2}(1+\sqrt{5})$ and $G_2=\frac{1}{2}(1-\sqrt{5})$. 

To proceed, let us expand the potential $U(x,y)$ in the vicinity of some saddle point $(x_s,y_s)$ and find the explicit form of the corresponding surface $U(x,y)$ near $(x_s,y_s)$. The Taylor expansion of $U(x,y)$ up to the second order reads: 
\be
U(x-x_s, y-y_s) \approx U(x_s,y_s) + U_{xx}\,(x-x_s)^2+2U_{xy}\,(x-x_s)(y-y_s)+U_{yy}\,(y-y_s)^2
\label{eq:saddle-tay1}
\ee
where the derivatives $U_{xx}, U_{xy}=U_{yx}, U_{yy}$ are taken at the point $(x_s,y_s)$. The first derivatives in the Taylor expansion \eq{eq:saddle-tay1} are nullified at the point $(x_s,y_s)$, and the condition $U_{xx}U_{yy} - U_{xy}^2<0$ ensures that the point $(x_s,y_s)$ is actually a saddle. Let us define the coefficients $U_{xx}=a_1$, $U_{x,y}=a_2$, $U_{yy}=a_3$. Note that the coefficients $a_1=a_1(k)$, $a_2=a_2(k)$ and $a_3=a_3(k)$ depend on $k$, where $k=0,1,...,\infty$ is the corresponding member in the Fibonacci sequence. In \fig{fig:saddle}a we depict the relief $U(x,y)$ where the white dots are saddle points with the following $(x,y)$ coordinates: $(0, 1), (\frac{1}{2}, \frac{1}{2}), (\frac{3}{5}, \frac{1}{5}), (\frac{8}{13}, \frac{1}{13}), (\frac{21}{34}, \frac{1}{34})$. The arc is the open geodesics which is parameterized by the equation $y(x)=\sqrt{\frac{5}{4} - \left(x + \frac{1}{2}\right)^2}$.

Let us consider the RG flow in the complex $z=x+iy$ plane in vicinity of saddle points $(x_s,y_s)$ of the surface $U(x,y)$. Introducing the coordinates $u=x-x_s$ and $v=y-y_s$ and separating real and imaginary parts, we may write down the system of nonlinear first-order differential equations describing the RG flow in complex plane $w=u+iv$ in the vicinity of the point $(x_s,y_s)$:
\be
\begin{cases}
\disp \frac{du}{d\ln \mu} = a_1 u^2 - a_3 v^2 \medskip \\ 
\disp \frac{dv}{d\ln \mu} = 2 a_2\, u v
\end{cases}
\label{eq:RG1}
\ee
where $\mu$ is the RG time. 

Equations \eq{eq:RG1} imply that the RG flow near the bifurcation points is fully determined by the topology of the Riemann surface $U(x,y)$. It is worth mentioning that our construction is consistent with ideas expressed in works \cite{kaplan1,lutken,carpentier}. In particular, in \cite{kaplan1} the connection between the RG flows and the topological structure of $\beta$-function has been discussed in the context of CFT/ADS$_2$ duality, while in \cite{lutken} and in \cite{carpentier} the equations for RG flows ideologically similar to \eq{eq:RG1} have been derived to describe the behavior of RG flows in the FQHE in the vicinity of critical points. The emergence of BKT fixed points in similar context has been also studied in \cite{fisher} for layered high-$T_c$ superconductors.

Dividing the first equation of \eq{eq:RG1} by the second one we can convert the system \eq{eq:RG1} into the following single equation
\be
\frac{du}{dv} = \frac{a_1}{2a_2}\frac{u}{v} - \frac{a_3}{2a_2}\frac{v}{u}
\label{eq:RG2}
\ee
Introducing the new function $h$ and writing $u = h v$, we arrive at the equation in which the variables can be separated:
\be
v\frac{dh}{dv} = \left(\frac{a_1}{2a_2}-1\right)h - \frac{a_3}{2a_2}h^{-1}
\label{eq:RG3}
\ee
Solving \eq{eq:RG3} we get
\be
\frac{a_2}{a_1 - 2 a_2} \ln\left(a_3 - (a_1 - 2 a_2) h^2\right) = \ln (G v)
\label{eq:RG4}
\ee
where $G$ remains invariant along the RG flow (i.e. $G$ does not depend on the scale $\mu$). Plugging the function $h=u/v$ in \eq{eq:RG3} and denoting $G^{a_1/a_2-2}$ by $\Delta$, we have
\be
a_3v^2-v^{a_1/a_2} \Delta = (a_1-2a_2)u^2
\label{eq:RG5}
\ee
Substituting $u(v)$ into the second equation in \eq{eq:RG1} and performing the integration, we obtain an non-explicit solution for $v(\mu)$
\begin{multline}
\frac{\sqrt{a_1-2 a_2}}{a_1 \sqrt{a_3 v^2-\Delta  v^{\frac{a_1}{a_2}}}} \Bigg(\left(a_1-2 a_2\right) a_3 v^2 \sqrt{1-\frac{\Delta  v^{\frac{a_1}{a_2}-2}}{a_3}} \, _2F_1\left(\frac{1}{2},\frac{a_2}{a_1-2 a_2};\frac{a_1-a_2}{a_1-2 a_2};\frac{v^{\frac{a_1}{a_2}-2} \Delta }{a_3}\right)+ \\ 2 a_2 \left(a_3 v^2-\Delta  v^{\frac{a_1}{a_2}}\Bigg)\right) = \ln \mu
\label{eq:RG6}
\end{multline}

Despite \eq{eq:RG6} looks rather complicated, it is essentially simplified in the limit of small $u$. Computing explicitly the shape $U(u,v)$ (recall that $u=x-x_s$ and $y=y-y_s$) in vicinity of the saddle point $(x_s(k), y_s(k))$, we see that with $k\to\infty$ the coefficient $a_1=U_{uu}$ tends to zero, while the coefficients $a_2=U_{v}$ and $a_3=U_{vv}$ remain finite. To demonstrate this, we have depicted in \fig{fig:saddle}(b)-(e) the potentials $U(x,y)$ in vicinity of four first terms of the Fibonacci series for $k=0,1,2,3$:
\be
\begin{cases}
U_0(u,v) = 0.768 + 0.029 u^2 & \mbox{for $x_s=0, y_s=1$}  \medskip \\
U_1(u,v) = 0.768 - 0.056 u^2 - 0.055 v^2 + 0.110 uv & \mbox{for $x_s=\tfrac{1}{2}, y_s=\tfrac{1}{2}$} \medskip \\
U_2(x,y) = 0.768 + 0.156 u^2 + 1.405 v^2 - 0.936 uv & \mbox{for $x_s=\tfrac{3}{5}, y_s=\tfrac{1}{5}$} \medskip \\
U_3(x,y) = 0.768 - 0.082 u^2 - 4.885 v^2 + 1.245 uv & \mbox{for $x_s=\tfrac{8}{13}, y_s=\tfrac{1}{13}$}
\end{cases}
\label{eq:potentials}
\ee
One sees from \eq{eq:potentials} that with increasing $k$ the coefficient $a_1(k)$ in front of the term $u^2$ relatively decreases.  Substituting $a_1=0$ (corresponding to $k\to \infty$) into \eq{eq:RG1} we get equations describing the RG flow in the $XY$-model in vicinity of the BKT transition. The critical scale (the correlation length) near the transition point is defined by the condition $\sqrt{-2a_2\Delta} \ln \mu_c \sim 1$ which implies the BKT dependence of the correlation length, $\mu_c$, on $\Delta$:
\be
\mu_c \sim e^{1/\sqrt{-2a_2 \Delta}}
\label{eq:bkt-corr}
\ee
One can see from \eq{eq:potentials} that the coefficient $a_2$ in front of the $uv$ term periodically changes the sign. So, one can expect the signature of the BKT-like transition \eq{eq:bkt-corr} when $a_2<0$. 

The physical meaning of encountered critical behavior could have the following interpretation. When the cylinder is squeezed along its principal axis, the corresponding lattice of repulsive particles experiences a set of successive  rearrangements (``bifurcations''). Each bifurcation is a collective effect that is accompanied by the melting of the lattice. Our analysis permits us to conjecture that some of these bifurcations in the strong compression limit have signatures of Berezinsky-Kosterliz-Thouless (BKT) transtions. 

\section{Riemann-Thomae function, Devil's staircase and long-range 1D lattice models on a ring}

Consider a one-dimensional system of $n$ particles positioned on a ring of $N$ sites. Particles interact via the repulsive pairwise long-range potential, $V(|i-j|)$, which depends on the distance between particles along the ring. In addition, there is an external field, $h$, acting on all particles. The corresponding Hamiltonian, $H$, reads
\be
H\{n_1,...,n_N\} = -h\sum_{i=1}^N n_i + \frac{1}{2}\sum_{i=1}^N\sum_{j\neq i}V(|i-j|) n_i n_j,
\label{eq:045}
\ee
where $n_i$ is the indicator function of a particle at a site $i$, i.e. $n_i=1$ if the particle is present at the site $i$, and $n_i=0$ if the site $i$ is empty. The Hamiltonian \eq{eq:045} corresponds to the discrete Coulomb gas on a ring considered in \cite{gaudin}.

The exact structure of the ground state of such a system for a fixed number of particles has been discovered independently by Hubbard \cite{Hubbard} and Pokrovsky and Uimin \cite{Pokrovsky}. Later, in \cite{Bak, Bak2} it has been demonstrated that ground states for a model with $h\neq 0$ form a complete Devil's staircase structure. Some mathematical aspects of the appearance of the Devil's staircase have been discussed in \cite{sinai}. The recursive algorithm for constructing the corresponding staircase is fairly simple and can be described as follows.
\begin{itemize}
\item First, we pick up a filling density $\rho_0 = 1/N$. This choice fixes the smallest step in density that we can detect. We also define, for completeness, $h_-(0)=0$.
\item On the next step we recursively determine $h_+(\rho_1)$ by the following equations
\be
\begin{cases}
h_+(\rho_k) = h_-(\rho_k) + \Delta h(\rho_k) \medskip \\
\rho_{k+1}  = \rho_k + \frac{1}{N} \medskip \\
h_-(\rho_{k+1})  = h_+(\rho_{k+1})
\end{cases}
\label{eq:046}
\ee
where the value $\Delta h$ is set by the sum
\be
\Delta h = 2\sum_{k=1}^{\infty}k N\Big(V(k N+1)+V(k N-1)-2V(k N)\Big).
\label{eq:047}
\ee
\item The algorithm runs $T$ times ($k=1,2,...,T$) until the requested density $\rho_T\equiv\rho$ is reached.
\end{itemize}
Two remarks concerning the above algorithm should be made. First of all, the potential $V$ should be convex. Secondly, we need to be careful to always write the running density $\rho_i$ as an irreducible fraction (the nominator and denominator should not have any common divisors). The denominator of this irreducible fraction is used to compute $\Delta h$ in \eq{eq:047}.

\begin{figure}[ht]
\centerline{\includegraphics[width=15cm]{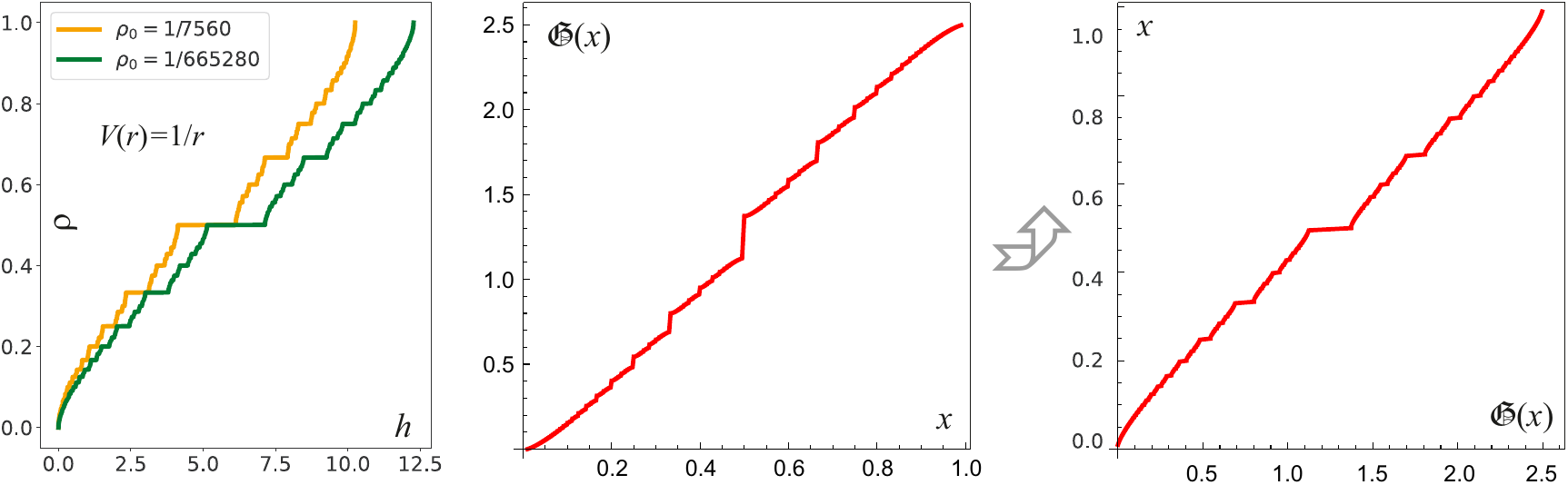}}
\caption{(a--b) Plot $\rho(h)$ for the potentials $V(r)=1/r$; (b--c) Integrated generalized Riemann-Thomae function $\mathfrak{G}(x)=G_2(x)$ defined in \eq{eq:019} in ordinary coordinates (b), and its reflected-rotated image (c).}
\label{fig:013}
\end{figure}

The densities $\rho(h)$ of systems of particles minimizing the energy described by the Hamiltonian \eq{eq:045} are computed by the described algorithm for two potentials, $V(r)=1/r$ and $V(r)=1/r^2$ on the lattice of $N=7560$ and $N=665280$ sites. The corresponding plots of Devil's staircases are depicted in \fig{fig:013}. It should be noted that the number of steps (plateaus) in staircases depends on the factorization of $N$. To see as many steps as possible, $N$ should be a ``superior highly composite number'' \cite{composite}. To the contrary, if $N$ is prime, no Devil's staircase structure emerges. From a physical point of view, jumps between plateaus are the phase transitions between the states minimizing the system energy at a varying external field, $h$, and the density $\rho(h)$ plays a role of an order parameter.

Comparing functions $\rho(h)$ in \fig{fig:010} and $\mathfrak{G}(x)=G_2(x)$ in \fig{fig:003}, it is eligible to ask the following questions:
\begin{itemize}
\item[(i)] Whether there is a coincidence of the function $h(\rho)$ (horizontally reflected) for some potential $V(r)=1/r^{\gamma}$ and the function $G_2(x)$ (subject to affine deformations);
\item[(ii)] If such a coincidence actually exists for some values of $\gamma$, what could be the physics behind it?
\end{itemize}

In \eq{eq:003} we have defined the generalized Riemann-Thomae (gRT) function $g_{\alpha}(x)$. Figures \fig{fig:002}a,b provide sample plots of $g_{\alpha}(x)$ for two values, $\alpha=0.41$ and $\alpha=2.76$ (for $n=100$). Let us consider now the function $g_{\alpha,\beta}(x)$ which extends the definition of $g_{\alpha}(x)$ as follows:
\be
g_{\alpha,\beta}(x) = \begin{cases} \disp \frac{1}{n^{\alpha}} + \frac{A}{n^{\beta}} & \mbox{if $x=\frac{m}{n}$, and $(m,n)$ coprime} \medskip \\
0 & \mbox{if $x$ is irrational} \end{cases}
\label{eq:048}
\ee
where $A$ is some scaling factor and $\alpha$ and $\beta$ are exponents such that $\alpha<\beta$.

\begin{figure}[ht]
\centerline{\includegraphics[width=16cm]{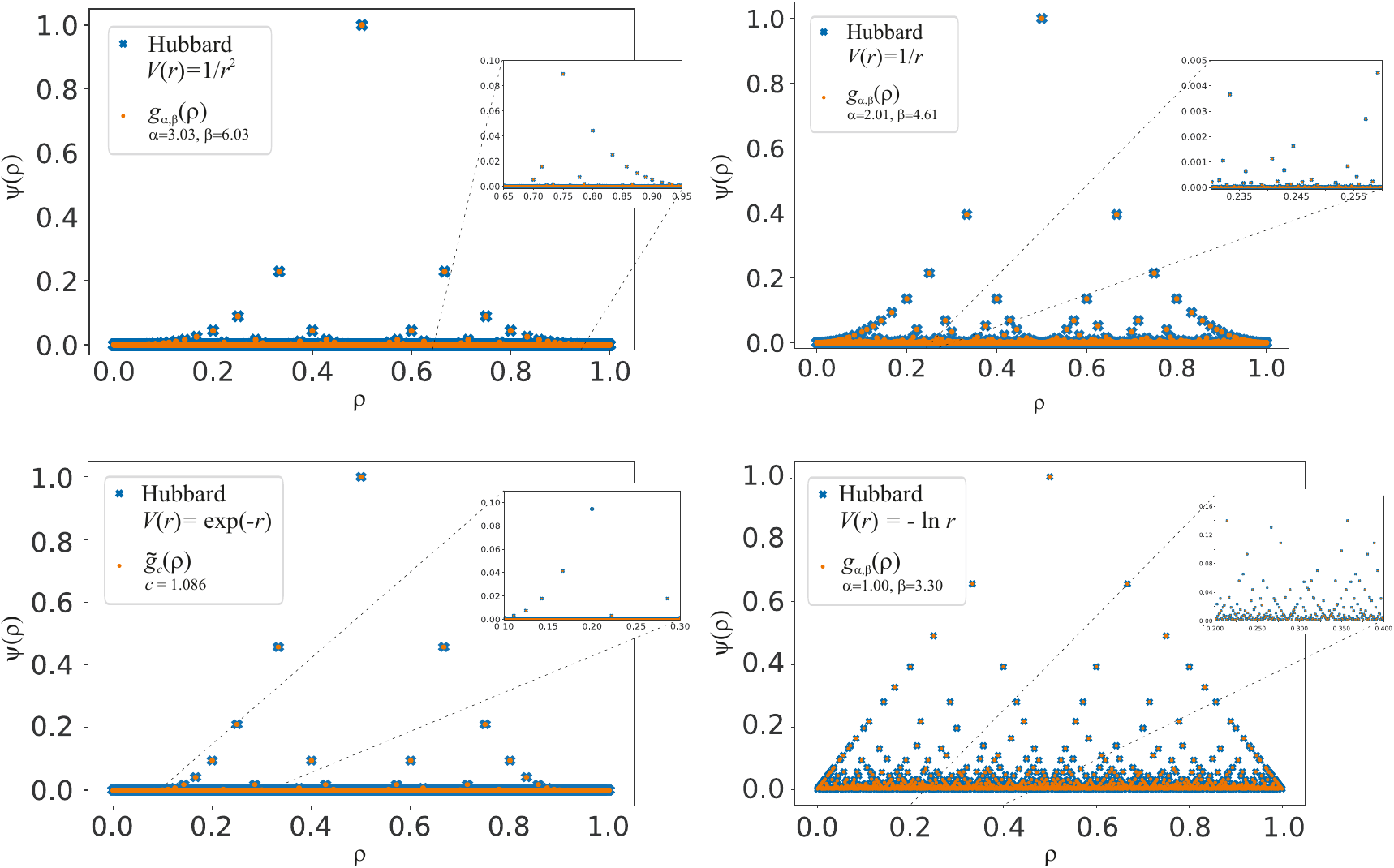}}
\caption{Comparison of the function $\psi(\rho)=\frac{dh(\rho)}{d\rho}$ with the function $g_{\alpha, \beta}(\rho)$ for three different potentials of the Hubbard model on a ring: (a) $V(r)=1/r^2$; (b) $V(r)=1/r$; (c) $V(r) = \exp(-r)$; (d) $V(r)=-\ln r$.}
\label{fig:014}
\end{figure}

Since the Devil's staircases shown in \fig{fig:010}a,b consists of two symmetric branches, in what follows, we will consider only one of them. Also, in view of further comparison with the Riemann-Thomae function $g(x)$, it is convenient to work with the derivative of the staircase. So, we consider the derivative 
\be
\psi(\rho)=\frac{d h(\rho)}{d\rho}
\label{eq:049}
\ee
of the ``inverted Devil's staircase'' $h(\rho)$. The question which we address is as follows: is it possible to find such values $\alpha$ and $\beta$ in the definition of the function $g_{\alpha,\beta}(x)$ (see \eq{eq:048}) that one can match $\psi(\rho)$ for any potential $V(r)$? The answer is positive and below we demonstrate an excellent agreement of the generalized Riemann-Thomae function $g_{\alpha,\beta}(x)$ with the function $\psi(\rho)$ for algebraically decaying potentials, $V(r)=1/r^{\gamma}$, where $\gamma=1,2,1/2$, as well as for $V(r) = \exp(-r)$ and $V(r)=-\ln r$. Note that since $x$ in the definition \eq{eq:001} lies on the segment $[0,1]$ and the density $\rho$ changes, by definition, within the interval $[0,1]$, we can identify $x$ with $\rho$. The plots for the potentials $V(r)=\left\{1/r^2; 1/r; \exp(-r); -\ln r \right\}$ are shown in \fig{fig:011}. The inserts demonstrate the magnification of the region near the horizontal axis. The parameters $\alpha$, $\beta$ and $A$ for various algebraically descending potentials defined in \eq{eq:048} are presented in Table \ref{tab:01} (to save the space in \fig{fig:014} we have not shown the plot for the potential $V(r) = 1/\sqrt{r}$).

\begin{table}[ht]
\begin{tabular}{lllll}
\hline \hline
& \hspace{0.5cm} $1/r^2$  & \hspace{1cm} $1/r$ & \hspace{1cm} $1/\sqrt{r}$ \hspace{0.6cm} $-\ln r$ \\ \hline
$\alpha$ & \hspace{0.5cm} 3.03 & \hspace{1cm} 2.01  & \hspace{1cm} 1.50 \hspace{1cm} 1.00 \\ \hline
$\beta$ & \hspace{0.5cm} 6.03 & \hspace{1cm} 4.61   & \hspace{1cm} 3.93 \hspace{1cm} 3.30 \\ \hline
$A$    & \hspace{0.5cm} 3.57 & \hspace{1cm} 1.21  & \hspace{1cm} 0.51 \hspace{1cm} 8.90 \\ \hline 
\end{tabular}
\caption{Values of optimal parameters for the generalized Riemann-Thomae function $g_{\alpha,\beta}(x)$ (see \eq{eq:048}) for three different potentials $\{1/r, 1/r^2, 1/\sqrt{r}, -\ln r\}$ (except $\exp(-r)$). In all cases we have set $\rho_0=1/7560$.}
\label{tab:01}
\end{table}

From the Table \ref{tab:01} we can conjecture the following relation between $\alpha$ in the definition of the generalized Riemann-Thomae function $g_{\alpha, \beta}$ and the exponent $\gamma$ in the definition of the algebraic potential $V(r)=1/r^{\gamma}$:
\be
\alpha = 1+\gamma
\label{eq:050}
\ee
The exponent $\beta$ is apparently non-universal and describes the finite-size corrections to the leading behavior of $g_{\alpha,\beta}(\rho)$ at small densities (i.e. when $\rho\to 0$).

The equation \eq{eq:050} is consistent with the expressions approximating the enveloping shapes of generalized Riemann-Thomae functions for a class of power-law and logarithmic potentials at $q\to\infty$ (see \eq{eq:047}):
\be
\Delta h(q) = \psi_{env}(q) = \begin{cases} 
q^{-(\gamma+1)} & \mbox{for $V(r)=r^{-\gamma}$, where $\gamma = \{1/2,\, 1,\, 2\}$ } \medskip \\ 
q^{-1} & \mbox{for $V(r)=-\ln(r)$} \medskip \\
q \exp(-q) & \mbox{for $V(r)=\exp(-r)$}
\end{cases}
\label{eq:051}
\ee 
This remark was pointed us by M. Gherardi \cite{marco}.

\subsection{Universality of the numerical parameters $\alpha$, $\beta$}

To find the values of the parameters $\alpha$ and $\beta$ in $g_{\alpha,\beta}(x)$ that best describe the Hubbard model on a ring, we should check if $\alpha$ and $\beta$ depend on the initial filling density $\rho_0=1/N$. Values presented in Table \ref{tab:01} correspond to $\rho_0=1/7560$. Here $7560$ has $64$ divisors and belongs to a group of highly composite numbers \cite{composite}. They are defined as natural numbers that have more divisors than all smaller numbers. 

\begin{figure}[ht]
\centerline{\includegraphics[width=16cm]{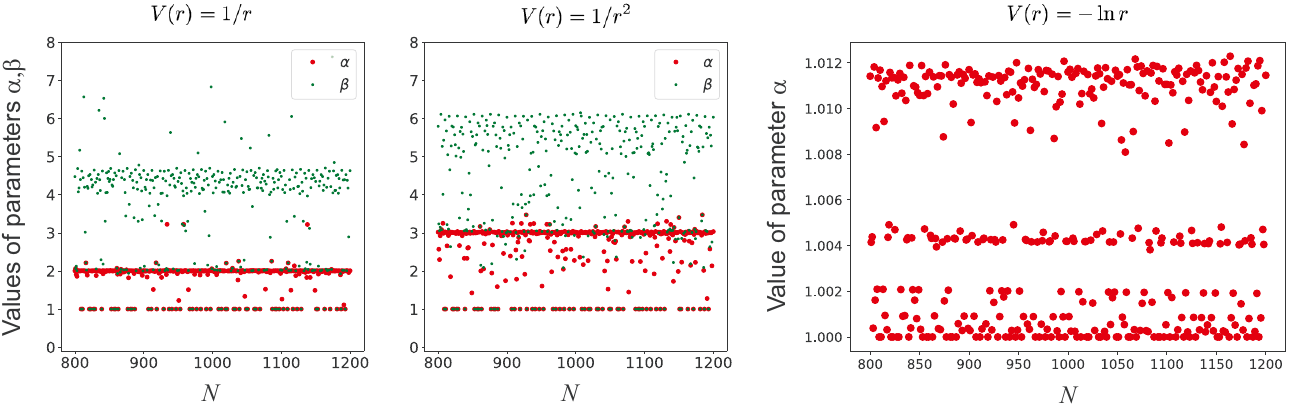}}
\caption{(a) Numerical values of $\alpha$ and $\beta$ given in \eq{eq:048} for different initial filling densities $\rho_0=1/N$. Left panel shows the data for $V(r)=1/r$ and the right one -- for $V(r)=1/r^2$; (b) Numerical values of $\alpha$ given in Eq. \ref{eq:048} for different initial filling densities $\rho_0=1/N$ and $V(r)=-\log(r)$.}
\label{fig:015}
\end{figure}

Since $7560$ belongs to a very specific group of numbers, we study the sensitivity of $\alpha$ and $\beta$ to $\rho_0$. To investigate the universality of these parameters, we numerically compute their values for different $N$. For the potential $V(r)=1/r$ and $V(r)=1/r^2$ the results are presented in \fig{fig:015}. Let us discuss in detail the case of $V(r)=1/r^2$. As expected, the exponent $\beta$ that controls the finite size corrections is non-universal and varies essentially. Values of $\alpha$ stay mostly near $\alpha=2$, with the exception of anomalous points that gather around $\alpha=1$. These exceptional points occur when $N$ is a prime number or has a small number of divisors. In this case, the devil's staircase has almost no jumps, and consequently, the derivative $\psi(\rho)$ is constant or has only a few jumps what makes the comparison with the generalized Thomae function senseless. We have also checked that all values of $N$ producing $\alpha=1$ and $\beta=1$ in the interval $N\in [853, 947]$ are prime numbers. Furthermore, these are also {\it all} prime numbers in this interval. We can also spot some points that are neither at $\alpha=1$ nor $\alpha=2$. We checked some of them, and found that they appear at $N$, that have a very low number of divisors, usually four. The results for $V(r)=1/r^2$ are presented in \fig{fig:015}. 

In the same manner, we analyzed optimal values of $\alpha$ and $\beta$ for other potentials, and found the same dependence of $\alpha$ on the divisibility of $N$. We can conclude that the parameter $\alpha$ is stable with respect to the choice of initial filling density $\rho_0(N)=1/N$. Exceptions are ``anomalous'' points that correspond to $N$ with a low number of divisors. The case of $V(r)=-\log(r)$ is slightly different since only one parameter suffices for describing the data. Nevertheless, we verified the universality of the obtained value $\alpha \approx 1$. Looking at \fig{fig:015}, we can see that regardless of $N$, the value of $\alpha$ is close to one.

\subsection{Fibonacci series in a Hubbard model on a ring}

\begin{figure}[ht]
\centerline{\includegraphics[width=14cm]{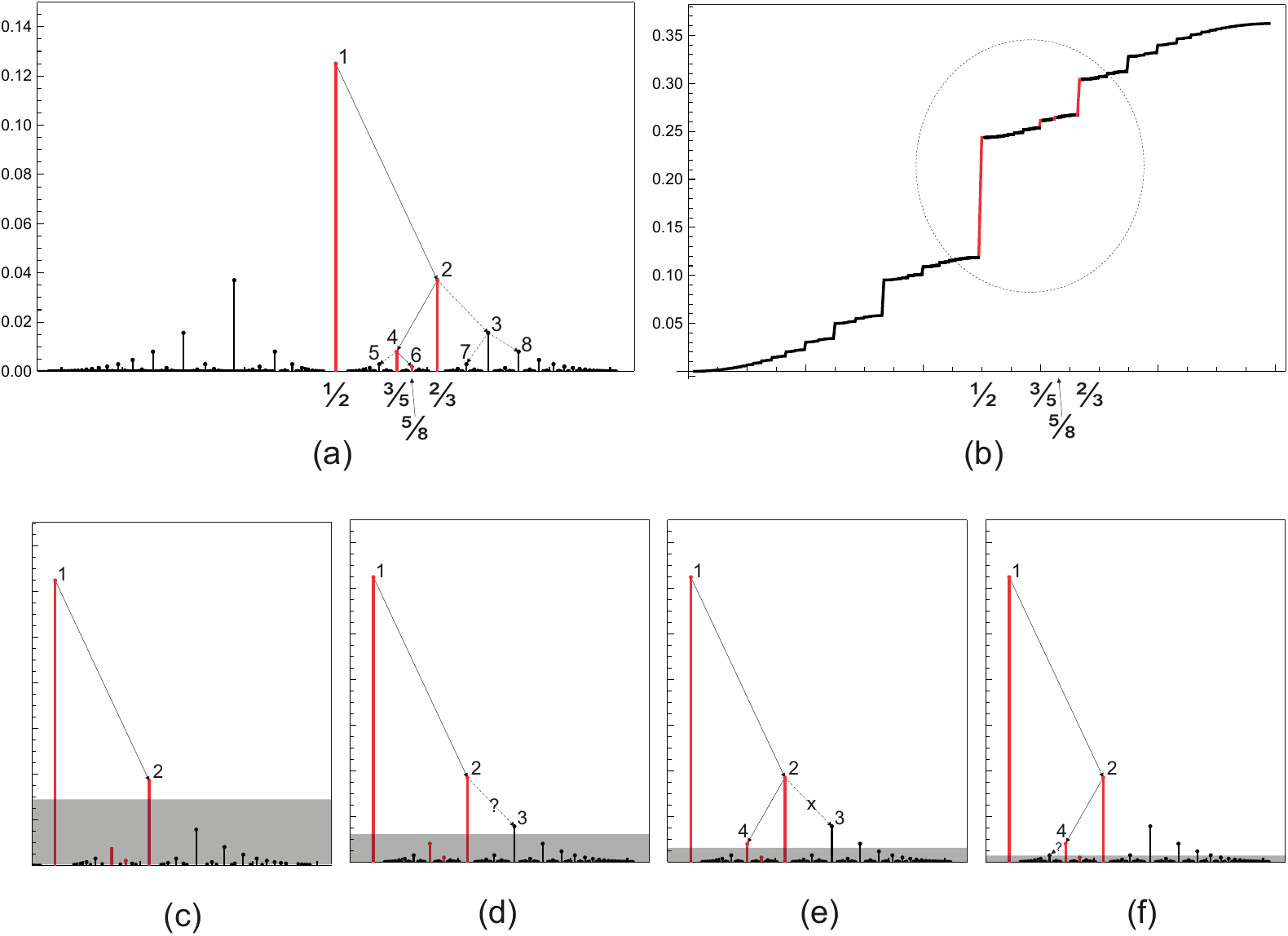}}
\caption{(a) Generalized Riemann-Thomae function emerging for the Hubbard model on a ring for the potential $V(r)=1/r^2$. The Fibonacci sequence $1-2-4-6-...$ is ``optimal'' with respect to other sequences from the point of view of local minimization of the relation \eq{eq:047}; (b) The associated Devil's staircase. Steps corresponding to the Fibonacci sequence are shown in red; (c)-(f) Sequential steps of the optimal path selection.}
\label{fig:fib3}
\end{figure}

Let us formulate the rules which select the Fibonacci sequence in the Hubbard model on a ring. Look at the forest of barriers in \fig{fig:014} and pay more detailed attention to the potential $V(r)=1/r^2$. The corresponding plots is shown in \fig{fig:fib3} for the set of potential barriers between different ground states (\fig{fig:fib3}a) and for the corresponding integrated function having a Devil's staircase structure (\fig{fig:fib3}b). It is worth noting that the rule formulated below works for any long-ranged potential producing the structure of the generalized Riemann-Thomae function.

So, we want to define a natural rule for the Hubbard model which selects in \fig{fig:fib3}a the sequence $1-2-4-6-...$ (corresponding to the Fibonacci ``zigzag'' sequence), but not $1-2-3-5...$, or not $1-2-3-8-...$. The idea of the rule is as follows. First, build a graph as a 3-branching tree on vertices that are close by heights. In \fig{fig:fib3} the vertex 2 has neighbors 3 and 4, the vertex 3 has neighbors 7 and 8, the vertex 4 has neighbors 5 and 6, etc. Now, we would like to select an ``optimal'' path on the constructed tree that corresponds to the Fibonacci sequence $1-2-4-6-...$. Note that in \eq{eq:047} the expression under the sum sign has the structure of the 2nd difference relation (the discrete version of the 2nd derivative) and is defined on each triple of adjacent points on the constructed tree. The prescription of path selection is simple: we locally minimize a 2nd difference relation for each triple of neighboring vertices. For example, comparing the subsequences $1-2-3$ and $1-2-4$ in Figs. \ref{fig:fib3}c-f we see that the 2nd difference for the triple $1-2-4$ is smaller than that of the triple $1-2-3$. So, we select the subsequence $1-2-4$. Coming to the point $4$ we should choose between $2-4-5$ and $2-4-6$. The heights of barriers are organized in such a way that the subsequence $2-3-6$ has smaller 2nd difference than that of the subsequence $2-4-5$, etc. Interestingly, the Fibonacci sequence survives for any symmetric potentials.

\section{Discussion}

Here we comment on a few points related to our study, restricting ourselves by qualitative arguments. We postpone more detailed analysis for a separate publication.

\subsection{Jack polynomials on the Devil's staircase}

The existence of the Devil's staircase structure in the fractional quantum Hall effect in the thin torus limit was the subject of the studies \cite{Tao1,Tao2}. In \cite{rotondo2016devil} it has been shown that there exists a precise mapping of FQHE in this limit onto the Hubbard model on a ring discussed above. In the thin torus limit the Hamiltonian of the FQHE for interacting fermions reads as
\be
H\{n_1,...,n_N\} = -h\sum_{i=1}^N n_i + \frac{1}{2}\sum_{i=1}^N\sum_{j\neq i}\frac{ n_i n_j}{k_i-k_j},
\label{hub}
\ee
where $k_i$ are momenta of the particles forming the lattice on the circle. The Devil's staircase structure for FQHE has been reformulated in terms of Jack polynomials in \cite{di2017jack} which looked rather surprisingly and the question ``What Jacks are doing on the Devil's staircase?'' seems eligible. Taking into account that the Jack polynomials are the wave functions of the Calogero many-body system, let us argue that the relations found in \cite{di2017jack,rotondo2016devil} for $V(r)=\frac{1}{r}$ can be understood via the chain of dualities known in the framework of the integrable many-body systems. 

First, recall the duality between the Calogero-Moser (CM) / Ruijsenaars-Schneider (RS) family of long-range many-body systems and the family of inhomogeneous twisted spin chains \cite{gorsky2014spectrum, gaiotto2013three, beketov2016trigonometric, bulycheva2014bps, zabrodin2017qkz}. This duality, for instance, provides the relation between the quantum inhomogeneous twisted XXZ chain and the trigonometric RS model \cite{beketov2016trigonometric}. Taking into account the realization of the FQHE on the torus in terms of the trigonometric RS model \cite{gorsky2002dualities} and combining this relation with the duality mentioned above, we arrive at the qualitative understanding how the Devil's staircase emerges in terms of the particular counting problem at the Calogero-Moser side.

To be more specific, consider the quantum trigonometric RS model {\it per se} whose wave functions are McDonald polynomials, and both, coordinates and momenta, live respectively on the circles $R_q$ and $R_p$. Geometrically the phase space of the trigonometric RS model with $N$ particles corresponds to the moduli space of the $SU(N)$ flat connections of the punctured torus \cite{gorsky1995relativistic}. The coupling constant corresponds to the operator inserted at one marked point. To get the system of particles, one performs the $T$-duality transformation for both torus cycles. Under such a transformation, the eigenvalues of holonomies over the cycles become the coordinates and momenta of particles. 

On the other hand, the quantum trigonometric RS model is dual to the quantum inhomogeneous twisted XXZ spin chain \cite{gorsky2022dualities, zabrodin2017qkz} via the quantum-quantum duality. The trigonometric $N$-body RS model is described by the Hamiltonian
\be
H_{trRS}= \sum_{j=1}^{N}e^{i\eta p_j} \prod_{j\neq k}\frac{\sin(R_q^{-1}(q_j -q_k -\eta \nu ))}
{\sin(R_q^{-1}(q_j-q_k))}
\ee
where momenta live on the circle of radius $R_p=\eta^{-1}$. The mapping between the trigonometric RS model and the spin chain goes as follows: (i) the coordinates, $q_i$, in the RS model are the inhomogeneities in the XXZ spin chain, (ii) the momenta $\Dot{q}_i$ are the non-local Hamiltonians, $H_i$, at the XXZ spin chain side, (iii)
the eigenvalues of the Lax operator at the RS side are the twists at the XXZ spin chain side, and (iv) the coupling constant at RS side is the Planck constant at the spin chain side. This correspondence is described in detail in \cite{gorsky2022dualities}.

The radius of ``coordinate circle'', $R_q$, is related to the anisotropy $\chi$ in XXZ chain and as the first step we can consider the limit $\chi \rightarrow \infty$ when the XXZ spin interaction term gets reduced to $S_i^z\, S_j^z$. Note that in this limit we have $R_q \to 0$ instead of more familiar limit $R_q \to \infty$, when the trigonometric RS model gets reduced to the rational RS model and the XXZ spin chain becomes the XXX chain. In our regime, to avoid fast oscillating behavior, particles tend to form the momentum lattice structure in the $R_q \rightarrow 0$ limit. To get the Hubbard-type Hamiltonian discussed above (see \eq{eq:045}), one has to make the second degeneration taking the semiclassical limit of the spin chain which transforms it to the Gaudin model. At the RS side this second degeneration corresponds to the $R_p \rightarrow \infty$ limit for the ``momentum circle''  when the relativistic RS model gets reduced to the non-relativistic CM model in the peculiar limit. Note that the considered limit of the RS model \cite{gorsky1995relativistic} corresponds to the $k\to \infty$ limit for the Kac-Moody level.

Taking the limits $R_q \to 0$ and $R_p \to \infty$ and transforming the trigonometric RS model to the Hubbard-like Hamiltonian, we have built the formal basis of our consideration. It is worth adding a more physical flavor to this construction. To this aim recall that the trigonometric RS model has been related to the FQHE on the torus \cite{gorsky2002dualities} by extending the approach developed in \cite{susskind2001quantum, polychronakos2001quantum}. Now we can provide a physical interpretation of considered limits for the radii $R_q$ and $R_p$. First, from the relation between the FQHE on the disc, the rational Calogero system supplemented with the oscillator potential $\omega \sum q_i^2$ can be mapped it to the trigonometric Sutherland model on the circle of radius $R_q$, where $\omega= \frac{1}{R_q}$, as it is shown in \cite{nekrasov1997duality}. Since $\omega$ is defined by the magnetic field \cite{polychronakos2001quantum}, the limit $R_q \to 0$ corresponds to the strong magnetic field. Second, in the limit $R_p \to \infty$ we arrive at the specific version of the trigonometric CM model, which means that we indeed find ourselves with the Jack polynomials for the ``momentum lattice'', as expected.

Let us complete the discussion by comparing the Devil's staircase structures at different extremities of our mapping. In the Hubbard model we focus at the chemical potential and the density of particles. In the spin chain before taking limit, these objects are involved into twist term in the non-local Hamiltonian ${\rm Tr}\,{\bf G}{\bf S}$ with the twist matrix $G$. In our case we can identify the twist with the diagonal $S^z$ and eigenvalue $h$, hence the term ${\rm Tr}\,{\bf G}{\bf S}$ gets reduced to the term $h n_i$ in the Hubbard Hamiltonian. The Hubbard Hamiltonian in terms of the non-local Gaudin Hamiltonians $H_i$ is the sum of individual terms $H =\sum_i ^N H_i$. In the Hubbard case we can recognize the Devil's staircase $\rho(h)$ for the density as the function 
of twist $h$, or the inverse function $h(\rho)$ -- see \cite{bak1982commensurate}.

What is the meaning of $\rho(h)$ function and Hubbard Hamiltonian at CM side? As we discussed above, the twist $h\leftrightarrow E_{cal}$ corresponds to the eigenvalue of the Lax operator and therefore to the eigenvalues of CM Hamiltonians. Since we have a single eigenvalue of the twist matrix, we consider the highly degenerate state. The $H_i$ Hamiltonians correspond to the momenta at the CM side, hence the total Hubbard Hamiltonian is nothing but the total momentum of CM particles $P=\sum_i p_i$. 

In the iterative procedure which yields the Devil's staircase we add at each step one additional particle demanding the total Hubbard energy to be constant. Being translated to the CM side it means that we add  particles keeping the total momentum fixed. It is worth mentioning the difference between the Devil's staircase structure in the CM model in the context of Fibonacci numbers and in Hubbard case. In the phyllotaxis problem we fix the energy $E=0$ and count the weighted degeneracy of this level. In the context of the Hubbard model we focus on the multiplicity of the $P=const$ state in a specific limit of RS system. Since the momentum of the RS particles is determined via the Bethe ansatz equations \cite{gorsky2014spectrum} one could a bit loosely say that the Devil's staircase structure emerges in the space of solutions to Bethe ansatz equations.

\subsection{On Fibonacci universality class at out-of-equilibrium}

It was suggested in \cite{popkov2015fibonacci} that critical exponents describing fluctuations in the non-equilibrium dynamics are of more generic nature than it is usually assumed. It was claimed that the Gauss and KPZ scalings are just two first representatives in the generic ``Fibonacci hierarchy''  with critical exponents $z_n= \frac{F_{n+1}}{F_n}$, where $F_n$ is the $n$th Finonacci number. This result has been obtained by analyzing the hydrodynamic equations with several conservation laws. As a toy example, the 3-species TASEP model has been discussed and the numerical simulations indeed exhibit the third critical exponent for the large-time asymptotics of the maximal value of the two-point correlator
\be
{\rm max}\, \left<\phi(x,t)\phi(0,0)\right>\propto t^{-1/z_3}
\ee
The existence of Fibonacci universality class is still under discussion. For instance, it was demonstrated rigorously that for the systems with the non-abelian global symmetries including the integrable spin models the large-time asymptotics of two point function enjoys the KPZ critical exponent \cite{ilievski2021superuniversality} generalizing the initial observation in \cite{ljubotina2017spin}.

It is eligible to ask a question of whether there is any relation between the Fibonacci hierarchy and our study. Let us speculate on the possible connection and assume that we got somehow the correlator of a particular operators in the modular domain, for instance the two-point correlator,
\be
\left<\phi(\tau)\phi(0)\right>  = f(\tau)
\label{eq:propagator}
\ee
where $\tau$ is a modular complex parameter, for which we interpret $\im\tau$ as a time variable. Such a viewpoint is valid at least in two situations. First, in the phyllotaxis problem Eq.\eq{eq:propagator} has a sense of the propagator in the modular domain with $\im \tau$ as an evolution parameter. Second, in the Whitham dynamics for the Seiberg-Witten solution, the Nekrasov partition function can be mapped via the AGT correspondence \cite{alday2010liouville} onto the conformal block in the Liouville or Toda models where the modular parameter corresponds to the insertion point for some vertex operator $\left<V_1(0)V_2(1)V_3(\infty)V_4(\tau)\right>$ and $\im\tau$ is the time for the Whitham dynamics.

Let us suppose a simple typical scaling behavior for a correlator \eq{eq:propagator} in the modular domain, say $f(\tau)\propto \exp(\tau)$ and consider the parameter $\tau= \theta + i\log t $ from the perspective discussed at the phyllotaxis side. In this representation at small values of $\im\, \tau$ one gets the critical exponents dictated by asymptotic values of $\theta$ for which we know that they are given by truncated continued fraction expansion \eq{eq:zigzag} for sequential quotients of Fibonacci numbers. This point of view supports
the idea of Fibonacci universality raised in \cite{popkov2015fibonacci}. Certainly, these arguments are very superficial and an accurate analysis is required, however it immediately poses the following challenging question: do we have non-equilibrium dynamical systems producing a ``Silver ratio universality class''?

\subsection{Dedekind in the proper place}

The Riemann-Thomae function has been discussed recently from an interesting perspective in the theory of massless free bosons and fermions on the circle of radius $R$ at finite temperature $T=\beta ^{-1}$ upon the Wick rotation \cite{chernodub2022fractal, ambrus2023rigidlyrotating}. The twisted boundary conditions involving both circles are imposed and for the rational $p/q$ the twist corresponds exactly to the $T_{p,q}$ torus knot for the closed space-time trajectory of a particle. The emergence of the RT function is not a surprise since the partition function of the 2D theory on the torus with the global $U(1)\times U(1)$ involves the Dedekind functions hence the free energy in the proper limit indeed has a RT structure leading to the Devil's staircase.

It was argued in \cite{chernodub2022fractal, ambrus2023rigidlyrotating} that for the torus knot boundary conditions the thermodynamic properties demonstrate interesting fractal behavior. It was suggested that the statistics of free particles with the torus knot boundary condition is level--dependent and the negative pressure regime can be found. Since the Dedekind enters the partition function of the massless scalar, the RT function defines the dependence of the pressure on the twist parameter in the limit $L=\frac{R}{\beta} \to \infty$.

It would be interesting to match the fractal thermodynamic properties for system with torus knot boundary condition and our RG approach. Indeed for the torus knot trajectories we have specific knot invariants  corresponding to particular multiplicities of states. Hence, their impact on the partition function is expected. It would be also interesting to recognize the possible fractal thermodynamics for the massless scalar on the mirror torus when the complex and Kahler structures get interchanged.

\section{Conclusion}

This work provides a modest attempt to add some flavor of universality to the interplay between the modular group acting in the parameter space of physical systems (spectra of random operators, phyllotaxis, Hubbard model on a ring) and the RG flows in the peculiar region of the fundamental domain of the modular $SL(2,R)$ group when the real part of the modular parameter tends to zero. We have argued that in this regime the systems possess the universality described by the generalized Riemann-Thomae (gRT) function, and the generalized Devil's staircase emerges. Using a natural regularization of the fractal gRT by the modular Dedekind $\eta$-function we were able to interpolate between ``neighboring'' fractal states and connect this interpolation with the RG flows on the modular group. Saying  differently we were looking for additional arguments supporting the universality of commensurability-incommensurability transitions. The problem can be reformulated as the derivation of the RG flow for the deformations of lattices of different nature via some disorder. We argued that the analysis of the lattice structure by studying the corresponding spectral properties of propagating probe could be very useful and in the ``thin torus limit'' the rearrangement (bifurcation) of highly squeezed lattice is a collective effect with a signature of the BKT transiton.

The limit $y\to 0$ of the modular parameter $\tau = \theta+iy$ we are focused at, has the clear physical interpretation. It corresponds to the situation when the disorder $y$ associated with the imaginary part of the modular parameter in some frame tends to zero, while the $\theta$-term associated with the real part of $\tau$ and serves as the chemical potential for some version of the topological charge, remains finite. The very notion of the disorder $y$ is model-dependent and in some systems it can be identified with the diffusion constant as for example in the Anderson transition problem, or with the magnetic coupling constant as in the Yang-Mills theory. The notion of the ``weak disorder'' is frame-dependent since the weak coupling limit in the magnetic frame corresponds to the strong coupling regime in the electric frame. We have argued that in this regime the modular the non-perturbative ``instanton'' renormalization dominates since the disorder is weak. In particular, our analysis suggests that one could expect the Devil's staircase in some version of multiplicities of BPS states near Argyres-Douglas point.

There are several issues that certainly deserve additional study. It would be interesting to find the place to gRT function in the group-theoretic framework. There are finite algebras that involve several parameters, such as Sklyanin algebra which has the $p$-adic and quantum  groups as the peculiar degenerations. The models which enjoy the devil staircase and generalized Devil's staircase have extended symmetries and it would be interesting to recognize the structures discussed at length of our paper in some limits of Sklyanin algebra. Another possible group-like structure concerns the algebra of BPS states which was identified as the hyperbolic Kac-Moody algebra \cite{harvey1996algebras}. The flow between the lattices in this framework correspond to the interpolation between hyperbolic algebras which are related with the Fibonacchi numbers  \cite{feingold1980hyperbolic, lechtenfeld2022hyperbolic}.

It would be also interesting to include into the framework of our study the structures associated with general $T_{n,m}$ torus knots and links related with the instanton counting via the instanton-torus knot duality. This should generalize the relation between invariants of $T_{2,n}$ knots and Fibonacci numbers. The last note concerns the resurgence theory (see \cite{aniceto2019primer} for the review) providing the interplay between the non-perturbative and perturbative contributions to different objects including $\beta$-function. 

\begin{acknowledgments}

We are grateful to K. Polovnikov for collaboration at the early stage of the work. We encourage the numerous discussions of different aspects of this work with V. Avetisov, M. Gherardi, P. Krapivsky, P. Rotondo, M. Tamm. A.G. thanks Nordita and IHES where the parts of the works have been done for the hospitality and support.
\end{acknowledgments}

\begin{appendix}

\section{Computation of the coefficient $C(y)$ for the regularized Riemann-Thomae function $g_2(x,y)$ in Eq. \eq{eq:017}}
\label{app:1}

The analytic structure of the function $f(x,y)=y^{1/4}|\eta(x,y)|$ has been discussed in \cite{krapiv} and in more detail in \cite{polov} in the context of the ultrametric landscape construction. The asymptotic behavior of the Dedekind $\eta$-function can be straightforwardly derived through the duality relation
\be
f\left(\left\{\frac{m}{k}\right\}, y\right) = f\left(\left\{\frac{n}{k}\right\},\frac{1}{k^2 y}
\right)
\label{eq:a1}
\ee
where $m n - k r = 1,\; \{k,m,n,r\}\in \mathbb{Z},\; (y>0)$ and $\left\{\frac{m}{k}\right\}, \left\{\frac{n}{k}\right\}$ denote fractional parts of corresponding quotients (see \cite{krapiv}). 

The constant $C(y)$ in the relation $g_2(x)=C(y) \ln (y^{1/4}|\eta(x,y)|)$ we obtain comparing the value $g_2\left(\frac{1}{2}\right)=\frac{1}{4}$ with the asymptotic expansion of the function $f\left(\frac{1}{2},y\right)$ at $y\to 0$. Using \eq{eq:a1} and choosing $\{m=1; k=2; n=3; r=1\}$ we get:
\be
f\left(\frac{1}{2}, y\right) = f\left(\frac{1}{2},\frac{1}{4y}\right)
\label{eq:a2}
\ee
Thus,
\be 
\left|\eta\left(\frac{1}{2},y\right)\right|=\frac{1}{\sqrt{2y}}\left|\eta\left(\frac{1}{2},\frac{1}{4y}\right)\right|
\label{eq:a3}
\ee
Remembering the relation between the Dedekind $\eta$-function and elliptic Jacobi function,
\be 
2\eta^3(z) =\left.\frac{d\theta_1\left(u,e^{i\pi z}\right)}{du}\right|_{u=0} \equiv \theta_1'(0,e^{i\pi z})
\label{eq:a4}
\ee
and the series representation of the Jacobi prime function
\be 
\theta_1'(0,e^{i\pi z}) = 2 e^{i\pi z/4} \sum_{k=0}^{\infty} (-1)^k(2k+1) e^{i\pi k(k+1) z}
\label{eq:a5}
\ee
we get the following asymptotic expression
\be 
\left|\eta\left(\frac{1}{2},\frac{1}{4y}\right)\right| = \left.\left|\frac{1}{2}\theta_1'\left(0,e^{i \pi/2-\pi/(4y)}\right)\right|^{1/3} \right|_{y\to 0} \approx e^{-\pi/(48y)}
\label{eq:a6}
\ee 
Substituting \eq{eq:a6} into expression for $f\left(\frac{1}{2},y\right)$, we get the following equation for the coefficient $C(y)$: 
\be 
C(y) \ln \left((4y)^{-1/4}e^{-\pi/(48y)}\right)=\frac{1}{4}
\label{eq:a7}
\ee
Hence,
\be 
C(y)=\left.\frac{12 y}{\pi + 12 y \ln (4y)}\right|_{y\to 0} \approx \frac{12 y}{\pi}
\label{eq:a8}
\ee
    
\end{appendix}

\bibliography{genpop.bib}

\begin{thebibliography}{96}%
\makeatletter
\providecommand \@ifxundefined [1]{%
 \@ifx{#1\undefined}
}%
\providecommand \@ifnum [1]{%
 \ifnum #1\expandafter \@firstoftwo
 \else \expandafter \@secondoftwo
 \fi
}%
\providecommand \@ifx [1]{%
 \ifx #1\expandafter \@firstoftwo
 \else \expandafter \@secondoftwo
 \fi
}%
\providecommand \natexlab [1]{#1}%
\providecommand \enquote  [1]{``#1''}%
\providecommand \bibnamefont  [1]{#1}%
\providecommand \bibfnamefont [1]{#1}%
\providecommand \citenamefont [1]{#1}%
\providecommand \href@noop [0]{\@secondoftwo}%
\providecommand \href [0]{\begingroup \@sanitize@url \@href}%
\providecommand \@href[1]{\@@startlink{#1}\@@href}%
\providecommand \@@href[1]{\endgroup#1\@@endlink}%
\providecommand \@sanitize@url [0]{\catcode `\\12\catcode `\$12\catcode
  `\&12\catcode `\#12\catcode `\^12\catcode `\_12\catcode `\%12\relax}%
\providecommand \@@startlink[1]{}%
\providecommand \@@endlink[0]{}%
\providecommand \url  [0]{\begingroup\@sanitize@url \@url }%
\providecommand \@url [1]{\endgroup\@href {#1}{\urlprefix }}%
\providecommand \urlprefix  [0]{URL }%
\providecommand \Eprint [0]{\href }%
\providecommand \doibase [0]{https://doi.org/}%
\providecommand \selectlanguage [0]{\@gobble}%
\providecommand \bibinfo  [0]{\@secondoftwo}%
\providecommand \bibfield  [0]{\@secondoftwo}%
\providecommand \translation [1]{[#1]}%
\providecommand \BibitemOpen [0]{}%
\providecommand \bibitemStop [0]{}%
\providecommand \bibitemNoStop [0]{.\EOS\space}%
\providecommand \EOS [0]{\spacefactor3000\relax}%
\providecommand \BibitemShut  [1]{\csname bibitem#1\endcsname}%
\let\auto@bib@innerbib\@empty
%</preamble>
\bibitem [{\citenamefont {Aubry}(1983)}]{aubry1983devil}%
  \BibitemOpen
  \bibfield  {author} {\bibinfo {author} {\bibfnamefont {S.}~\bibnamefont
  {Aubry}},\ }\bibfield  {title} {\bibinfo {title} {Devil's staircase and order
  without periodicity in classical condensed matter},\ }\href@noop {}
  {\bibfield  {journal} {\bibinfo  {journal} {Journal de Physique}\ }\textbf
  {\bibinfo {volume} {44}},\ \bibinfo {pages} {147} (\bibinfo {year}
  {1983})}\BibitemShut {NoStop}%
\bibitem [{\citenamefont {Bak}(1982{\natexlab{a}})}]{bak1982commensurate}%
  \BibitemOpen
  \bibfield  {author} {\bibinfo {author} {\bibfnamefont {P.}~\bibnamefont
  {Bak}},\ }\bibfield  {title} {\bibinfo {title} {Commensurate phases,
  incommensurate phases and the devil's staircase},\ }\href@noop {} {\bibfield
  {journal} {\bibinfo  {journal} {Reports on Progress in Physics}\ }\textbf
  {\bibinfo {volume} {45}},\ \bibinfo {pages} {587} (\bibinfo {year}
  {1982}{\natexlab{a}})}\BibitemShut {NoStop}%
\bibitem [{\citenamefont {Bergholtz}\ \emph {et~al.}(2007)\citenamefont
  {Bergholtz}, \citenamefont {Hansson}, \citenamefont {Hermanns},\ and\
  \citenamefont {Karlhede}}]{Tao1}%
  \BibitemOpen
  \bibfield  {author} {\bibinfo {author} {\bibfnamefont {E.~J.}\ \bibnamefont
  {Bergholtz}}, \bibinfo {author} {\bibfnamefont {T.~H.}\ \bibnamefont
  {Hansson}}, \bibinfo {author} {\bibfnamefont {M.}~\bibnamefont {Hermanns}},\
  and\ \bibinfo {author} {\bibfnamefont {A.}~\bibnamefont {Karlhede}},\
  }\bibfield  {title} {\bibinfo {title} {Microscopic theory of the quantum hall
  hierarchy},\ }\href {https://doi.org/10.1103/PhysRevLett.99.256803}
  {\bibfield  {journal} {\bibinfo  {journal} {Phys. Rev. Lett.}\ }\textbf
  {\bibinfo {volume} {99}},\ \bibinfo {pages} {256803} (\bibinfo {year}
  {2007})}\BibitemShut {NoStop}%
\bibitem [{\citenamefont {Bergholtz}\ and\ \citenamefont
  {Karlhede}(2008)}]{Tao2}%
  \BibitemOpen
  \bibfield  {author} {\bibinfo {author} {\bibfnamefont {E.~J.}\ \bibnamefont
  {Bergholtz}}\ and\ \bibinfo {author} {\bibfnamefont {A.}~\bibnamefont
  {Karlhede}},\ }\bibfield  {title} {\bibinfo {title} {Quantum hall system in
  tao-thouless limit},\ }\href {https://doi.org/10.1103/PhysRevB.77.155308}
  {\bibfield  {journal} {\bibinfo  {journal} {Phys. Rev. B}\ }\textbf {\bibinfo
  {volume} {77}},\ \bibinfo {pages} {155308} (\bibinfo {year}
  {2008})}\BibitemShut {NoStop}%
\bibitem [{\citenamefont {Lundholm}(2017)}]{lundholm}%
  \BibitemOpen
  \bibfield  {author} {\bibinfo {author} {\bibfnamefont {D.}~\bibnamefont
  {Lundholm}},\ }\bibfield  {title} {\bibinfo {title} {Many-anyon trial
  states},\ }\href {https://doi.org/10.1103/PhysRevA.96.012116} {\bibfield
  {journal} {\bibinfo  {journal} {Phys. Rev. A}\ }\textbf {\bibinfo {volume}
  {96}},\ \bibinfo {pages} {012116} (\bibinfo {year} {2017})}\BibitemShut
  {NoStop}%
\bibitem [{\citenamefont {Planat}\ and\ \citenamefont {Eckert}(2000)}]{planat}%
  \BibitemOpen
  \bibfield  {author} {\bibinfo {author} {\bibfnamefont {M.}~\bibnamefont
  {Planat}}\ and\ \bibinfo {author} {\bibfnamefont {C.}~\bibnamefont
  {Eckert}},\ }\bibfield  {title} {\bibinfo {title} {On the frequency and
  amplitude spectrum and the fluctuations at the output of a communication
  receiver},\ }\href {https://doi.org/10.1109/58.869063} {\bibfield  {journal}
  {\bibinfo  {journal} {IEEE Transactions on Ultrasonics, Ferroelectrics, and
  Frequency Control}\ }\textbf {\bibinfo {volume} {47}},\ \bibinfo {pages}
  {1173} (\bibinfo {year} {2000})}\BibitemShut {NoStop}%
\bibitem [{\citenamefont {Trifonov}\ \emph {et~al.}(2011)\citenamefont
  {Trifonov}, \citenamefont {Pascualucci}, \citenamefont {Dalla-Favera},\ and\
  \citenamefont {Rabadan}}]{dna}%
  \BibitemOpen
  \bibfield  {author} {\bibinfo {author} {\bibfnamefont {V.}~\bibnamefont
  {Trifonov}}, \bibinfo {author} {\bibfnamefont {L.}~\bibnamefont
  {Pascualucci}}, \bibinfo {author} {\bibfnamefont {R.}~\bibnamefont
  {Dalla-Favera}},\ and\ \bibinfo {author} {\bibfnamefont {R.}~\bibnamefont
  {Rabadan}},\ }\bibfield  {title} {\bibinfo {title} {Fractal-like
  distributions over the rational numbers in high-throughput biological and
  clinical data},\ }\href {https://doi.org/https://doi.org/10.1038/srep00191}
  {\bibfield  {journal} {\bibinfo  {journal} {Sci. Rep.}\ }\textbf {\bibinfo
  {volume} {1}},\ \bibinfo {pages} {191} (\bibinfo {year} {2011})}\BibitemShut
  {NoStop}%
\bibitem [{\citenamefont {Middendorf}\ \emph {et~al.}(2005)\citenamefont
  {Middendorf}, \citenamefont {Ziv},\ and\ \citenamefont
  {Wiggins}}]{drosophilla}%
  \BibitemOpen
  \bibfield  {author} {\bibinfo {author} {\bibfnamefont {M.}~\bibnamefont
  {Middendorf}}, \bibinfo {author} {\bibfnamefont {E.}~\bibnamefont {Ziv}},\
  and\ \bibinfo {author} {\bibfnamefont {C.}~\bibnamefont {Wiggins}},\
  }\bibfield  {title} {\bibinfo {title} {Inferring network mechanisms: The
  drosophila melanogaster protein interaction network},\ }\href@noop {}
  {\bibfield  {journal} {\bibinfo  {journal} {PNAS}\ }\textbf {\bibinfo
  {volume} {102}},\ \bibinfo {pages} {3192} (\bibinfo {year}
  {2005})}\BibitemShut {NoStop}%
\bibitem [{\citenamefont {Altshuler}\ and\ \citenamefont
  {Kravtsov}(2023)}]{altshuler2023random}%
  \BibitemOpen
  \bibfield  {author} {\bibinfo {author} {\bibfnamefont {B.}~\bibnamefont
  {Altshuler}}\ and\ \bibinfo {author} {\bibfnamefont {V.}~\bibnamefont
  {Kravtsov}},\ }\bibfield  {title} {\bibinfo {title} {Random cantor sets and
  mini-bands in local spectrum of quantum systems},\ }\href@noop {} {\bibfield
  {journal} {\bibinfo  {journal} {arXiv preprint arXiv:2301.12279}\ } (\bibinfo
  {year} {2023})}\BibitemShut {NoStop}%
\bibitem [{\citenamefont {Dzyaloshinskij}\ and\ \citenamefont
  {Krichever}(1982)}]{dzyaloshinskij1982commensurability}%
  \BibitemOpen
  \bibfield  {author} {\bibinfo {author} {\bibfnamefont {I.}~\bibnamefont
  {Dzyaloshinskij}}\ and\ \bibinfo {author} {\bibfnamefont {I.}~\bibnamefont
  {Krichever}},\ }\bibfield  {title} {\bibinfo {title} {Commensurability
  effects in the discrete peierls model},\ }\href@noop {} {\bibfield  {journal}
  {\bibinfo  {journal} {Zh. Ehksp. Teor. Fiz}\ }\textbf {\bibinfo {volume}
  {83}},\ \bibinfo {pages} {1576} (\bibinfo {year} {1982})}\BibitemShut
  {NoStop}%
\bibitem [{\citenamefont {Brazovskii}\ \emph {et~al.}(1982)\citenamefont
  {Brazovskii}, \citenamefont {Dzyaloshinskii},\ and\ \citenamefont
  {Krichever}}]{brazovskii1982exactly}%
  \BibitemOpen
  \bibfield  {author} {\bibinfo {author} {\bibfnamefont {S.}~\bibnamefont
  {Brazovskii}}, \bibinfo {author} {\bibfnamefont {I.}~\bibnamefont
  {Dzyaloshinskii}},\ and\ \bibinfo {author} {\bibfnamefont {I.}~\bibnamefont
  {Krichever}},\ }\bibfield  {title} {\bibinfo {title} {Exactly soluble peierls
  models},\ }\href@noop {} {\bibfield  {journal} {\bibinfo  {journal} {Physics
  Letters A}\ }\textbf {\bibinfo {volume} {91}},\ \bibinfo {pages} {40}
  (\bibinfo {year} {1982})}\BibitemShut {NoStop}%
\bibitem [{\citenamefont {Gukov}(2017)}]{gukov2017rg}%
  \BibitemOpen
  \bibfield  {author} {\bibinfo {author} {\bibfnamefont {S.}~\bibnamefont
  {Gukov}},\ }\bibfield  {title} {\bibinfo {title} {Rg flows and
  bifurcations},\ }\href@noop {} {\bibfield  {journal} {\bibinfo  {journal}
  {Nuclear Physics B}\ }\textbf {\bibinfo {volume} {919}},\ \bibinfo {pages}
  {583} (\bibinfo {year} {2017})}\BibitemShut {NoStop}%
\bibitem [{\citenamefont {Jepsen}\ and\ \citenamefont
  {Popov}(2021)}]{jepsen2021homoclinic}%
  \BibitemOpen
  \bibfield  {author} {\bibinfo {author} {\bibfnamefont {C.~B.}\ \bibnamefont
  {Jepsen}}\ and\ \bibinfo {author} {\bibfnamefont {F.~K.}\ \bibnamefont
  {Popov}},\ }\bibfield  {title} {\bibinfo {title} {Homoclinic renormalization
  group flows, or when relevant operators become irrelevant},\ }\href@noop {}
  {\bibfield  {journal} {\bibinfo  {journal} {Physical Review Letters}\
  }\textbf {\bibinfo {volume} {127}},\ \bibinfo {pages} {141602} (\bibinfo
  {year} {2021})}\BibitemShut {NoStop}%
\bibitem [{\citenamefont {Bosschaert}\ \emph {et~al.}(2022)\citenamefont
  {Bosschaert}, \citenamefont {Jepsen},\ and\ \citenamefont
  {Popov}}]{bosschaert2022chaotic}%
  \BibitemOpen
  \bibfield  {author} {\bibinfo {author} {\bibfnamefont {M.~M.}\ \bibnamefont
  {Bosschaert}}, \bibinfo {author} {\bibfnamefont {C.~B.}\ \bibnamefont
  {Jepsen}},\ and\ \bibinfo {author} {\bibfnamefont {F.~K.}\ \bibnamefont
  {Popov}},\ }\bibfield  {title} {\bibinfo {title} {Chaotic rg flow in tensor
  models},\ }\href@noop {} {\bibfield  {journal} {\bibinfo  {journal} {Physical
  Review D}\ }\textbf {\bibinfo {volume} {105}},\ \bibinfo {pages} {065021}
  (\bibinfo {year} {2022})}\BibitemShut {NoStop}%
\bibitem [{\citenamefont {Wilkinson}(1984)}]{wilkinson1984critical}%
  \BibitemOpen
  \bibfield  {author} {\bibinfo {author} {\bibfnamefont {M.}~\bibnamefont
  {Wilkinson}},\ }\bibfield  {title} {\bibinfo {title} {Critical properties of
  electron eigenstates in incommensurate systems},\ }\href@noop {} {\bibfield
  {journal} {\bibinfo  {journal} {Proceedings of the Royal Society of London.
  A. Mathematical and Physical Sciences}\ }\textbf {\bibinfo {volume} {391}},\
  \bibinfo {pages} {305} (\bibinfo {year} {1984})}\BibitemShut {NoStop}%
\bibitem [{\citenamefont {Wilkinson}(1987)}]{wilkinson1987exact}%
  \BibitemOpen
  \bibfield  {author} {\bibinfo {author} {\bibfnamefont {M.}~\bibnamefont
  {Wilkinson}},\ }\bibfield  {title} {\bibinfo {title} {An exact
  renormalisation group for bloch electrons in a magnetic field},\ }\href@noop
  {} {\bibfield  {journal} {\bibinfo  {journal} {Journal of Physics A:
  Mathematical and General}\ }\textbf {\bibinfo {volume} {20}},\ \bibinfo
  {pages} {4337} (\bibinfo {year} {1987})}\BibitemShut {NoStop}%
\bibitem [{\citenamefont {Altland}\ \emph {et~al.}(2015)\citenamefont
  {Altland}, \citenamefont {Bagrets},\ and\ \citenamefont
  {Kamenev}}]{Altland2015topology}%
  \BibitemOpen
  \bibfield  {author} {\bibinfo {author} {\bibfnamefont {A.}~\bibnamefont
  {Altland}}, \bibinfo {author} {\bibfnamefont {D.}~\bibnamefont {Bagrets}},\
  and\ \bibinfo {author} {\bibfnamefont {A.}~\bibnamefont {Kamenev}},\
  }\bibfield  {title} {\bibinfo {title} {Topology versus {Anderson}
  localization: Nonperturbative solutions in one dimension},\ }\href
  {https://doi.org/10.1103/PhysRevB.91.085429} {\bibfield  {journal} {\bibinfo
  {journal} {Phys. Rev. B}\ }\textbf {\bibinfo {volume} {91}},\ \bibinfo
  {pages} {085429} (\bibinfo {year} {2015})}\BibitemShut {NoStop}%
\bibitem [{\citenamefont {Pruisken}(1984)}]{pruisken1984localization}%
  \BibitemOpen
  \bibfield  {author} {\bibinfo {author} {\bibfnamefont {A.~M.}\ \bibnamefont
  {Pruisken}},\ }\bibfield  {title} {\bibinfo {title} {On localization in the
  theory of the quantized hall effect: A two-dimensional realization of the
  $\theta$-vacuum},\ }\href@noop {} {\bibfield  {journal} {\bibinfo  {journal}
  {Nuclear Physics B}\ }\textbf {\bibinfo {volume} {235}},\ \bibinfo {pages}
  {277} (\bibinfo {year} {1984})}\BibitemShut {NoStop}%
\bibitem [{\citenamefont {Levine}\ \emph {et~al.}(1984)\citenamefont {Levine},
  \citenamefont {Libby},\ and\ \citenamefont {Pruisken}}]{levine1984theory}%
  \BibitemOpen
  \bibfield  {author} {\bibinfo {author} {\bibfnamefont {H.}~\bibnamefont
  {Levine}}, \bibinfo {author} {\bibfnamefont {S.}~\bibnamefont {Libby}},\ and\
  \bibinfo {author} {\bibfnamefont {A.}~\bibnamefont {Pruisken}},\ }\bibfield
  {title} {\bibinfo {title} {Theory of the quantum hall effect (i)-(iii)},\
  }\href@noop {} {\bibfield  {journal} {\bibinfo  {journal} {Nuclear Physics
  B}\ }\textbf {\bibinfo {volume} {240}},\ \bibinfo {pages} {30} (\bibinfo
  {year} {1984})}\BibitemShut {NoStop}%
\bibitem [{\citenamefont {Montonen}\ and\ \citenamefont
  {Olive}(1977)}]{montonen1977magnetic}%
  \BibitemOpen
  \bibfield  {author} {\bibinfo {author} {\bibfnamefont {C.}~\bibnamefont
  {Montonen}}\ and\ \bibinfo {author} {\bibfnamefont {D.}~\bibnamefont
  {Olive}},\ }\bibfield  {title} {\bibinfo {title} {Magnetic monopoles as gauge
  particles?},\ }\href@noop {} {\bibfield  {journal} {\bibinfo  {journal}
  {Physics Letters B}\ }\textbf {\bibinfo {volume} {72}},\ \bibinfo {pages}
  {117} (\bibinfo {year} {1977})}\BibitemShut {NoStop}%
\bibitem [{\citenamefont {Cardy}\ and\ \citenamefont
  {Rabinovici}(1982)}]{cardy1982phase}%
  \BibitemOpen
  \bibfield  {author} {\bibinfo {author} {\bibfnamefont {J.~L.}\ \bibnamefont
  {Cardy}}\ and\ \bibinfo {author} {\bibfnamefont {E.}~\bibnamefont
  {Rabinovici}},\ }\bibfield  {title} {\bibinfo {title} {Phase structure of zp
  models in the presence of a $\theta$ parameter},\ }\href@noop {} {\bibfield
  {journal} {\bibinfo  {journal} {Nuclear Physics B}\ }\textbf {\bibinfo
  {volume} {205}},\ \bibinfo {pages} {1} (\bibinfo {year} {1982})}\BibitemShut
  {NoStop}%
\bibitem [{\citenamefont {Avetisov}\ \emph {et~al.}(2015)\citenamefont
  {Avetisov}, \citenamefont {Krapivsky},\ and\ \citenamefont
  {Nechaev}}]{krapiv}%
  \BibitemOpen
  \bibfield  {author} {\bibinfo {author} {\bibfnamefont {V.}~\bibnamefont
  {Avetisov}}, \bibinfo {author} {\bibfnamefont {P.~L.}\ \bibnamefont
  {Krapivsky}},\ and\ \bibinfo {author} {\bibfnamefont {S.}~\bibnamefont
  {Nechaev}},\ }\bibfield  {title} {\bibinfo {title} {Native ultrametricity of
  sparse random ensembles},\ }\href
  {https://doi.org/10.1088/1751-8113/49/3/035101} {\bibfield  {journal}
  {\bibinfo  {journal} {J. Phys. A: Math. Theor.}\ }\textbf {\bibinfo {volume}
  {49}},\ \bibinfo {pages} {035101} (\bibinfo {year} {2015})}\BibitemShut
  {NoStop}%
\bibitem [{\citenamefont {Nechaev}\ and\ \citenamefont
  {Polovnikov}(2018)}]{polov}%
  \BibitemOpen
  \bibfield  {author} {\bibinfo {author} {\bibfnamefont {S.~K.}\ \bibnamefont
  {Nechaev}}\ and\ \bibinfo {author} {\bibfnamefont {K.}~\bibnamefont
  {Polovnikov}},\ }\bibfield  {title} {\bibinfo {title} {Rare-event statistics
  and modular invariance},\ }\href
  {https://doi.org/10.3367/UFNe.2017.01.038106} {\bibfield  {journal} {\bibinfo
   {journal} {Phys. Usp.}\ }\textbf {\bibinfo {volume} {61}},\ \bibinfo {pages}
  {99} (\bibinfo {year} {2018})}\BibitemShut {NoStop}%
\bibitem [{\citenamefont {Helfand}\ and\ \citenamefont
  {Pearson}(1983)}]{primitive1}%
  \BibitemOpen
  \bibfield  {author} {\bibinfo {author} {\bibfnamefont {E.}~\bibnamefont
  {Helfand}}\ and\ \bibinfo {author} {\bibfnamefont {D.~S.}\ \bibnamefont
  {Pearson}},\ }\bibfield  {title} {\bibinfo {title} {Statistics of the
  entanglement of polymers: Unentangled loops and primitive paths},\
  }\href@noop {} {\bibfield  {journal} {\bibinfo  {journal} {J. Chem. Phys.}\
  }\textbf {\bibinfo {volume} {79}},\ \bibinfo {pages} {2054} (\bibinfo {year}
  {1983})}\BibitemShut {NoStop}%
\bibitem [{\citenamefont {Nechaev}(1988)}]{nech1}%
  \BibitemOpen
  \bibfield  {author} {\bibinfo {author} {\bibfnamefont {S.~K.}\ \bibnamefont
  {Nechaev}},\ }\bibfield  {title} {\bibinfo {title} {Topological properties of
  a two-dimensional polymer chain in the lattice of obstacles},\ }\href
  {https://doi.org/10.1088/0305-4470/21/18/018} {\bibfield  {journal} {\bibinfo
   {journal} {Journal of Physics A: Mathematical and General}\ }\textbf
  {\bibinfo {volume} {21}},\ \bibinfo {pages} {3659} (\bibinfo {year}
  {1988})}\BibitemShut {NoStop}%
\bibitem [{\citenamefont {Nechaev}(1999)}]{nechaev-houches}%
  \BibitemOpen
  \bibfield  {author} {\bibinfo {author} {\bibfnamefont {S.}~\bibnamefont
  {Nechaev}},\ }\bibfield  {title} {\bibinfo {title} {Statistics of knots and
  entangled random walks},\ }in\ \href@noop {} {\emph {\bibinfo {booktitle}
  {Aspects topologiques de la physique en basse dimension. Topological aspects
  of low dimensional systems}}},\ \bibinfo {editor} {edited by\ \bibinfo
  {editor} {\bibfnamefont {A.}~\bibnamefont {Comtet}}, \bibinfo {editor}
  {\bibfnamefont {T.}~\bibnamefont {Jolic{\oe}ur}}, \bibinfo {editor}
  {\bibfnamefont {S.}~\bibnamefont {Ouvry}},\ and\ \bibinfo {editor}
  {\bibfnamefont {F.}~\bibnamefont {David}}}\ (\bibinfo  {publisher} {Springer
  Berlin Heidelberg},\ \bibinfo {address} {Berlin, Heidelberg},\ \bibinfo
  {year} {1999})\ pp.\ \bibinfo {pages} {643--733}\BibitemShut {NoStop}%
\bibitem [{\citenamefont {Khokhlov}\ and\ \citenamefont
  {Nechaev}(1985)}]{primitive2}%
  \BibitemOpen
  \bibfield  {author} {\bibinfo {author} {\bibfnamefont {A.}~\bibnamefont
  {Khokhlov}}\ and\ \bibinfo {author} {\bibfnamefont {S.}~\bibnamefont
  {Nechaev}},\ }\bibfield  {title} {\bibinfo {title} {Polymer chain in an array
  of obstacles},\ }\href
  {https://doi.org/https://doi.org/10.1016/0375-9601(85)90678-4} {\bibfield
  {journal} {\bibinfo  {journal} {Physics Letters A}\ }\textbf {\bibinfo
  {volume} {112}},\ \bibinfo {pages} {156} (\bibinfo {year}
  {1985})}\BibitemShut {NoStop}%
\bibitem [{\citenamefont {Nechaev}(1998)}]{nech-UFN}%
  \BibitemOpen
  \bibfield  {author} {\bibinfo {author} {\bibfnamefont {S.~K.}\ \bibnamefont
  {Nechaev}},\ }\bibfield  {title} {\bibinfo {title} {Problems of probabilistic
  topology: the statistics of knots and non-commutative random walks},\ }\href
  {https://doi.org/10.1070/PU1998v041n04ABEH000382} {\bibfield  {journal}
  {\bibinfo  {journal} {Phys. Usp.}\ }\textbf {\bibinfo {volume} {41}},\
  \bibinfo {pages} {313} (\bibinfo {year} {1998})}\BibitemShut {NoStop}%
\bibitem [{\citenamefont {Bulycheva}\ and\ \citenamefont
  {Gorsky}(2014{\natexlab{a}})}]{bulycheva2014limit}%
  \BibitemOpen
  \bibfield  {author} {\bibinfo {author} {\bibfnamefont {K.}~\bibnamefont
  {Bulycheva}}\ and\ \bibinfo {author} {\bibfnamefont {A.}~\bibnamefont
  {Gorsky}},\ }\bibfield  {title} {\bibinfo {title} {Limit cycles in
  renormalization group dynamics},\ }\href@noop {} {\bibfield  {journal}
  {\bibinfo  {journal} {Physics-Uspekhi}\ }\textbf {\bibinfo {volume} {57}},\
  \bibinfo {pages} {171} (\bibinfo {year} {2014}{\natexlab{a}})}\BibitemShut
  {NoStop}%
\bibitem [{\citenamefont {Seiberg}\ and\ \citenamefont
  {Witten}(1994)}]{seiberg1994electric}%
  \BibitemOpen
  \bibfield  {author} {\bibinfo {author} {\bibfnamefont {N.}~\bibnamefont
  {Seiberg}}\ and\ \bibinfo {author} {\bibfnamefont {E.}~\bibnamefont
  {Witten}},\ }\bibfield  {title} {\bibinfo {title} {Electric-magnetic duality,
  monopole condensation, and confinement in n= 2 supersymmetric yang-mills
  theory},\ }\href@noop {} {\bibfield  {journal} {\bibinfo  {journal} {Nuclear
  Physics B}\ }\textbf {\bibinfo {volume} {426}},\ \bibinfo {pages} {19}
  (\bibinfo {year} {1994})}\BibitemShut {NoStop}%
\bibitem [{\citenamefont {Nekrasov}(2003)}]{nekrasov2003seiberg}%
  \BibitemOpen
  \bibfield  {author} {\bibinfo {author} {\bibfnamefont {N.~A.}\ \bibnamefont
  {Nekrasov}},\ }\bibfield  {title} {\bibinfo {title} {Seiberg-witten
  prepotential from instanton counting},\ }\href@noop {} {\bibfield  {journal}
  {\bibinfo  {journal} {Advances in Theoretical and Mathematical Physics}\
  }\textbf {\bibinfo {volume} {7}},\ \bibinfo {pages} {831} (\bibinfo {year}
  {2003})}\BibitemShut {NoStop}%
\bibitem [{\citenamefont {Gorsky}\ \emph {et~al.}(1995)\citenamefont {Gorsky},
  \citenamefont {Krichever}, \citenamefont {Marshakov}, \citenamefont
  {Mironov},\ and\ \citenamefont {Morozov}}]{gorsky1995integrability}%
  \BibitemOpen
  \bibfield  {author} {\bibinfo {author} {\bibfnamefont {A.}~\bibnamefont
  {Gorsky}}, \bibinfo {author} {\bibfnamefont {I.}~\bibnamefont {Krichever}},
  \bibinfo {author} {\bibfnamefont {A.}~\bibnamefont {Marshakov}}, \bibinfo
  {author} {\bibfnamefont {A.}~\bibnamefont {Mironov}},\ and\ \bibinfo {author}
  {\bibfnamefont {A.}~\bibnamefont {Morozov}},\ }\bibfield  {title} {\bibinfo
  {title} {Integrability and seiberg-witten exact solution},\ }\href@noop {}
  {\bibfield  {journal} {\bibinfo  {journal} {Physics Letters B}\ }\textbf
  {\bibinfo {volume} {355}},\ \bibinfo {pages} {466} (\bibinfo {year}
  {1995})}\BibitemShut {NoStop}%
\bibitem [{\citenamefont {Martinec}\ and\ \citenamefont
  {Warner}(1996)}]{martinec1996integrable}%
  \BibitemOpen
  \bibfield  {author} {\bibinfo {author} {\bibfnamefont {E.~J.}\ \bibnamefont
  {Martinec}}\ and\ \bibinfo {author} {\bibfnamefont {N.~P.}\ \bibnamefont
  {Warner}},\ }\bibfield  {title} {\bibinfo {title} {Integrable systems and
  supersymmetric gauge theory},\ }\href@noop {} {\bibfield  {journal} {\bibinfo
   {journal} {Nuclear Physics B}\ }\textbf {\bibinfo {volume} {459}},\ \bibinfo
  {pages} {97} (\bibinfo {year} {1996})}\BibitemShut {NoStop}%
\bibitem [{\citenamefont {Donagi}\ and\ \citenamefont
  {Witten}(1996)}]{donagi1996supersymmetric}%
  \BibitemOpen
  \bibfield  {author} {\bibinfo {author} {\bibfnamefont {R.}~\bibnamefont
  {Donagi}}\ and\ \bibinfo {author} {\bibfnamefont {E.}~\bibnamefont
  {Witten}},\ }\bibfield  {title} {\bibinfo {title} {Supersymmetric yang-mills
  theory and integrable systems},\ }\href@noop {} {\bibfield  {journal}
  {\bibinfo  {journal} {Nuclear Physics B}\ }\textbf {\bibinfo {volume}
  {460}},\ \bibinfo {pages} {299} (\bibinfo {year} {1996})}\BibitemShut
  {NoStop}%
\bibitem [{\citenamefont {Nekrasov}(2019)}]{nekrasov2019bps}%
  \BibitemOpen
  \bibfield  {author} {\bibinfo {author} {\bibfnamefont {N.}~\bibnamefont
  {Nekrasov}},\ }\bibfield  {title} {\bibinfo {title} {Bps/cft correspondence
  iv: sigma models and defects in gauge theory},\ }\href@noop {} {\bibfield
  {journal} {\bibinfo  {journal} {Letters in Mathematical Physics}\ }\textbf
  {\bibinfo {volume} {109}},\ \bibinfo {pages} {579} (\bibinfo {year}
  {2019})}\BibitemShut {NoStop}%
\bibitem [{\citenamefont {Alday}\ \emph {et~al.}(2010)\citenamefont {Alday},
  \citenamefont {Gaiotto},\ and\ \citenamefont
  {Tachikawa}}]{alday2010liouville}%
  \BibitemOpen
  \bibfield  {author} {\bibinfo {author} {\bibfnamefont {L.~F.}\ \bibnamefont
  {Alday}}, \bibinfo {author} {\bibfnamefont {D.}~\bibnamefont {Gaiotto}},\
  and\ \bibinfo {author} {\bibfnamefont {Y.}~\bibnamefont {Tachikawa}},\
  }\bibfield  {title} {\bibinfo {title} {Liouville correlation functions from
  four-dimensional gauge theories},\ }\href@noop {} {\bibfield  {journal}
  {\bibinfo  {journal} {Letters in Mathematical Physics}\ }\textbf {\bibinfo
  {volume} {91}},\ \bibinfo {pages} {167} (\bibinfo {year} {2010})}\BibitemShut
  {NoStop}%
\bibitem [{\citenamefont {Beanland}\ \emph {et~al.}(2009)\citenamefont
  {Beanland}, \citenamefont {Roberts},\ and\ \citenamefont {Stevenson}}]{RT}%
  \BibitemOpen
  \bibfield  {author} {\bibinfo {author} {\bibfnamefont {K.}~\bibnamefont
  {Beanland}}, \bibinfo {author} {\bibfnamefont {J.~W.}\ \bibnamefont
  {Roberts}},\ and\ \bibinfo {author} {\bibfnamefont {C.}~\bibnamefont
  {Stevenson}},\ }\bibfield  {title} {\bibinfo {title} {Modifications of
  thomae's function and differentiability},\ }\href
  {https://doi.org/10.1080/00029890.2009.11920968} {\bibfield  {journal}
  {\bibinfo  {journal} {The American Mathematical Monthly}\ }\textbf {\bibinfo
  {volume} {116}},\ \bibinfo {pages} {531} (\bibinfo {year} {2009})},\ \Eprint
  {https://arxiv.org/abs/https://doi.org/10.1080/00029890.2009.11920968}
  {https://doi.org/10.1080/00029890.2009.11920968} \BibitemShut {NoStop}%
\bibitem [{\citenamefont {Vandervelde}(2009)}]{euclid}%
  \BibitemOpen
  \bibfield  {author} {\bibinfo {author} {\bibfnamefont {S.}~\bibnamefont
  {Vandervelde}},\ }\href@noop {} {\emph {\bibinfo {title} {Chapter 9: Sneaky
  segments. Circle in a Box}}}\ (\bibinfo  {publisher} {MSRI Mathematical
  Circles Library. Mathematical Sciences Research Institute and American
  Mathematical Society},\ \bibinfo {year} {2009})\ p.\ \bibinfo {pages}
  {101–106}\BibitemShut {NoStop}%
\bibitem [{\citenamefont {O'Sullivan}(2018)}]{eisen}%
  \BibitemOpen
  \bibfield  {author} {\bibinfo {author} {\bibfnamefont {C.}~\bibnamefont
  {O'Sullivan}},\ }\bibfield  {title} {\bibinfo {title} {Formulas for
  non-holomorphic eisenstein series and for the riemann zeta function at odd
  integers},\ }\href {https://doi.org/10.1007/s40993-018-0129-7} {\bibfield
  {journal} {\bibinfo  {journal} {Research in Number Theory}\ }\textbf
  {\bibinfo {volume} {4}},\ \bibinfo {pages} {36} (\bibinfo {year}
  {2018})}\BibitemShut {NoStop}%
\bibitem [{\citenamefont {Ribeiro}\ and\ \citenamefont
  {Yakubovich}(2022)}]{epstein}%
  \BibitemOpen
  \bibfield  {author} {\bibinfo {author} {\bibfnamefont {P.}~\bibnamefont
  {Ribeiro}}\ and\ \bibinfo {author} {\bibfnamefont {S.}~\bibnamefont
  {Yakubovich}},\ }\href@noop {} {\bibinfo {title} {On the epstein zeta
  function and the zeros of a class of dirichlet series}} (\bibinfo {year}
  {2022}),\ \Eprint {https://arxiv.org/abs/2112.10561} {arXiv:2112.10561
  [math.NT]} \BibitemShut {NoStop}%
\bibitem [{\citenamefont {Siegel}\ and\ \citenamefont
  {Raghavan}(1961)}]{siegel}%
  \BibitemOpen
  \bibfield  {author} {\bibinfo {author} {\bibfnamefont {C.~L.}\ \bibnamefont
  {Siegel}}\ and\ \bibinfo {author} {\bibfnamefont {S.}~\bibnamefont
  {Raghavan}},\ }\href@noop {} {\emph {\bibinfo {title} {Lectures on advanced
  analytic number theory}}}\ (\bibinfo  {publisher} {Tata Institute of
  Fundamental Research, Mumbai, India},\ \bibinfo {year} {1961})\BibitemShut
  {NoStop}%
\bibitem [{\citenamefont {Motohashi}(1968)}]{motohashi}%
  \BibitemOpen
  \bibfield  {author} {\bibinfo {author} {\bibfnamefont {Y.}~\bibnamefont
  {Motohashi}},\ }\bibfield  {title} {\bibinfo {title} {A new proof of the
  limit formula of kronecker},\ }\href {https://doi.org/10.3792/pja/1195521077}
  {\bibfield  {journal} {\bibinfo  {journal} {Proceedings of the Japan
  Academy}\ }\textbf {\bibinfo {volume} {44}},\ \bibinfo {pages} {614 }
  (\bibinfo {year} {1968})}\BibitemShut {NoStop}%
\bibitem [{\citenamefont {Apostol}(1990)}]{dedekind}%
  \BibitemOpen
  \bibfield  {author} {\bibinfo {author} {\bibfnamefont {T.~M.}\ \bibnamefont
  {Apostol}},\ }\href@noop {} {\emph {\bibinfo {title} {Modular functions and
  Dirichlet Series in Number Theory. Chapter 3}}},\ Vol.~\bibinfo {volume}
  {41}\ (\bibinfo  {publisher} {Springer-Verlag},\ \bibinfo {year}
  {1990})\BibitemShut {NoStop}%
\bibitem [{\citenamefont {Fleron}(1994)}]{devil}%
  \BibitemOpen
  \bibfield  {author} {\bibinfo {author} {\bibfnamefont {J.~F.}\ \bibnamefont
  {Fleron}},\ }\bibfield  {title} {\bibinfo {title} {A note on the history of
  the cantor set and cantor function},\ }\href
  {https://doi.org/10.1080/0025570X.1994.11996201} {\bibfield  {journal}
  {\bibinfo  {journal} {Mathematics Magazine}\ }\textbf {\bibinfo {volume}
  {67}},\ \bibinfo {pages} {136} (\bibinfo {year} {1994})},\ \Eprint
  {https://arxiv.org/abs/https://doi.org/10.1080/0025570X.1994.11996201}
  {https://doi.org/10.1080/0025570X.1994.11996201} \BibitemShut {NoStop}%
\bibitem [{\citenamefont {Gorsky}\ and\ \citenamefont
  {Milekhin}(2015)}]{gorsky2015rg}%
  \BibitemOpen
  \bibfield  {author} {\bibinfo {author} {\bibfnamefont {A.}~\bibnamefont
  {Gorsky}}\ and\ \bibinfo {author} {\bibfnamefont {A.}~\bibnamefont
  {Milekhin}},\ }\bibfield  {title} {\bibinfo {title} {Rg-whitham dynamics and
  complex hamiltonian systems},\ }\href@noop {} {\bibfield  {journal} {\bibinfo
   {journal} {Nuclear Physics B}\ }\textbf {\bibinfo {volume} {895}},\ \bibinfo
  {pages} {33} (\bibinfo {year} {2015})}\BibitemShut {NoStop}%
\bibitem [{\citenamefont {Tom\'as}(2014)}]{farey}%
  \BibitemOpen
  \bibfield  {author} {\bibinfo {author} {\bibfnamefont {R.}~\bibnamefont
  {Tom\'as}},\ }\bibfield  {title} {\bibinfo {title} {From farey sequences to
  resonance diagrams},\ }\href {https://doi.org/10.1103/PhysRevSTAB.17.014001}
  {\bibfield  {journal} {\bibinfo  {journal} {Phys. Rev. ST Accel. Beams}\
  }\textbf {\bibinfo {volume} {17}},\ \bibinfo {pages} {014001} (\bibinfo
  {year} {2014})}\BibitemShut {NoStop}%
\bibitem [{\citenamefont {Northshield}(2015)}]{ford}%
  \BibitemOpen
  \bibfield  {author} {\bibinfo {author} {\bibfnamefont {S.}~\bibnamefont
  {Northshield}},\ }\href@noop {} {\bibinfo {title} {Ford circles and spheres}}
  (\bibinfo {year} {2015}),\ \Eprint {https://arxiv.org/abs/1503.00813}
  {arXiv:1503.00813 [math.NT]} \BibitemShut {NoStop}%
\bibitem [{\citenamefont {Georgelin}\ \emph {et~al.}(1997)\citenamefont
  {Georgelin}, \citenamefont {Masson},\ and\ \citenamefont {Wallet}}]{wallett}%
  \BibitemOpen
  \bibfield  {author} {\bibinfo {author} {\bibfnamefont {Y.}~\bibnamefont
  {Georgelin}}, \bibinfo {author} {\bibfnamefont {T.}~\bibnamefont {Masson}},\
  and\ \bibinfo {author} {\bibfnamefont {J.-C.}\ \bibnamefont {Wallet}},\
  }\bibfield  {title} {\bibinfo {title} {Modular groups, visibility diagram and
  quantum hall effect},\ }\href {https://doi.org/10.1088/0305-4470/30/14/017}
  {\bibfield  {journal} {\bibinfo  {journal} {J. Phys. A: Math. Gen.}\ }\textbf
  {\bibinfo {volume} {30}},\ \bibinfo {pages} {5065} (\bibinfo {year}
  {1997})}\BibitemShut {NoStop}%
\bibitem [{\citenamefont {Livio}(2008)}]{phyllotaxis}%
  \BibitemOpen
  \bibfield  {author} {\bibinfo {author} {\bibfnamefont {M.}~\bibnamefont
  {Livio}},\ }\href@noop {} {\emph {\bibinfo {title} {The Golden Ratio: The
  Story of PHI, the World's Most Astonishing Number}}}\ (\bibinfo  {publisher}
  {Broadway Books, New York},\ \bibinfo {year} {2008})\BibitemShut {NoStop}%
\bibitem [{\citenamefont {Rothen}\ and\ \citenamefont
  {Koch}(1989{\natexlab{a}})}]{phyllo2}%
  \BibitemOpen
  \bibfield  {author} {\bibinfo {author} {\bibfnamefont {F.}~\bibnamefont
  {Rothen}}\ and\ \bibinfo {author} {\bibfnamefont {A.-J.}\ \bibnamefont
  {Koch}},\ }\bibfield  {title} {\bibinfo {title} {Phyllotaxis, or the
  properties of spiral lattices. - i. shape invariance under compression},\
  }\href {https://doi.org/10.1051/jphys:01989005006063300} {\bibfield
  {journal} {\bibinfo  {journal} {J. Phys. France}\ }\textbf {\bibinfo {volume}
  {50}},\ \bibinfo {pages} {633} (\bibinfo {year}
  {1989}{\natexlab{a}})}\BibitemShut {NoStop}%
\bibitem [{\citenamefont {Rothen}\ and\ \citenamefont
  {Koch}(1989{\natexlab{b}})}]{phyllo3}%
  \BibitemOpen
  \bibfield  {author} {\bibinfo {author} {\bibfnamefont {F.}~\bibnamefont
  {Rothen}}\ and\ \bibinfo {author} {\bibfnamefont {A.~J.}\ \bibnamefont
  {Koch}},\ }\bibfield  {title} {\bibinfo {title} {Phyllotaxis or the
  properties of spiral lattices. - ii. packing of circles along logarithmic
  spirals},\ }\href@noop {} {\bibfield  {journal} {\bibinfo  {journal} {Journal
  De Physique}\ }\textbf {\bibinfo {volume} {50}},\ \bibinfo {pages} {1603}
  (\bibinfo {year} {1989}{\natexlab{b}})}\BibitemShut {NoStop}%
\bibitem [{\citenamefont {Kunz}\ and\ \citenamefont {Rothen}(1992)}]{phyllo4}%
  \BibitemOpen
  \bibfield  {author} {\bibinfo {author} {\bibfnamefont {M.}~\bibnamefont
  {Kunz}}\ and\ \bibinfo {author} {\bibfnamefont {F.}~\bibnamefont {Rothen}},\
  }\bibfield  {title} {\bibinfo {title} {Phyllotaxis or the properties of
  spiral lattices. iii. an algebraic model of morphogenesis},\ }\href
  {https://doi.org/10.1051/jp1:1992273} {\bibfield  {journal} {\bibinfo
  {journal} {J. Phys. I France}\ }\textbf {\bibinfo {volume} {2}},\ \bibinfo
  {pages} {2131} (\bibinfo {year} {1992})}\BibitemShut {NoStop}%
\bibitem [{\citenamefont {Lee}\ and\ \citenamefont {Levitov}(1998)}]{levitov}%
  \BibitemOpen
  \bibfield  {author} {\bibinfo {author} {\bibfnamefont {H.-W.}\ \bibnamefont
  {Lee}}\ and\ \bibinfo {author} {\bibfnamefont {L.}~\bibnamefont {Levitov}},\
  }\href@noop {} {\emph {\bibinfo {title} {Universality in Phyllotaxis: a
  Mechanical Theory}}}\ (\bibinfo  {publisher} {World Scientific},\ \bibinfo
  {year} {1998})\BibitemShut {NoStop}%
\bibitem [{\citenamefont {Levitov}(1991{\natexlab{a}})}]{levitov1}%
  \BibitemOpen
  \bibfield  {author} {\bibinfo {author} {\bibfnamefont {L.}~\bibnamefont
  {Levitov}},\ }\bibfield  {title} {\bibinfo {title} {Energetic approach to
  phyllotaxis},\ }\href@noop {} {\bibfield  {journal} {\bibinfo  {journal}
  {Europhysics Letters}\ }\textbf {\bibinfo {volume} {14}},\ \bibinfo {pages}
  {533} (\bibinfo {year} {1991}{\natexlab{a}})}\BibitemShut {NoStop}%
\bibitem [{\citenamefont {Levitov}(1991{\natexlab{b}})}]{levitov3}%
  \BibitemOpen
  \bibfield  {author} {\bibinfo {author} {\bibfnamefont {L.~S.}\ \bibnamefont
  {Levitov}},\ }\bibfield  {title} {\bibinfo {title} {Phyllotaxis of flux
  lattices in layered superconductors},\ }\href@noop {} {\bibfield  {journal}
  {\bibinfo  {journal} {Physical review letters}\ }\textbf {\bibinfo {volume}
  {66}},\ \bibinfo {pages} {224} (\bibinfo {year}
  {1991}{\natexlab{b}})}\BibitemShut {NoStop}%
\bibitem [{\citenamefont {Douady}\ and\ \citenamefont
  {Couder}(1992)}]{rotating}%
  \BibitemOpen
  \bibfield  {author} {\bibinfo {author} {\bibfnamefont {S.}~\bibnamefont
  {Douady}}\ and\ \bibinfo {author} {\bibfnamefont {Y.}~\bibnamefont
  {Couder}},\ }\bibfield  {title} {\bibinfo {title} {Phyllotaxis as a physical
  self-organized growth process},\ }\href
  {https://doi.org/10.1103/PhysRevLett.68.2098} {\bibfield  {journal} {\bibinfo
   {journal} {Phys. Rev. Lett.}\ }\textbf {\bibinfo {volume} {68}},\ \bibinfo
  {pages} {2098} (\bibinfo {year} {1992})}\BibitemShut {NoStop}%
\bibitem [{\citenamefont {Nisoli}\ \emph {et~al.}(2009)\citenamefont {Nisoli},
  \citenamefont {Gabor}, \citenamefont {Lammert}, \citenamefont {Maynard},\
  and\ \citenamefont {Crespi}}]{cactus}%
  \BibitemOpen
  \bibfield  {author} {\bibinfo {author} {\bibfnamefont {C.}~\bibnamefont
  {Nisoli}}, \bibinfo {author} {\bibfnamefont {N.~M.}\ \bibnamefont {Gabor}},
  \bibinfo {author} {\bibfnamefont {P.~E.}\ \bibnamefont {Lammert}}, \bibinfo
  {author} {\bibfnamefont {J.~D.}\ \bibnamefont {Maynard}},\ and\ \bibinfo
  {author} {\bibfnamefont {V.~H.}\ \bibnamefont {Crespi}},\ }\bibfield  {title}
  {\bibinfo {title} {Static and dynamical phyllotaxis in a magnetic cactus},\
  }\href {https://doi.org/10.1103/PhysRevLett.102.186103} {\bibfield  {journal}
  {\bibinfo  {journal} {Phys. Rev. Lett.}\ }\textbf {\bibinfo {volume} {102}},\
  \bibinfo {pages} {186103} (\bibinfo {year} {2009})}\BibitemShut {NoStop}%
\bibitem [{\citenamefont {Abrikosov}(1957)}]{abrikosov}%
  \BibitemOpen
  \bibfield  {author} {\bibinfo {author} {\bibfnamefont {A.}~\bibnamefont
  {Abrikosov}},\ }\bibfield  {title} {\bibinfo {title} {The magnetic properties
  of superconducting alloys},\ }\href
  {https://doi.org/https://doi.org/10.1016/0022-3697(57)90083-5} {\bibfield
  {journal} {\bibinfo  {journal} {Journal of Physics and Chemistry of Solids}\
  }\textbf {\bibinfo {volume} {2}},\ \bibinfo {pages} {199} (\bibinfo {year}
  {1957})}\BibitemShut {NoStop}%
\bibitem [{\citenamefont {Rammal}\ \emph {et~al.}(1986)\citenamefont {Rammal},
  \citenamefont {Toulouse},\ and\ \citenamefont {Virasoro}}]{ultra1}%
  \BibitemOpen
  \bibfield  {author} {\bibinfo {author} {\bibfnamefont {R.}~\bibnamefont
  {Rammal}}, \bibinfo {author} {\bibfnamefont {G.}~\bibnamefont {Toulouse}},\
  and\ \bibinfo {author} {\bibfnamefont {M.~A.}\ \bibnamefont {Virasoro}},\
  }\bibfield  {title} {\bibinfo {title} {Ultrametricity for physicists},\
  }\href {https://doi.org/10.1103/RevModPhys.58.765} {\bibfield  {journal}
  {\bibinfo  {journal} {Rev. Mod. Phys.}\ }\textbf {\bibinfo {volume} {58}},\
  \bibinfo {pages} {765} (\bibinfo {year} {1986})}\BibitemShut {NoStop}%
\bibitem [{\citenamefont {Mezard}\ \emph {et~al.}(1986)\citenamefont {Mezard},
  \citenamefont {Parisi},\ and\ \citenamefont {Virasoro}}]{mez}%
  \BibitemOpen
  \bibfield  {author} {\bibinfo {author} {\bibfnamefont {M.}~\bibnamefont
  {Mezard}}, \bibinfo {author} {\bibfnamefont {G.}~\bibnamefont {Parisi}},\
  and\ \bibinfo {author} {\bibfnamefont {M.}~\bibnamefont {Virasoro}},\
  }\href@noop {} {\emph {\bibinfo {title} {Spin glass theory and beyond}}}\
  (\bibinfo  {publisher} {World Scientific},\ \bibinfo {year}
  {1986})\BibitemShut {NoStop}%
\bibitem [{\citenamefont {Rajput}(2021)}]{silver}%
  \BibitemOpen
  \bibfield  {author} {\bibinfo {author} {\bibfnamefont {D.~C.}\ \bibnamefont
  {Rajput}},\ }\bibfield  {title} {\bibinfo {title} {Golden ratio},\ }\href
  {https://doi.org/10.24297/jam.v20i.8945} {\bibfield  {journal} {\bibinfo
  {journal} {Journal of Advances in Mathematics}\ }\textbf {\bibinfo {volume}
  {20}},\ \bibinfo {pages} {19–42} (\bibinfo {year} {2021})}\BibitemShut
  {NoStop}%
\bibitem [{\citenamefont {Nechaev}\ and\ \citenamefont
  {Polovnikov}(2016)}]{conf}%
  \BibitemOpen
  \bibfield  {author} {\bibinfo {author} {\bibfnamefont {S.~K.}\ \bibnamefont
  {Nechaev}}\ and\ \bibinfo {author} {\bibfnamefont {K.~E.}\ \bibnamefont
  {Polovnikov}},\ }\bibfield  {title} {\bibinfo {title} {From geometric optics
  to plants: the eikonal equation for buckling},\ }\href@noop {} {\bibfield
  {journal} {\bibinfo  {journal} {Soft matter}\ }\textbf {\bibinfo {volume}
  {13}},\ \bibinfo {pages} {1420} (\bibinfo {year} {2016})}\BibitemShut
  {NoStop}%
\bibitem [{\citenamefont {Kaplan}\ \emph {et~al.}(2009)\citenamefont {Kaplan},
  \citenamefont {Lee}, \citenamefont {Son},\ and\ \citenamefont
  {Stephanov}}]{kaplan1}%
  \BibitemOpen
  \bibfield  {author} {\bibinfo {author} {\bibfnamefont {D.~B.}\ \bibnamefont
  {Kaplan}}, \bibinfo {author} {\bibfnamefont {J.-W.}\ \bibnamefont {Lee}},
  \bibinfo {author} {\bibfnamefont {D.~T.}\ \bibnamefont {Son}},\ and\ \bibinfo
  {author} {\bibfnamefont {M.~A.}\ \bibnamefont {Stephanov}},\ }\bibfield
  {title} {\bibinfo {title} {Conformality lost},\ }\href
  {https://doi.org/10.1103/PhysRevD.80.125005} {\bibfield  {journal} {\bibinfo
  {journal} {Phys. Rev. D}\ }\textbf {\bibinfo {volume} {80}},\ \bibinfo
  {pages} {125005} (\bibinfo {year} {2009})}\BibitemShut {NoStop}%
\bibitem [{\citenamefont {Lutken}\ and\ \citenamefont {Ross}(2009)}]{lutken}%
  \BibitemOpen
  \bibfield  {author} {\bibinfo {author} {\bibfnamefont {C.~A.}\ \bibnamefont
  {Lutken}}\ and\ \bibinfo {author} {\bibfnamefont {G.~G.}\ \bibnamefont
  {Ross}},\ }\href@noop {} {\bibinfo {title} {Implications of experimental
  probes of the rg-flow in quantum hall systems}} (\bibinfo {year} {2009}),\
  \Eprint {https://arxiv.org/abs/0906.5551} {arXiv:0906.5551 [cond-mat.other]}
  \BibitemShut {NoStop}%
\bibitem [{\citenamefont {Carpentier}(1999)}]{carpentier}%
  \BibitemOpen
  \bibfield  {author} {\bibinfo {author} {\bibfnamefont {D.}~\bibnamefont
  {Carpentier}},\ }\bibfield  {title} {\bibinfo {title} {Renormalization of
  modular invariant coulomb gas and sine-gordon theories, and the quantum hall
  flow diagram},\ }\href {https://doi.org/10.1088/0305-4470/32/21/301}
  {\bibfield  {journal} {\bibinfo  {journal} {Journal of Physics A:
  Mathematical and General}\ }\textbf {\bibinfo {volume} {32}},\ \bibinfo
  {pages} {3865} (\bibinfo {year} {1999})}\BibitemShut {NoStop}%
\bibitem [{\citenamefont {Fischer}(1993)}]{fisher}%
  \BibitemOpen
  \bibfield  {author} {\bibinfo {author} {\bibfnamefont {K.}~\bibnamefont
  {Fischer}},\ }\bibfield  {title} {\bibinfo {title} {Kosterlitz-thouless
  transition in layered high-tc superconductors},\ }\href
  {https://doi.org/https://doi.org/10.1016/0921-4534(93)90023-J} {\bibfield
  {journal} {\bibinfo  {journal} {Physica C: Superconductivity}\ }\textbf
  {\bibinfo {volume} {210}},\ \bibinfo {pages} {179} (\bibinfo {year}
  {1993})}\BibitemShut {NoStop}%
\bibitem [{\citenamefont {Gaudin}(1973)}]{gaudin}%
  \BibitemOpen
  \bibfield  {author} {\bibinfo {author} {\bibfnamefont {M.}~\bibnamefont
  {Gaudin}},\ }\bibfield  {title} {\bibinfo {title} {Gaz coulombien discret \`a
  une dimension},\ }\href {https://doi.org/10.1051/jphys:01973003407051100}
  {\bibfield  {journal} {\bibinfo  {journal} {J. Phys. France}\ }\textbf
  {\bibinfo {volume} {34}},\ \bibinfo {pages} {511} (\bibinfo {year}
  {1973})}\BibitemShut {NoStop}%
\bibitem [{\citenamefont {Hubbard}(1978)}]{Hubbard}%
  \BibitemOpen
  \bibfield  {author} {\bibinfo {author} {\bibfnamefont {J.}~\bibnamefont
  {Hubbard}},\ }\bibfield  {title} {\bibinfo {title} {Generalized wigner
  lattices in one dimension and some applications to tetracyanoquinodimethane
  (tcnq) salts},\ }\href {https://doi.org/10.1103/PhysRevB.17.494} {\bibfield
  {journal} {\bibinfo  {journal} {Phys. Rev. B}\ }\textbf {\bibinfo {volume}
  {17}},\ \bibinfo {pages} {494} (\bibinfo {year} {1978})}\BibitemShut
  {NoStop}%
\bibitem [{\citenamefont {Pokrovsky}\ and\ \citenamefont
  {Uimin}(1978)}]{Pokrovsky}%
  \BibitemOpen
  \bibfield  {author} {\bibinfo {author} {\bibfnamefont {V.~L.}\ \bibnamefont
  {Pokrovsky}}\ and\ \bibinfo {author} {\bibfnamefont {G.~V.}\ \bibnamefont
  {Uimin}},\ }\bibfield  {title} {\bibinfo {title} {On the properties of
  monolayers of adsorbed atoms},\ }\href
  {https://doi.org/10.1088/0022-3719/11/16/022} {\bibfield  {journal} {\bibinfo
   {journal} {Journal of Physics C: Solid State Physics}\ }\textbf {\bibinfo
  {volume} {11}},\ \bibinfo {pages} {3535} (\bibinfo {year}
  {1978})}\BibitemShut {NoStop}%
\bibitem [{\citenamefont {Bak}\ and\ \citenamefont {Bruinsma}(1982)}]{Bak}%
  \BibitemOpen
  \bibfield  {author} {\bibinfo {author} {\bibfnamefont {P.}~\bibnamefont
  {Bak}}\ and\ \bibinfo {author} {\bibfnamefont {R.}~\bibnamefont {Bruinsma}},\
  }\bibfield  {title} {\bibinfo {title} {One-dimensional ising model and the
  complete devil's staircase},\ }\href
  {https://doi.org/10.1103/PhysRevLett.49.249} {\bibfield  {journal} {\bibinfo
  {journal} {Phys. Rev. Lett.}\ }\textbf {\bibinfo {volume} {49}},\ \bibinfo
  {pages} {249} (\bibinfo {year} {1982})}\BibitemShut {NoStop}%
\bibitem [{\citenamefont {Bak}(1982{\natexlab{b}})}]{Bak2}%
  \BibitemOpen
  \bibfield  {author} {\bibinfo {author} {\bibfnamefont {P.}~\bibnamefont
  {Bak}},\ }\bibfield  {title} {\bibinfo {title} {Commensurate phases,
  incommensurate phases and the devil's staircase},\ }\href
  {https://doi.org/10.1088/0034-4885/45/6/001} {\bibfield  {journal} {\bibinfo
  {journal} {Reports on Progress in Physics}\ }\textbf {\bibinfo {volume}
  {45}},\ \bibinfo {pages} {587} (\bibinfo {year}
  {1982}{\natexlab{b}})}\BibitemShut {NoStop}%
\bibitem [{\citenamefont {Burkov}\ and\ \citenamefont {Sinai}(1983)}]{sinai}%
  \BibitemOpen
  \bibfield  {author} {\bibinfo {author} {\bibfnamefont {S.~E.}\ \bibnamefont
  {Burkov}}\ and\ \bibinfo {author} {\bibfnamefont {Y.~G.}\ \bibnamefont
  {Sinai}},\ }\bibfield  {title} {\bibinfo {title} {Phase diagrams of
  one-dimensional lattice models with long-range antiferromagnetic
  interaction},\ }\href {https://doi.org/10.1070/RM1983v038n04ABEH004211}
  {\bibfield  {journal} {\bibinfo  {journal} {Russian Mathematical Surveys}\
  }\textbf {\bibinfo {volume} {38}},\ \bibinfo {pages} {235} (\bibinfo {year}
  {1983})}\BibitemShut {NoStop}%
\bibitem [{\citenamefont {Nicolas}\ and\ \citenamefont
  {Robin}(1997)}]{composite}%
  \BibitemOpen
  \bibfield  {author} {\bibinfo {author} {\bibfnamefont {J.-L.}\ \bibnamefont
  {Nicolas}}\ and\ \bibinfo {author} {\bibfnamefont {G.}~\bibnamefont
  {Robin}},\ }\bibfield  {title} {\bibinfo {title} {Highly composite numbers by
  srinivasa ramanujan},\ }\href {https://doi.org/10.1023/A:1009764017495}
  {\bibfield  {journal} {\bibinfo  {journal} {The Ramanujan Journal}\ }\textbf
  {\bibinfo {volume} {1}},\ \bibinfo {pages} {119} (\bibinfo {year}
  {1997})}\BibitemShut {NoStop}%
\bibitem [{\citenamefont {Gherardi}(2022)}]{marco}%
  \BibitemOpen
  \bibfield  {author} {\bibinfo {author} {\bibfnamefont {M.}~\bibnamefont
  {Gherardi}},\ }\href@noop {} {\bibfield  {journal} {\bibinfo  {journal}
  {privae communication}\ } (\bibinfo {year} {2022})}\BibitemShut {NoStop}%
\bibitem [{\citenamefont {Rotondo}\ \emph {et~al.}(2016)\citenamefont
  {Rotondo}, \citenamefont {Molinari}, \citenamefont {Ratti},\ and\
  \citenamefont {Gherardi}}]{rotondo2016devil}%
  \BibitemOpen
  \bibfield  {author} {\bibinfo {author} {\bibfnamefont {P.}~\bibnamefont
  {Rotondo}}, \bibinfo {author} {\bibfnamefont {L.~G.}\ \bibnamefont
  {Molinari}}, \bibinfo {author} {\bibfnamefont {P.}~\bibnamefont {Ratti}},\
  and\ \bibinfo {author} {\bibfnamefont {M.}~\bibnamefont {Gherardi}},\
  }\bibfield  {title} {\bibinfo {title} {Devil’s staircase phase diagram of
  the fractional quantum hall effect in the thin-torus limit},\ }\href@noop {}
  {\bibfield  {journal} {\bibinfo  {journal} {Physical Review Letters}\
  }\textbf {\bibinfo {volume} {116}},\ \bibinfo {pages} {256803} (\bibinfo
  {year} {2016})}\BibitemShut {NoStop}%
\bibitem [{\citenamefont {Di~Gioacchino}\ \emph {et~al.}(2017)\citenamefont
  {Di~Gioacchino}, \citenamefont {Gherardi}, \citenamefont {Molinari},\ and\
  \citenamefont {Rotondo}}]{di2017jack}%
  \BibitemOpen
  \bibfield  {author} {\bibinfo {author} {\bibfnamefont {A.}~\bibnamefont
  {Di~Gioacchino}}, \bibinfo {author} {\bibfnamefont {M.}~\bibnamefont
  {Gherardi}}, \bibinfo {author} {\bibfnamefont {L.~G.}\ \bibnamefont
  {Molinari}},\ and\ \bibinfo {author} {\bibfnamefont {P.}~\bibnamefont
  {Rotondo}},\ }\bibfield  {title} {\bibinfo {title} {Jack on a devil’s
  staircase},\ }in\ \href@noop {} {\emph {\bibinfo {booktitle} {Congress of the
  Department of Physics}}}\ (\bibinfo {organization} {Springer},\ \bibinfo
  {year} {2017})\ pp.\ \bibinfo {pages} {193--207}\BibitemShut {NoStop}%
\bibitem [{\citenamefont {Gorsky}\ \emph {et~al.}(2014)\citenamefont {Gorsky},
  \citenamefont {Zabrodin},\ and\ \citenamefont {Zotov}}]{gorsky2014spectrum}%
  \BibitemOpen
  \bibfield  {author} {\bibinfo {author} {\bibfnamefont {A.}~\bibnamefont
  {Gorsky}}, \bibinfo {author} {\bibfnamefont {A.}~\bibnamefont {Zabrodin}},\
  and\ \bibinfo {author} {\bibfnamefont {A.}~\bibnamefont {Zotov}},\ }\bibfield
   {title} {\bibinfo {title} {Spectrum of quantum transfer matrices via
  classical many-body systems},\ }\href@noop {} {\bibfield  {journal} {\bibinfo
   {journal} {Journal of High Energy Physics}\ }\textbf {\bibinfo {volume}
  {2014}},\ \bibinfo {pages} {1} (\bibinfo {year} {2014})}\BibitemShut
  {NoStop}%
\bibitem [{\citenamefont {Gaiotto}\ and\ \citenamefont
  {Koroteev}(2013)}]{gaiotto2013three}%
  \BibitemOpen
  \bibfield  {author} {\bibinfo {author} {\bibfnamefont {D.}~\bibnamefont
  {Gaiotto}}\ and\ \bibinfo {author} {\bibfnamefont {P.}~\bibnamefont
  {Koroteev}},\ }\bibfield  {title} {\bibinfo {title} {On three dimensional
  quiver gauge theories and integrability},\ }\href@noop {} {\bibfield
  {journal} {\bibinfo  {journal} {Journal of High Energy Physics}\ }\textbf
  {\bibinfo {volume} {2013}},\ \bibinfo {pages} {1} (\bibinfo {year}
  {2013})}\BibitemShut {NoStop}%
\bibitem [{\citenamefont {Beketov}\ \emph {et~al.}(2016)\citenamefont
  {Beketov}, \citenamefont {Liashyk}, \citenamefont {Zabrodin},\ and\
  \citenamefont {Zotov}}]{beketov2016trigonometric}%
  \BibitemOpen
  \bibfield  {author} {\bibinfo {author} {\bibfnamefont {M.}~\bibnamefont
  {Beketov}}, \bibinfo {author} {\bibfnamefont {A.}~\bibnamefont {Liashyk}},
  \bibinfo {author} {\bibfnamefont {A.}~\bibnamefont {Zabrodin}},\ and\
  \bibinfo {author} {\bibfnamefont {A.}~\bibnamefont {Zotov}},\ }\bibfield
  {title} {\bibinfo {title} {Trigonometric version of quantum--classical
  duality in integrable systems},\ }\href@noop {} {\bibfield  {journal}
  {\bibinfo  {journal} {Nuclear Physics B}\ }\textbf {\bibinfo {volume}
  {903}},\ \bibinfo {pages} {150} (\bibinfo {year} {2016})}\BibitemShut
  {NoStop}%
\bibitem [{\citenamefont {Bulycheva}\ and\ \citenamefont
  {Gorsky}(2014{\natexlab{b}})}]{bulycheva2014bps}%
  \BibitemOpen
  \bibfield  {author} {\bibinfo {author} {\bibfnamefont {K.}~\bibnamefont
  {Bulycheva}}\ and\ \bibinfo {author} {\bibfnamefont {A.}~\bibnamefont
  {Gorsky}},\ }\bibfield  {title} {\bibinfo {title} {Bps states in the
  Ω-background and torus knots},\ }\href@noop {} {\bibfield  {journal}
  {\bibinfo  {journal} {Journal of High Energy Physics}\ }\textbf {\bibinfo
  {volume} {2014}},\ \bibinfo {pages} {1} (\bibinfo {year}
  {2014}{\natexlab{b}})}\BibitemShut {NoStop}%
\bibitem [{\citenamefont {Zabrodin}\ and\ \citenamefont
  {Zotov}(2017)}]{zabrodin2017qkz}%
  \BibitemOpen
  \bibfield  {author} {\bibinfo {author} {\bibfnamefont {A.}~\bibnamefont
  {Zabrodin}}\ and\ \bibinfo {author} {\bibfnamefont {A.}~\bibnamefont
  {Zotov}},\ }\bibfield  {title} {\bibinfo {title} {Qkz--ruijsenaars
  correspondence revisited},\ }\href@noop {} {\bibfield  {journal} {\bibinfo
  {journal} {Nuclear Physics B}\ }\textbf {\bibinfo {volume} {922}},\ \bibinfo
  {pages} {113} (\bibinfo {year} {2017})}\BibitemShut {NoStop}%
\bibitem [{\citenamefont {Gorsky}\ \emph {et~al.}(2002)\citenamefont {Gorsky},
  \citenamefont {Kogan},\ and\ \citenamefont
  {Korthals-Altes}}]{gorsky2002dualities}%
  \BibitemOpen
  \bibfield  {author} {\bibinfo {author} {\bibfnamefont {A.}~\bibnamefont
  {Gorsky}}, \bibinfo {author} {\bibfnamefont {I.~I.}\ \bibnamefont {Kogan}},\
  and\ \bibinfo {author} {\bibfnamefont {C.}~\bibnamefont {Korthals-Altes}},\
  }\bibfield  {title} {\bibinfo {title} {Dualities in quantum hall system and
  noncommutative chern-simons theory},\ }\href@noop {} {\bibfield  {journal}
  {\bibinfo  {journal} {Journal of High Energy Physics}\ }\textbf {\bibinfo
  {volume} {2002}},\ \bibinfo {pages} {002} (\bibinfo {year}
  {2002})}\BibitemShut {NoStop}%
\bibitem [{\citenamefont {Gorsky}\ and\ \citenamefont
  {Nekrasov}(1995)}]{gorsky1995relativistic}%
  \BibitemOpen
  \bibfield  {author} {\bibinfo {author} {\bibfnamefont {A.}~\bibnamefont
  {Gorsky}}\ and\ \bibinfo {author} {\bibfnamefont {N.}~\bibnamefont
  {Nekrasov}},\ }\bibfield  {title} {\bibinfo {title} {Relativistic
  calogero-moser model as gauged wzw theory},\ }\href@noop {} {\bibfield
  {journal} {\bibinfo  {journal} {Nuclear Physics B}\ }\textbf {\bibinfo
  {volume} {436}},\ \bibinfo {pages} {582} (\bibinfo {year}
  {1995})}\BibitemShut {NoStop}%
\bibitem [{\citenamefont {Gorsky}\ \emph {et~al.}(2022)\citenamefont {Gorsky},
  \citenamefont {Vasilyev},\ and\ \citenamefont {Zotov}}]{gorsky2022dualities}%
  \BibitemOpen
  \bibfield  {author} {\bibinfo {author} {\bibfnamefont {A.}~\bibnamefont
  {Gorsky}}, \bibinfo {author} {\bibfnamefont {M.}~\bibnamefont {Vasilyev}},\
  and\ \bibinfo {author} {\bibfnamefont {A.}~\bibnamefont {Zotov}},\ }\bibfield
   {title} {\bibinfo {title} {Dualities in quantum integrable many-body systems
  and integrable probabilities. part i},\ }\href@noop {} {\bibfield  {journal}
  {\bibinfo  {journal} {Journal of High Energy Physics}\ }\textbf {\bibinfo
  {volume} {2022}},\ \bibinfo {pages} {1} (\bibinfo {year} {2022})}\BibitemShut
  {NoStop}%
\bibitem [{\citenamefont {Susskind}(2001)}]{susskind2001quantum}%
  \BibitemOpen
  \bibfield  {author} {\bibinfo {author} {\bibfnamefont {L.}~\bibnamefont
  {Susskind}},\ }\bibfield  {title} {\bibinfo {title} {The quantum hall fluid
  and non-commutative chern simons theory},\ }\href@noop {} {\bibfield
  {journal} {\bibinfo  {journal} {arXiv preprint hep-th/0101029}\ } (\bibinfo
  {year} {2001})}\BibitemShut {NoStop}%
\bibitem [{\citenamefont {Polychronakos}(2001)}]{polychronakos2001quantum}%
  \BibitemOpen
  \bibfield  {author} {\bibinfo {author} {\bibfnamefont {A.~P.}\ \bibnamefont
  {Polychronakos}},\ }\bibfield  {title} {\bibinfo {title} {Quantum hall states
  as matrix chern-simons theory},\ }\href@noop {} {\bibfield  {journal}
  {\bibinfo  {journal} {Journal of High Energy Physics}\ }\textbf {\bibinfo
  {volume} {2001}},\ \bibinfo {pages} {011} (\bibinfo {year}
  {2001})}\BibitemShut {NoStop}%
\bibitem [{\citenamefont {Nekrasov}(1997)}]{nekrasov1997duality}%
  \BibitemOpen
  \bibfield  {author} {\bibinfo {author} {\bibfnamefont {N.}~\bibnamefont
  {Nekrasov}},\ }\bibfield  {title} {\bibinfo {title} {On a duality in
  calogero-moser-sutherland systems},\ }\href@noop {} {\bibfield  {journal}
  {\bibinfo  {journal} {arXiv preprint hep-th/9707111}\ } (\bibinfo {year}
  {1997})}\BibitemShut {NoStop}%
\bibitem [{\citenamefont {Popkov}\ \emph {et~al.}(2015)\citenamefont {Popkov},
  \citenamefont {Schadschneider}, \citenamefont {Schmidt},\ and\ \citenamefont
  {Sch{\"u}tz}}]{popkov2015fibonacci}%
  \BibitemOpen
  \bibfield  {author} {\bibinfo {author} {\bibfnamefont {V.}~\bibnamefont
  {Popkov}}, \bibinfo {author} {\bibfnamefont {A.}~\bibnamefont
  {Schadschneider}}, \bibinfo {author} {\bibfnamefont {J.}~\bibnamefont
  {Schmidt}},\ and\ \bibinfo {author} {\bibfnamefont {G.~M.}\ \bibnamefont
  {Sch{\"u}tz}},\ }\bibfield  {title} {\bibinfo {title} {Fibonacci family of
  dynamical universality classes},\ }\href@noop {} {\bibfield  {journal}
  {\bibinfo  {journal} {Proceedings of the National Academy of Sciences}\
  }\textbf {\bibinfo {volume} {112}},\ \bibinfo {pages} {12645} (\bibinfo
  {year} {2015})}\BibitemShut {NoStop}%
\bibitem [{\citenamefont {Ilievski}\ \emph {et~al.}(2021)\citenamefont
  {Ilievski}, \citenamefont {De~Nardis}, \citenamefont {Gopalakrishnan},
  \citenamefont {Vasseur},\ and\ \citenamefont
  {Ware}}]{ilievski2021superuniversality}%
  \BibitemOpen
  \bibfield  {author} {\bibinfo {author} {\bibfnamefont {E.}~\bibnamefont
  {Ilievski}}, \bibinfo {author} {\bibfnamefont {J.}~\bibnamefont {De~Nardis}},
  \bibinfo {author} {\bibfnamefont {S.}~\bibnamefont {Gopalakrishnan}},
  \bibinfo {author} {\bibfnamefont {R.}~\bibnamefont {Vasseur}},\ and\ \bibinfo
  {author} {\bibfnamefont {B.}~\bibnamefont {Ware}},\ }\bibfield  {title}
  {\bibinfo {title} {Superuniversality of superdiffusion},\ }\href@noop {}
  {\bibfield  {journal} {\bibinfo  {journal} {Physical Review X}\ }\textbf
  {\bibinfo {volume} {11}},\ \bibinfo {pages} {031023} (\bibinfo {year}
  {2021})}\BibitemShut {NoStop}%
\bibitem [{\citenamefont {Ljubotina}\ \emph {et~al.}(2017)\citenamefont
  {Ljubotina}, \citenamefont {{\v{Z}}nidari{\v{c}}},\ and\ \citenamefont
  {Prosen}}]{ljubotina2017spin}%
  \BibitemOpen
  \bibfield  {author} {\bibinfo {author} {\bibfnamefont {M.}~\bibnamefont
  {Ljubotina}}, \bibinfo {author} {\bibfnamefont {M.}~\bibnamefont
  {{\v{Z}}nidari{\v{c}}}},\ and\ \bibinfo {author} {\bibfnamefont
  {T.}~\bibnamefont {Prosen}},\ }\bibfield  {title} {\bibinfo {title} {Spin
  diffusion from an inhomogeneous quench in an integrable system},\ }\href@noop
  {} {\bibfield  {journal} {\bibinfo  {journal} {Nature communications}\
  }\textbf {\bibinfo {volume} {8}},\ \bibinfo {pages} {16117} (\bibinfo {year}
  {2017})}\BibitemShut {NoStop}%
\bibitem [{\citenamefont {Chernodub}(2022)}]{chernodub2022fractal}%
  \BibitemOpen
  \bibfield  {author} {\bibinfo {author} {\bibfnamefont {M.~N.}\ \bibnamefont
  {Chernodub}},\ }\bibfield  {title} {\bibinfo {title} {Fractal thermodynamics
  and ninionic statistics of coherent rotational states: realization via
  imaginary angular rotation in imaginary time formalism},\ }\href@noop {}
  {\bibfield  {journal} {\bibinfo  {journal} {arXiv preprint arXiv:2210.05651}\
  } (\bibinfo {year} {2022})}\BibitemShut {NoStop}%
\bibitem [{\citenamefont {Ambrus}\ and\ \citenamefont
  {Chernodub}(2023)}]{ambrus2023rigidlyrotating}%
  \BibitemOpen
  \bibfield  {author} {\bibinfo {author} {\bibfnamefont {V.~E.}\ \bibnamefont
  {Ambrus}}\ and\ \bibinfo {author} {\bibfnamefont {M.~N.}\ \bibnamefont
  {Chernodub}},\ }\href@noop {} {\bibinfo {title} {Rigidly-rotating scalar
  fields: between real divergence and imaginary fractalization}} (\bibinfo
  {year} {2023}),\ \Eprint {https://arxiv.org/abs/2304.05998} {arXiv:2304.05998
  [hep-th]} \BibitemShut {NoStop}%
\bibitem [{\citenamefont {Harvey}\ and\ \citenamefont
  {Moore}(1996)}]{harvey1996algebras}%
  \BibitemOpen
  \bibfield  {author} {\bibinfo {author} {\bibfnamefont {J.~A.}\ \bibnamefont
  {Harvey}}\ and\ \bibinfo {author} {\bibfnamefont {G.}~\bibnamefont {Moore}},\
  }\bibfield  {title} {\bibinfo {title} {Algebras, bps states, and strings},\
  }\href@noop {} {\bibfield  {journal} {\bibinfo  {journal} {Nuclear Physics
  B}\ }\textbf {\bibinfo {volume} {463}},\ \bibinfo {pages} {315} (\bibinfo
  {year} {1996})}\BibitemShut {NoStop}%
\bibitem [{\citenamefont {Feingold}(1980)}]{feingold1980hyperbolic}%
  \BibitemOpen
  \bibfield  {author} {\bibinfo {author} {\bibfnamefont {A.~J.}\ \bibnamefont
  {Feingold}},\ }\bibfield  {title} {\bibinfo {title} {A hyperbolic gcm lie
  algebra and the fibonacci numbers},\ }\href@noop {} {\bibfield  {journal}
  {\bibinfo  {journal} {Proceedings of the American Mathematical Society}\
  }\textbf {\bibinfo {volume} {80}},\ \bibinfo {pages} {379} (\bibinfo {year}
  {1980})}\BibitemShut {NoStop}%
\bibitem [{\citenamefont {Lechtenfeld}\ and\ \citenamefont
  {Zagier}(2022)}]{lechtenfeld2022hyperbolic}%
  \BibitemOpen
  \bibfield  {author} {\bibinfo {author} {\bibfnamefont {O.}~\bibnamefont
  {Lechtenfeld}}\ and\ \bibinfo {author} {\bibfnamefont {D.}~\bibnamefont
  {Zagier}},\ }\bibfield  {title} {\bibinfo {title} {A hyperbolic kac-moody
  calogero model},\ }\href@noop {} {\bibfield  {journal} {\bibinfo  {journal}
  {arXiv preprint arXiv:2203.06519}\ } (\bibinfo {year} {2022})}\BibitemShut
  {NoStop}%
\bibitem [{\citenamefont {Aniceto}\ \emph {et~al.}(2019)\citenamefont
  {Aniceto}, \citenamefont {Ba{\c{s}}ar},\ and\ \citenamefont
  {Schiappa}}]{aniceto2019primer}%
  \BibitemOpen
  \bibfield  {author} {\bibinfo {author} {\bibfnamefont {I.}~\bibnamefont
  {Aniceto}}, \bibinfo {author} {\bibfnamefont {G.}~\bibnamefont
  {Ba{\c{s}}ar}},\ and\ \bibinfo {author} {\bibfnamefont {R.}~\bibnamefont
  {Schiappa}},\ }\bibfield  {title} {\bibinfo {title} {A primer on resurgent
  transseries and their asymptotics},\ }\href@noop {} {\bibfield  {journal}
  {\bibinfo  {journal} {Physics Reports}\ }\textbf {\bibinfo {volume} {809}},\
  \bibinfo {pages} {1} (\bibinfo {year} {2019})}\BibitemShut {NoStop}%
\end{thebibliography}%

\end{document}